\documentclass[times,twocolumn,final]{elsarticle}

\usepackage{medima}
\usepackage{framed,multirow}

\usepackage{amssymb}
\usepackage{latexsym}

\usepackage{url}
\usepackage[dvipsnames]{xcolor}

\usepackage[colorlinks]{hyperref}

\usepackage{bbding}
\usepackage{soul}
\usepackage{bibunits}
\usepackage{bm}
\usepackage{natbib}
\usepackage{array}
\usepackage{subfigure}
\usepackage{pdflscape}
\usepackage{makecell}
\usepackage{framed,multirow}
\usepackage{threeparttable}
\usepackage{amssymb}
\usepackage{pifont}
\usepackage{algorithm}
\usepackage{layouts}
\usepackage{listings}
\usepackage{colortbl}
\usepackage{booktabs}
\usepackage{amsmath}
\usepackage{pythonhighlight}
\makeatletter
\newcommand\footnoteref[1]{\protected@xdef\@thefnmark{\ref{#1}}\@footnotemark}
\makeatother

\newcolumntype{P}[1]{>{\centering\arraybackslash}p{#1}}
\newlength\savewidth
\newcommand{\eg}{\mbox{e.g.,\ }}
\newcommand{\etal}{\mbox{et al.}}
\newcommand{\ie}{\mbox{i.e.,\ }}

\def\arrvline{\hfil\kern\arraycolsep\vline\kern-\arraycolsep\hfilneg}

\definecolor{Highlight}{HTML}{39b54a}  

\definecolor{iblue}{rgb}{0.06, 0.75, 1.0}
\definecolor{igray}{rgb}{0.00, 0.00, 0.00}

\newcommand{\ourmodel}{{\fontfamily{ppl}\selectfont Universal Model}}

\newcommand{\major}{\textcolor[RGB]{0, 0, 0}}
\newcommand{\significant}{\textcolor[RGB]{239,134,131}}
\definecolor{significant}{RGB}{239,134,131}

\definecolor{newcolor}{rgb}{.8,.349,.1}

\journal{Medical Image Analysis}

\begin{document}
	
	\verso{Jie Liu \textit{et~al.}}
	
	\begin{frontmatter}
 
		\title{Universal and Extensible Language-Vision Models for Organ Segmentation and Tumor Detection from Abdominal Computed Tomography}
		
		\author[1]{Jie \snm{Liu}}
		\author[2]{Yixiao \snm{Zhang}}
		\author[3]{Kang \snm{Wang}}
		\author[3]{Mehmet Can \snm{Yavuz}}
		\author[4]{Xiaoxi \snm{Chen}}
		\author[5]{Yixuan \snm{Yuan}}
		\author[1]{Haoliang \snm{Li}}
		\author[3]{Yang \snm{Yang}}
		\author[2]{Alan \snm{Yuille}}
		\author[6]{Yucheng \snm{Tang}\corref{cor1}}
		\author[2]{Zongwei \snm{Zhou}\corref{cor1}}
		\cortext[cor1]{Corresponding authors: Yucheng Tang (\href{mailto:yuchengt@nvidia.com}{\textsc{yuchengt@nvidia.com}}) and Zongwei Zhou (\href{mailto:zzhou82@jh.edu}{\textsc{zzhou82@jh.edu}})}
		
		\address[1]{City University of Hong Kong}
		\address[2]{Johns Hopkins University}
		\address[3]{University of California, San Francisco}
		\address[4]{University of Illinois Urbana-Champaign}
		\address[5]{Chinese University of Hong Kong}
		\address[6]{NVIDIA}
		
		\received{1 May 2013}
		\finalform{10 May 2013}
		\accepted{13 May 2013}
		\availableonline{15 May 2013}
		\communicated{S. Sarkar}
		
		\begin{abstract}
			The advancement of artificial intelligence (AI) for organ segmentation and tumor detection is propelled by the growing availability of computed tomography (CT) datasets with detailed, per-voxel annotations. However, these AI models often struggle with \ul{flexibility} for partially annotated datasets and \ul{extensibility} for new classes due to limitations in the one-hot encoding, architectural design, and learning scheme. To overcome these limitations, we propose a universal, extensible framework enabling a single model, termed \ourmodel, to deal with multiple public datasets and adapt to new classes (e.g., organs/tumors). Firstly, we introduce a novel language-driven parameter generator that leverages language embeddings from large language models, enriching semantic encoding compared with one-hot encoding. Secondly, the conventional output layers are replaced with lightweight, class-specific heads, allowing \ourmodel\ to simultaneously segment 25 organs and six types of tumors and ease the addition of new classes. We train our \ourmodel\ on 3,410 CT volumes assembled from 14 publicly available datasets \major{and then test it on 6,173 CT volumes from four external datasets.} \ourmodel\ achieves \major{first place on six CT tasks in the Medical Segmentation Decathlon (MSD) public leaderboard} and leading performance on the Beyond The Cranial Vault (BTCV) dataset. In summary, \ourmodel\ exhibits remarkable computational efficiency (6$\times$ faster than other dataset-specific models), demonstrates strong generalization across different hospitals, transfers well to numerous downstream tasks, and more importantly, facilitates the extensibility to new classes while alleviating the catastrophic forgetting of previously learned classes. Codes, models, and datasets are available at \href{https://github.com/ljwztc/CLIP-Driven-Universal-Model}{https://github.com/ljwztc/CLIP-Driven-Universal-Model}
		\end{abstract}
		
		
	\end{frontmatter}
	
	\section{Introduction}\label{sec:introduction}
	
	Computer tomography (CT) is a widely used and powerful tool for disease diagnosis and treatment planning \citep{mattikalli2022universal,zhou2022interpreting,qu2023annotating,chen2024learning}. In routine clinical workflow, radiologists analyze hundreds of 2D slices in a single CT volume to find and interpret diagnostic information, which can be tedious and prone to misdiagnosis \citep{zhou2021towards}. Medical image segmentation offers a promising solution, improving diagnostic efficiency and quality by automatically identifying organs, delineating their boundaries, and highlighting abnormalities \citep{liu2021graph,hu2023label,chen2023cancerunit,chen2024towards,lai2024pixel}. 
	
	The progress of medical image segmentation relies heavily on specialized datasets. These include organ/tumor-specific datasets, e.g., LiTS (Liver Tumor Segmentation) \citep{bilic2019liver}, KiTS (Kidney Tumor Segmentation) \citep{heller2019kits19}, and MSD (Medical Segmentation Decathlon) \citep{simpson2019large}, to abdominal multi-organ annotation datasets like BTCV (Beyond The Cranial Vault) \citep{landman2015miccai}, AMOS (Abdominal Multi-organ Segmentation) \citep{ji2022amos}, and AbdomenAtlas \citep{li2024well}. Additionally, \citet{wasserthal2023totalsegmentator} have proposed to provide a holistic anatomical view of the whole body instead of the specific body region, which aims to capture the comprehensive anatomical structure of the human body.
	
	Considering the highly complex nature of anatomical structures in the human body and the need for fine-grained clinical requirements, we can expect the emergence of new organ/tumor annotations \citep{jaus2023towards}, such as vermiform appendix and spleen tumors. Moreover, there is a growing demand for more detailed anatomy annotations, such as distinguishing between the right lobe liver and left lobe liver \citep{germain2014liver}. While liver annotation is common in current datasets, the partition of liver lobes remains under-investigated \citep{bilic2023liver}, leading to potential overlapping annotations between liver lobes and overall liver in the future. The latest endeavors to meet these emerging requirements involve re-annotating existing datasets with human-in-the-loop and retraining models accordingly \citep{qu2023annotating,wasserthal2023totalsegmentator,jaus2023towards}. However, this approach incurs significant annotation costs, particularly for 3D medical imaging, as well as substantial computational consumption for retraining models from scratch \citep{zhang2023continual,zhang2024leveraging}. Therefore, it is important to explore a new framework that can effectively handle new organ/tumor annotations while mitigating the computational burden associated with retraining models.
	
	Several challenges must be addressed when designing such a continual multi-organ segmentation and tumor detection framework. The first challenge involves the encoding orthogonality problem. Existing methods commonly employ one-hot encoding, ignoring the semantic relationship between organs and tumors. For instance, when utilizing one-hot encoding for liver [1,0,0], pancreas [0,1,0], and pancreas tumor [0,0,1], there is no semantic distinction among these classes. 
	A possible solution is the few-hot encoding~\citep{shi2021marginal}, where the liver, pancreas, and pancreas tumor can be encoded as [1,0,0], [0,1,0], and [0,1,1], respectively. Although few-hot labels could indicate that pancreatic tumors are part of the pancreas, the relationship between organs remains orthogonal. In addition, these encoding approaches become challenging to extend when adding new organs into the current framework, as the entire encoding scheme needs to be modified to accommodate the new classes. 
	
	The second challenge lies in the label taxonomy inconsistency, where publicly available datasets usually contain annotations of different classes. This disparity in label spaces restricts the traditional segmentation head, which employs a pre-defined kernel number and Softmax activation function, to produce single-class predictions for each voxel based on fixed classes \citep{liu2022universal,ma2021abdomenct,schoppe2020deep}. Consequently, it lacks the flexibility to accommodate new organ/tumor annotations and handle detailed anatomy annotations. For example, ``Right Lobe Liver'' is part of the ``Liver'' and ``Kidney Tumor'' represents a sub-volume of the ``Kidney'', which means that certain voxels do not exclusively belong to a single specific class as per clinical requirements. One trivial solution is to replace the segmentation head and retrain from scratch, but this would forget the learned knowledge and cause computational consumption. This motivates us to design a novel framework for segmentation heads that allows for flexible prediction and extensibility to new classes.
	
	The third challenge pertains to the design of a dynamic network architecture for continual learning. When it comes to annotating new organs/tumors, the medical segmentation model should be able to extend its segmentation capabilities without accessing previous data, in order to comply with medical privacy regulations. Existing works typically freeze the encoder and decoder, and add an additional decoder when learning new classes \citep{ji2023continual}, leading to tremendous memory costs for network parameters. Furthermore, the newly added decoder lacks awareness of the relationship between the newly added organs/tumors and existing classes, resulting in limited utilization of existing knowledge. Hence, our objective is to design a network architecture that minimizes the addition of parameters during different continual learning steps and ensures awareness of the relationship between newly added organs/tumors and existing classes.
	
	To overcome these challenges, we propose a universal and extensible framework that offers a promising solution to multi-organ segmentation and tumor detection. Firstly, we propose the language-driven parameter generator (LPG) to employ the language embeddings, learned from large language models, as a novel semantic encoding method. This enables the capture of semantic relationships between organs and tumors, facilitating the learning of structured features. Secondly, by replacing the conventional output layer with a lightweight class-specific segment head (CSH) integrated with language embeddings, our model achieves precise segmentation of 25 organs and 6 tumor types. The framework design allows for efficient extensibility, enabling the addition of new classes without catastrophic forgetting. To evaluate the effectiveness of our proposed framework, we conducted extensive experiments on a comprehensive dataset assembly. The training set comprised 3,410 CT volumes from 14 different datasets, and the model was evaluated on 6,162 external CT volumes from three additional datasets. Our framework achieved \major{first place on six CT tasks in the Medical Segmentation Decathlon (MSD) public leaderboard} and demonstrated state-of-the-art results on the Beyond The Cranial Vault (BTCV) dataset. Furthermore, the proposed \ourmodel\ exhibited strong generalization to CT volumes from different centers, superior transfer learning capabilities across various downstream segmentation tasks, and outstanding continual learning capacity on two extension benchmarks. It also demonstrated computational efficiency, being six times faster than dataset-specific models.

	\section{Related Work}
	\label{sec:related_work}
	\subsection{Organ Segmentation and Tumor Detection}
	Organ segmentation and tumor detection, which are essential components of radiotherapy, offer quantitative insights for diagnosing and categorizing diseases. Currently, deep learning techniques have gained considerable popularity in this domain due to their ability to extract data-driven features and facilitate end-to-end training. Among these techniques, U-Net~\citep{ronneberger2015u} and its variations~\citep{zhou2019unet++,liang2021incorporating,oktay2018attention,isensee2021nnu} have emerged as prominent approaches, exhibiting promising outcomes.
	In addition to these convolutional neural networks (CNN)-based methods, generative adversarial network (GAN)-based approaches were also proposed to achieve high accuracy in segmentation tasks by employing adversarial losses \citep{cai2019end,gao2021focusnetv2,mahmood2019deep}. Nevertheless, training GAN networks is challenging and time-consuming since the generator must attain Nash equilibrium with the discriminator.
	Recently, transformer-based models \citep{zhou2021nnformer,hatamizadeh2022unetr,tang2022self,chen2021transunet,he2023swinunetr} use attention mechanism to learn the long-range relationship with stronger modeling capacities. Transformer-based approaches can capture long-range dependencies and achieve better performance than CNNs in many tasks.
	
	These works often simplify multi-organ segmentation and tumor detection, focusing on specific organs and tumors \citep{chen2023cancerunit,ronneberger2015u,zhou2019unet++,isensee2021nnu,liang2021incorporating,li2023early}, which have been verified to be effective for segmenting the organs and localizing the tumors. However, these works ignore the correlation between organs and tumors, specifically the prior anatomical information. Unlike these works, \ourmodel\ advances a single framework that utilizes the large language model embedding to capture the semantic relationship between organs and tumors, addressing both tasks with extensibility. Moreover, we demonstrate our work on publicly available datasets, which is beneficial to reproducibility.
	
	\subsection{Large Language Vision Model}
	With the widespread success of large models in the field of language processing~\citep{brown2020language,devlin2018bert}, there have been significant advancements in the application of language image models to visual recognition problems \citep{luddecke2022image,rao2022denseclip,wang2022cris,park2022per,xie2022clims}. Some language image models adopt BERT-like architecture \citep{yan2022clinical,conneau2019cross} or adopt contrastive learning paradigm \citep{radford2021learning,wang2022medclip} to deal with vision and language data. While these work focus on pre-trained vision language models, another line of works focus on integrating the pre-trained language model into visual recognition problems. These models are capable of image captioning \citep{hu2022scaling}, visual question answering \citep{eslami2023pubmedclip}, pathology image analysis \citep{huang2023visual}, and so on. \citet{qin2022medical} suggested that a language model could be used for detection tasks in the medical domain with carefully designed medical prompts. 
	Grounded in these findings, we are among the first to introduce large language embedding to voxel-level semantic understanding medical tasks, i.e., segmentation, in which we underline the importance of the semantic relationship between anatomical structures. 
	
	\subsection{Incremental Learning}
	Incremental learning aims to acquire knowledge about new classes in multiple stages without experiencing catastrophic forgetting \citep{lewandowsky1995catastrophic}, even when previous data is not accessible due to medical privacy regulations. This ability to extend a model to new classes without relying on previous data is crucial for organ segmentation and tumor detection.
	In the medical field, there are only a few existing methods for continual segmentation. For instance, \citet{ozdemir2018learn,ozdemir2019extending} extended the distillation loss to address medical image segmentation. Another study by~\citet{liu2022learning} introduced a memory module to store prototypical representations of different organ categories. However, these methods rely on a limited number of representative exemplars, which may not be practically available. In a recent work by~\citet{ji2023continual}, the focus lies on network extension to tackle the forgetting problem. This is achieved by freezing the encoder and decoder and adding additional decoders when learning new classes. Although these approaches have successfully mitigated the forgetting issue, they incur significant memory costs in terms of network parameters. In contrast, \ourmodel\ proposes a novel continual multi-organ and tumor segmentation method that overcomes the forgetting problem while imposing minimal memory and computation overhead.

	\subsection{Learning with Integrated Datasets}
	\label{sec:review_partial}
	Due to financial constraints and clinical limitations, publicly available datasets for abdominal imaging primarily focus on a limited number of organs and tumors~\citep{landman2017multiatlas,ma2021abdomenct,luo2021word,ji2022amos}.
	For example, the AbdomenCT-1K dataset is designed for segmenting four organs~\citep{ma2021abdomenct}, the WORD dataset for segmenting 16 organs~\citep{luo2021word}, and the TotalSegmentor dataset for segmenting 104 anatomical structures~\citep{wasserthal2022totalsegmentator}. To enhance representation capabilities, some studies have proposed simultaneous integration of multiple datasets \citep{zhou2019prior,fang2020multi,zhang2021dodnet,zhang2022merging}. However, integrating data can introduce challenges due to inconsistent label taxonomy, leading to the issue of partial labeling. \major{Several approaches have been explored to leverage partial labels in organ segmentation and tumor detection~\citep{liu2022universal,chen2021deep,bai2022end,zlocha2019universal,yan2019mulan,liu2022improving,naga2022universal,yan2020universal,mattikalli2022universal,xie2023learning,liu2024cosst,wu2022tgnet}. For example, DoDNet~\citep{zhang2021dodnet} and TransDoDNet~\citep{xie2023learning} used one-hot embedding to generate task-specific output. COSST~\citep{liu2024cosst} utilized ground truth and pseudo-labels with self-training. TGNet~\citep{wu2022tgnet}) proposed task-guided attention modules and residual blocks to extract task-related features. Nonetheless, these studies have certain limitations.
		Firstly, the integrated datasets in these studies were relatively small-scale\footnote{{\citet{zhou2019prior} assembled 150 CT volumes from 4 datasets; \citet{fang2020multi} assembled 548 CT volumes from 4 datasets; \citet{zhang2021dodnet,xie2023learning} assembled 1,155 CT volumes from 7 datasets.}}, and the potential benefits of dataset integration were not convincingly demonstrated. Their performance was comparable to that of models trained on specific datasets, and lacked evaluation on official benchmarks.
		Secondly, the used one-hot embedding resulted in the disregard of the semantic relationship between organs and tumors.}
	
	\subsection{Our Previous Work}
	\label{sec:our_previous_work}
	
	We first presented CLIP-Driven Universal Model in our ICCV 2023 paper~\citep{liu2023clip} and our MICCAI 2023 paper~\citep{zhang2023continual}, capable of segmenting 25 organs and seven types of tumors within CT volumes. 
	\ourmodel\ has since been quickly adopted by the research community, either as a strong baseline for comparison~\citep{jiang2023zept,chen2023versatile,gao2023training,ye2023uniseg,ulrich2023multitalent}, or as a source of inspiration for developing newer unified architectures~\citep{liu20233d,ye2023continual,zhang2023spatially,zeng2023segment,silva2023towards}; it has also been integrated into multiple open-source software, such as MONAI platform\footnote{MONAI: \href{https://monai.io}{https://monai.io}} at NVIDIA and Chimera/ChimeraX Group\footnote{UCSF ChimeraX: \href{https://www.cgl.ucsf.edu/chimerax/}{https://www.cgl.ucsf.edu/chimerax/}} at UCSF. To further strengthen \ourmodel\ on our own, this paper presents \textbf{six extensions} to our previous work:
		\begin{enumerate}
			\item We design a new language-driven parameter generator with an independent multi-layer perceptron to avoid the entangling among various classes (\S\ref{sec:lpg}).
			\item We investigate the generalization capabilities of \ourmodel\ on four external datasets without additional fine-tuning or adaptation (\figureautorefname~\ref{fig:fig_generalization} (a)).
			\item We benchmark various partially labeled methods utilizing multiple datasets to evaluate the effectiveness of our model on the BTCV and MOTS datasets (Tables~\ref{tab:btcv_benchmark} and \ref{tab:mots}).
			\item We further conduct case studies on challenging tumors, comparing the performance of AI with human experts (\figureautorefname~\ref{fig:pseudo_truth_evaluation}).
			\item We perform inference in a clinical study setting to evaluate the model's capability in detecting small tumors that may have been overlooked during radiological evaluation (\S\ref{sec:generalizability}).
			\item We analyze the model's transferability in segmenting sub-classes of pancreatic tumors, which represents a more challenging task requiring fine-grained discrimination (\tableautorefname~\ref{tab:transfer_learning}).
	\end{enumerate}
 
	\begin{figure*}[t]
		\centerline{\includegraphics[width=1.0\linewidth]{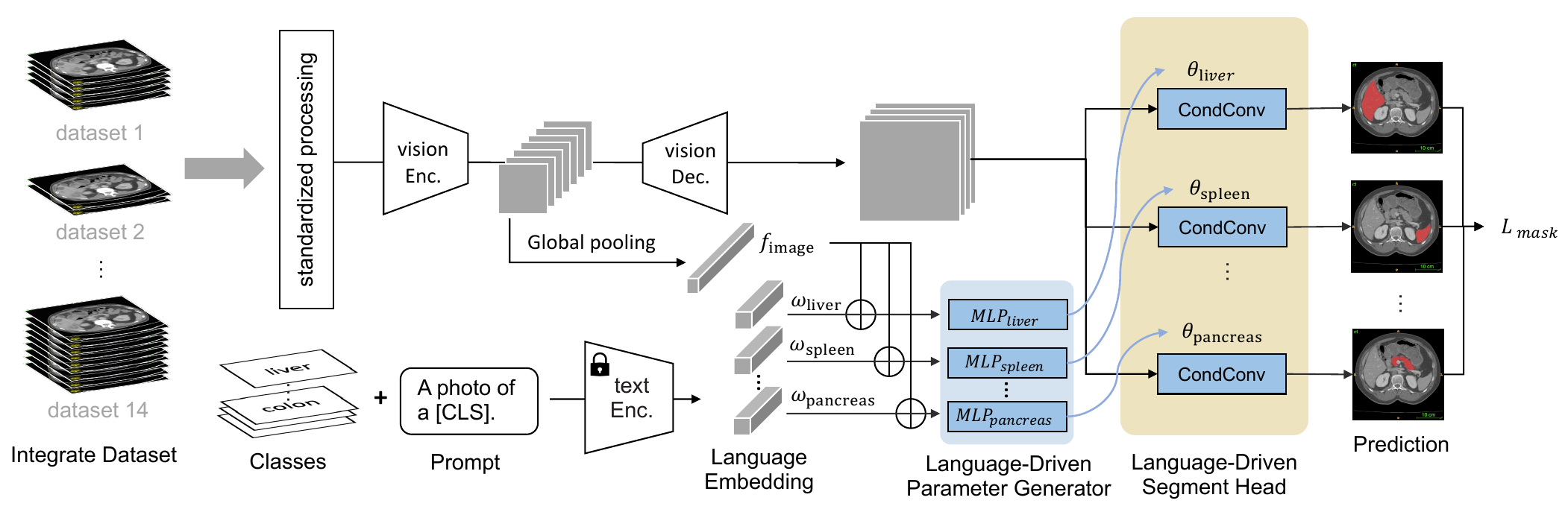}}
		\caption{
			\textbf{\textit{Overview.}} We have developed the continual CLIP-Driven Universal Model from an assembly of 14 public datasets of 3,410 CT volumes. In total, 25 organs and 6 types of tumors are partially labeled. To deal with partial labels, \ourmodel\ consists of a language branch and a vision branch (\S\ref{sec:universal_model}). The official test set of MSD and BTCV are used to benchmark the performance of organ segmentation (\S\ref{sec:strong_challenge_ranking}) and tumor detection (\S\ref{sec:high_specificity}). 3D-IRCADb, TotalSegmentator and a large-scale private dataset, consisting of 5,038 CT volumes with 21 annotated organs, are used for independent, external validation of model generalizability (\S\ref{sec:generalizability}) and transferability (\S\ref{sec:transfer_learning}). \major{The LPG module utilizes separate MLPs for each organ to overcome the entanglement issue present in the ICCV version~\citep{liu2023clip}, which relied on a single MLP.}
		}
		\label{fig:method}
	\end{figure*}
	
	\section{Methodology}
	\subsection{Problem Definition}
	Let us consider a set of partially labeled datasets $\{\mathcal{D}_1, \mathcal{D}_2, ..., \mathcal{D}_N\}$ with corresponding organs/tumors label space $\{\mathcal{L}_1, \mathcal{L}_2, ..., \mathcal{L}_N\}$, we learn a unified multi-organ segmentation model to segment tumors/organs in $\mathcal{L} = \bigcup^N_i \mathcal{L}_i$, where $\Vert \mathcal{L} \Vert \geq \Vert \mathcal{L}_i \Vert$ for all $i \in \{1,2,...,N\}$. When it comes a new dataset $\mathcal{D}_{N+1}$ with organs/tumors label space $\mathcal{L}_{N+1}$, $\mathcal{L}_{N+1} \setminus \mathcal{L} \neq \varnothing$, all previous training data are not accessible. The model is required to predict the accumulated label space $\mathcal{L} \bigcup \mathcal{L}_{N+1}$ for all seen datasets $\{\mathcal{D}_1, \mathcal{D}_2, ..., \mathcal{D}_N, \mathcal{D}_{N+1}\}$. 
 
	\subsection{CLIP-Driven Universal Model}
	\label{sec:universal_model}
	As illustrate in \figureautorefname~\ref{fig:method}, we present the framework of CLIP-Driven Universal Model for multi-organ segmentation and tumor detection with an integrated dataset, which consists of two main branches: the language branch and the vision branch. Given a 3D CT volume $\mathbf{X} \in\mathcal{D}_n$ , along with a partial label $\mathbf{Y}$, the language branch firstly generates the language embedding for each organ/tumor $\text{cls} \in \mathcal{L}_n$. To ensure that each parameter is aware of the anatomical relationship among classes, the language-driven parameter generator (LPG) consists of independent multi-layer perceptron (MLP) that take language embedding as input and generate a parameter $\theta$ for each class. On the other hand, the vision branch focuses on extracting the feature embedding $\bm{F}_E$ for the CT volume $\mathbf{X}$. Subsequently, the class-specific segment head (CSH) takes the generated parameter $\theta$ as parameter and $\bm{F}_D$ as input to predict the binary mask for each organ/tumor. During the continual learning process, the LPG module adopts another MLP to generate the new parameter and predict the new organ/tumor mask, minimizing the potential disruption caused by the introduction of new classes on the old classes.

	\subsubsection{Backbone Network}
	For each 3D CT volume, isotropic spacing and uniformed intensity scale are utilized to reduce the domain gap among various datasets\footnote{The variations in acquisition parameters, reconstruction kernels, contrast enhancements,  intensity variation and so on would lead to potential domain gap for different datasets \citep{orbes2019multi,yan2020mri,guo2021multi}.} The standardized and normalized CT volumes are subjected to the vision encoder and vision decoder. The choice of backbone architecture for this process is flexible and depends on the specific requirements of the task. Transformer-based architectures such as Swin UNETR or CNN-based architectures like U-Net are viable options, as long as they can effectively learn general representations for organ segmentation and tumor detection. The vision encoder extracts image features denoted as $\bm{F}_E$, then the vision decoder further refines the image features as $\bm{F}_D$. These features capture relevant information about the anatomical structures and potential abnormalities present in the volumes.
	
	\subsubsection{Language-driven Parameter Generator}\label{sec:lpg}
	To capture the semantic relations among organs/tumors, \ourmodel\ utilizes the embedding of large language model, \eg text encoder in CLIP \citep{radford2021learning}\footnote{CLIP (Contrastive Language–Image Pre-training), which has been trained on a vast dataset consisting of 400 million image-text pairs, including medical images and text \citep{chambon2022adapting}, exploits the semantic relationship between images and language.} and BioBERT \citep{yasunaga2022linkbert}, to encode different classes since they can learn the semantic relationships present in abundant language data. To generate the language embedding $\bm{w}_{cls}$ for each class $\text{cls} \in \mathcal{L}_n$, the medical prompt is utilized, \eg ``a computerized tomography of a [CLS]'', where [CLS] is the organ/tumor name of class $cls$. Subsequently, a global average pooling (GAP) operation is applied to the encoded image feature $\bm{F}_E$ to extract image-specific information, resulting in a global feature $\bm{f} = \text{GAP}(\bm{F}_E)$.
	The obtained global feature $\bm{f}$ and the language embedding are concatenated and fed into an independent multi-layer perceptron (MLP) to generate the corresponding parameter $\bm{\theta}_{cls}$,
	\begin{align}
		\boldsymbol{\theta}_{cls}= \text{MLP}_{cls}(\bm{w}_{cls} \oplus \bm{f}),
		\label{equ:lpg}
	\end{align}
	where $\oplus$ denotes the concatenation. In this way, \ourmodel\ exploits the semantic relationships among classes to address the label orthogonality problem, enhancing its ability to perform organ segmentation and tumor detection tasks.

	\subsubsection{Class-specific Segment Head}
	To address the label taxonomy inconsistency and enable efficient dynamic learning for different classes, we advance the Class-specific Segmentation Head for \ourmodel\ framework. To achieve this, we employ conditional convolution, which involves three sequential convolutional layers with $1\times1\times1$ kernels and a non-linear function. The purpose of this approach is to generate a binary mask for each class. The first two layers consist of 8 channels, while the last layer has 1 channel.
	To initiate the process, we divide the parameter $\bm{\theta}_{cls}$ into three parts, namely $\bm{\theta}_{{cls}_1}$, $\bm{\theta}_{{cls}_2}$, and $\bm{\theta}_{{cls}_3}$. These parts are then assigned as parameters for the conditional convolution kernel. Next, we input the decoded feature $\bm{F}_D$ into the conditional convolution to obtain the binary mask,
	\begin{align}
		\bm{P}_{cls}=\text{Sigmoid} ( \phi( \phi(\bm{F}_D * \boldsymbol{\theta}_{cls_1} ) * \boldsymbol{\theta}_{cls_2} ) * \boldsymbol{\theta}_{cls_3} ),
		\label{equ:lsh}
	\end{align}
	where $\phi(\cdot)$ represents a non-linear function, and the symbol * denotes the convolution operation. It is important to note that for each class $cls$, we generate the prediction using the ``one vs. all'' approach, employing the Sigmoid function instead of Softmax. This strategy allows for multi-label prediction, accommodating scenarios where a pixel may belong to multiple classes simultaneously (e.g., a tumor on an organ).

	\begin{table*}[h]
		\caption{
			\textbf{\textit{The information for an assembly of datasets.}} We have developed \ourmodel\ from an assembly of 1--14 public datasets. The official test and validation sets of Medical Segmentation Decathlon (MSD) and Beyond the Cranial Vault (BTCV) are used to benchmark the performance of organ segmentation (\S\ref{sec:strong_challenge_ranking}) and tumor detection (\S\ref{sec:high_specificity}). 3D-IRCADb (15), TotalSegmentator (16) and a large-scale private dataset (17), consisting of 5,038 CT volumes with 21 annotated organs, are used for independent evaluation of model generalizability (\S\ref{sec:generalizability}) and transferability (\S\ref{sec:transfer_learning}).
		}\vspace{2px}
		\centering
		\scriptsize
		\begin{tabular}{p{0.18\linewidth}P{0.05\linewidth}P{0.05\linewidth}p{0.62\linewidth}}
			\toprule
			Datasets (Year) & \# Targets & \# Volumes & Annotated Organs or Tumors \\
			\midrule
			1. Pancreas-CT \citeyearpar{roth2015deeporgan} & 1 & 82 & {Pancreas} \\
			2. LiTS \citeyearpar{bilic2019liver} & 2 & 201 & {Liver, Liver Tumor$^*$} \\
			3. KiTS \citeyearpar{heller2020international} & 2 & 300 & {Kidney, Kidney Tumor$^*$} \\
			4. AbdomenCT-1K \citeyearpar{ma2021abdomenct} & 4 & 1000 & {Spleen, Kidney, Liver, Pancreas} \\
			5. CT-ORG \citeyearpar{rister2020ct} & 4 & 140 & {Lung, Liver, Kidneys and Bladder} \\
			6. CHAOS \citeyearpar{valindria2018multi} & 4 & 40 & {Liver, Left Kidney, Right Kidney, Spl} \\
			7-11. MSD CT Tasks \citeyearpar{antonelli2021medical} & 9 & 947 & {Spl, Liver and Tumor$^*$, Lung Tumor$^*$, Colon Tumor$^*$, Pan and Tumor$^*$, Hepatic Vessel and Tumor$^*$} \\
			12. BTCV \citeyearpar{landman2015miccai} & 13 & 50 & {Spl, RKid, LKid, Gall, Eso, Liv, Sto, Aor, IVC, R\&SVeins, Pan, RAG, LAG} \\
			13. AMOS22 \citeyearpar{ji2022amos} & 15 & 500 & {Spl, RKid, LKid, Gall, Eso, Liv, Sto, Aor, IVC, Pan, RAG, LAG, Duo, Bla, Pro/UTE} \\
			14. WORD \citeyearpar{luo2021word} & 16 & 150 & {Spl, RKid, LKid, Gall, Eso, Liv, Sto, Pan, RAG, Duo, Col, Int, Rec, Bla, LFH, RFH} \\
			\midrule
			15. 3D-IRCADb \citeyearpar{soler20103d} & 13 & 20 & {Liv, Liv Cyst, RLung, LLung, Venous, PVein, Aor, Spl, RKid, LKid, Gall, IVC} \\
			\cmidrule{4-4}
			\multirow{5}{*}{16. TotalSegmentator \citeyearpar{wasserthal2022totalsegmentator}}  & \multirow{5}{*}{104} & \multirow{5}{*}{1,024} & { Clavicula, Humerus, Scapula, Rib 1-12, Vertebrae C1-7, Vertebrae T1-9, Vertebrae L1-5, Hip, Sacrum, Femur, Aorta, Pulmonary Artery, Right Ventricle, Right Atrium, Left Atrium, Left Ventricle, Myocardium, PVein, SVein, IVC, Iliac Artery, Iliac Vena, Brain, Trachea, Lung Upper Lobe, Lung Middle Lobe, Lung Lower Lobe, AG, Spl, Liv, Gall, Pan, Kid, Eso, Sto, Duo, Small Bowel, Colon, Bla, Autochthon, Iliopsoas, Gluteus Minimus, Gluteus Medius, Gluteus Maximus} \\
			\cmidrule{4-4}
			\multirow{2}{*}{17. JHH \citeyearpar{xia2022felix}} & \multirow{2}{*}{21} & \multirow{2}{*}{5,038} & Aor, AG, CBD, Celiac AA, Colon, duo, Gall, IVC, Lkid, RKid, Liv, Pan, Pan Duct, SMA, Small bowel, Spl, Sto, Veins, Kid LtRV, Kid RtRV, CBD Stent, PDAC$^*$, PanNET$^*$, Pancreatic Cyst$^*$\\
			\bottomrule
		\end{tabular}
		\label{tab:public_dataset}
	\end{table*}
	
	\subsubsection{Optimization}
	The integrated dataset contains only partial annotations, resulting in some other organs being annotated as background. To mitigate this conflict, we employ the binary cross entropy (BCE) and binary dice loss functions for supervision. In order to generate the binary ground truth $\mathcal{M}_{cls}$ for each class within $\mathbf{L}_n$, we define the following formulation:
	\begin{align}
		\mathbf{M}_{cls}^{i} =
		\begin{cases}
			1, & \text{if } \mathbf{Y}^{i} = cls \\
			0, & \text{otherwise},
		\end{cases}
	\end{align}
	where $i$ represents the voxel index. Subsequently, we calculate the BCE loss and dice loss summation for the classes contained in $\mathcal{L}_n$, and back-propagate the accurate supervision to update the entire framework. The formulation for this process is as follows:
	\begin{align}
		\mathbf{L}_{mask} = & \sum_{cls} Dice(\bm{P}_{cls}, \mathbf{M}_{cls}) + BCE(\bm{P}_{cls}, \mathbf{M}_{cls}), \text{for}~cls \in \mathcal{L}_n.
	\end{align}
	The application of masked back-propagation effectively addresses the issue of label inconsistency in the context of the partial label problem.
	
	\subsection{Extended to Novel Classes}\label{sec:method_continual_learning}
	For the new CT volume $\{\mathbf{X}, \mathbf{Y}\}$ in $\mathcal{D}_{N+1}$, there are new classes $cls \in \mathcal{L}_{N+1} \setminus \mathcal{L}$. We add additional MLPs in LPG to learn the parameter for CSH and segment the new binary result for each new class. The feed forward pipeline follows the Equation \ref{equ:lpg} and \ref{equ:lsh}. The separate heads allow independent probability prediction for newly introduced and previously learned classes, therefore minimizing the impact of new classes on old ones during continual learning.
	
	In the context of continual organ segmentation, a major challenge arises from the model's inability to access previous datasets, which often leads to forgetting of previously learned classes. To address this issue and preserve the existing knowledge, we propose a method of generating soft pseudo annotations for the old classes on newly-arrived data. Specifically, we leverage the output prediction from model denoted as $\hat{\mathbf{Y}}$, which includes the old classes $\mathcal{L}$. This prediction serves as the pseudo label for the current step's old classes. For the new classes, we continue to use the ground truth label. Formally, the binary ground truth $\mathbf{M}_{cls}$ for each class in continual learning can be expressed as:
	\begin{align}
		\mathbf{M}_{cls} &= 
		\begin{cases}
			\mathbf{Y}_{cls}   & \text{ if } cls \in \mathcal{L}_{N+1} \setminus \mathcal{L} \\
			\hat{\mathbf{Y}}_{cls}   & \text{ if } cls \in \mathcal{L},
		\end{cases}
	\end{align}
	where $\mathbf{Y}_{cls}$ represents the ground truth label for class $cls$. By adopting this approach, our aim is to retain the previous knowledge and prevent the model from forgetting the information learned from the old classes while simultaneously learning the new classes.

	\begin{table}[t]
		\centering
		\scriptsize
		\caption{ \textit{The unification of the label taxonomy.}
			The corresponding template for this taxonomy is presented, where ind. denotes the index of the class.
		}\vspace{2px}
		\begin{tabular}{p{0.23\linewidth}P{0.035\linewidth}|p{0.18\linewidth}P{0.035\linewidth}|p{0.19\linewidth}P{0.035\linewidth}}
			\toprule
			Class & Ind. & Class & Ind. & Class & Ind. \\
			\midrule
			spleen & 1 & R kidney & 2 & L kidney & 3 \\
			gall bladder & 4 & esophagus & 5 & liver & 6 \\
			stomach & 7 &  aorta & 8 & postcava & 9 \\
			portal~\&~spenic vein & 10 & pancreas & 11 & R adrenal gland & 12 \\
			L adrenal gland & 13 & duodenum & 14 & hepatic vessel & 15 \\
			R lung & 16 & L lung & 17 & colon & 18 \\
			intestine & 19 & rectum & 20 & bladder & 21 \\
			prostate/uterus & 22 & L head of femur & 23 & R head of femur & 24 \\
			celiac truck & 25 \\
			\midrule
			kidney tumor & 26 & liver tumor  & 27 & pancreatic tumor & 28 \\
			hepatic vessel tumor & 29 & lung tumor & 30 & colon tumor & 31 \\
			kidney cyst & 32 \\
			\bottomrule
		\end{tabular}
		\label{tab:label_index}
	\end{table}

	\section{Experiment I: Integrated Learning}
	\label{sec:integrated}
	
	\subsection{Experimental Settings}
	
	\subsubsection{Datasets}
	The assembly of datasets for this research comprises a total of 14 publicly available datasets, consisting of 3,410 CT volumes, which were used for AI development---2,100 for training and 1,310 for validation. Additionally, 2 public datasets, 2 private datasets were employed for testing the developed model. The details of these datasets are summarized in \tableautorefname~\ref{tab:public_dataset}.
	To preprocess the datasets, we initially map them to the standard index template, as shown in Table \ref{tab:label_index}. For datasets such as KiTS, WORD, AbdomenCT-1K, and CT-ORG, where the left and right organs are not distinguished, we utilize a script to split the organs (Kidney, Adrenal Gland, and Lung) into left and right parts. Additionally, we account for the inclusion relation, such as considering organ tumors as part of the respective organs and the hepatic vessel being inside the liver. Since we represent each organ segmentation result as a binary mask, we can independently organize the segmentation ground truth for the overlapped organs in a binary mask format. This pipeline allows for efficient handling of the segmentation ground truth.
	
	\major{In addition to 16 public datasets, we also used two proprietary datasets collected from Johns Hopkins Hospital (JHH) and University of California, San Francisco (UCSF).} JHH includes 5,038 CT volumes with annotations for 21 organs. Each case in this dataset was scanned using contrast-enhanced CT in both venous and arterial phases, employing Siemens MDCT scanners. The JHH dataset is particularly employed to investigate the extensibility of new organ classes. \major{The UCSF dataset comprises outpatient contrast-enhanced abdominal CT exams acquired at the portal venous phase from a single day (3/22/2024).   These volumes were performed from multiple sites at UCSF, comprising a cohort of 11 patients with known liver metastases. These patients had varying numbers (ranging from 2 to 50) and sizes (ranging from 0.5 cm to 4.6 cm) of liver metastases.}

	\subsubsection{Implementation Details}
	\label{sec:implementation}
	\smallskip\noindent\textbf{Optimization.} The training of \ourmodel\ employs the AdamW optimizer, along with a warm-up cosine scheduler lasting for 50 epochs. For the segmentation experiments, a batch size of 6 per GPU is utilized, with a patch size of $96\times96\times96$. The initial learning rate is set to $4e^{-4}$, with a momentum of 0.9 and a decay rate of $1e^{-5}$ on a multi-GPU setup consisting of 4 GPUs using Distributed Data Parallel (DDP). The implementation of the framework is based on MONAI 0.9.0. To ensure the robustness of the results, a five-fold cross-validation strategy is employed. The best model is selected in each fold based on the evaluation of the validation metrics. The training process is conducted on eight NVIDIA RTX A5000 cards.

	\smallskip\noindent\textbf{Data Augmentation.} Data augmentation techniques are applied to enhance the diversity and generalization capability of the training dataset. In our implementation, we utilize the MONAI library \citep{cardoso2022monai} in Python. \major{The uniform sampling strategy is utilized to ensure that volumes from each dataset have an equal probability of being selected.} The preprocessing steps involve several operations to standardize the input CT volumes. Firstly, the orientation of the volumes is adjusted to conform to specified axcodes. Additionally, isotropic spacing is employed to reslice each volume, ensuring a consistent voxel size of $1.5 \times 1.5 \times 1.5mm^3$. To normalize the intensity values, we truncate the range to $[-175, 250]$ and perform linear normalization to map the intensity values to the interval $[0, 1]$. \major{As we focus on the relevant structures within the medical images, we perform cropping to extract the foreground objects with a foreground-to-background patch ratio of 2:1.} Specifically, during the training phase, random fixed-sized regions of $96 \times 96 \times 96$ are cropped, with the center of the patch selected based on a predefined ratio. The selection process takes into account whether the center voxel corresponds to a foreground or background region. Also, we introduce random rotations of the input patches by 90 degrees and apply intensity shifts with a probability of $0.1$ and $0.2$, respectively. It is worth noting that mirroring augmentation is not utilized to avoid confusion between organ structures in the right and left parts of the images.

	\smallskip\noindent\textbf{Network Structures.}
	\major{We use the pre-trained text encoder with masked self-attention transformer architecture from the CLIP framework as the text branch}\footnote{\href{https://github.com/openai/CLIP}{https://github.com/openai/CLIP}}. We can extract and store the text features to reduce overhead brought by the text encoder in the training and inference stage since the CLIP embedding only depends on the dictionary, which is fixed. For the vision branch, we employ Swin UNETR as our vision encoder. The Swin UNETR architecture consists of four attention stages, each composed of two transformer blocks, as well as five convolution stages utilizing a CNN-based structure. To reduce the resolution by a factor of 2 in the attention stage, a patch merging layer is utilized. Specifically, in Stage 1, a linear embedding layer and transformer blocks are employed to maintain the number of tokens at $\frac{H}{2} \times \frac{W}{2} \times \frac{D}{2}$. Following this, a patch merging layer groups patches with a resolution of $2 \times 2 \times 2$ and concatenates them, resulting in a 4C-dimensional feature embedding. Subsequently, a linear layer is used to down-sample the resolution by reducing the dimension to 2C. This process continues in stages 2, 3, and 4, as outlined in \citet{tang2022self}.
	The text-based controller, on the other hand, is implemented as a single convolutional layer. It takes as input the CLIP embedding and the global pooling feature from the last convolution stages in the vision encoder. This design choice enables efficient integration of text and vision information within the framework.

	\begin{table*}[t]
		\caption{\major{\textit{Leaderboard performance of CT tasks on MSD.} The results are evaluated in the server on the MSD competition test dataset. The dice similarity coefficient (DSC) and normalized surface distance (NSD) metrics are obtained from the \href{https://decathlon-10.grand-challenge.org/evaluation/challenge/leaderboard/}{MSD public leaderboard}. The results of MRI-related tasks were generated by nnU-Net \citep{isensee2021nnu}.}}\vspace{2px}
		\scriptsize
		\begin{tabular}{p{0.23\linewidth}|P{0.038\linewidth}P{0.038\linewidth}P{0.038\linewidth}|P{0.038\linewidth}P{0.038\linewidth}P{0.038\linewidth}|P{0.038\linewidth}P{0.038\linewidth}P{0.038\linewidth}|P{0.038\linewidth}P{0.038\linewidth}P{0.038\linewidth}}
			\toprule
			& \multicolumn{6}{c|}{Task03 Liver} & \multicolumn{6}{c}{Task07 Pancreas}  \\
			\cline{2-13}
			&  \multicolumn{3}{c|}{DSC} &  \multicolumn{3}{c|}{NSD} &  \multicolumn{3}{c|}{DSC} &  \multicolumn{3}{c}{NSD} 
			\\
			Method & Organ & Tumor &Avg. & Organ  & Tumor &Avg. & Organ  & Tumor  & Avg.  & Organ & Tumor & Avg. \\ 
			\midrule
			Kim~\etal~\citep{kim2019scalable}  & 94.25 & 72.96   & 83.61  & 96.76 & 88.58 & 92.67 & 80.61    & 51.75  & 66.18 & 95.83   & 73.09  & 84.46 \\
			Trans VW~\citep{haghighi2021transferable}  & 95.18 & 76.90 & 86.04   & 97.86  & {92.03} & 94.95 & 81.42    & 51.08  & 66.25 & 96.07 & 70.13   & 83.10 \\
			C2FNAS\citep{yu2020c2fnas}  & 94.98 & 72.89   & 83.94  & 98.38 & 89.15 & 93.77  & 80.76    & 54.41  & 67.59 & 96.16   & 75.58  & 85.87 \\
			Models Gen.~\citep{zhou2021models}   & 95.72 & {77.50}   & {86.61}  & 98.48 & 91.92 & {95.20} & 81.36    & 50.36  & 65.86 & 96.16   & 70.02  & 83.09 \\
			nnU-Net~\citep{isensee2021nnu}  & \textbf{95.75} & 75.97   & 85.86  & 98.55 & 90.65 & 94.60 &  81.64    & 52.78  & 67.21 & 96.14   & 71.47  & 83.81 \\
			DiNTS~\citep{he2021dints}   & 95.35 & 74.62   & 84.99  & \textbf{98.69} & 91.02 & 94.86  & 81.02    & 55.35  & 68.19 & 96.26   & 75.90  & 86.08 \\
			Swin UNETR~\citep{tang2022self}   & 95.35 & 75.68 & 85.52 & 98.34 & 91.59 & 94.97 &  {81.85} &{58.21} & {70.71} & {96.57} &{79.10} & {87.84} \\
			\midrule
			\ourmodel\ & 95.42 & \textbf{79.35} & \textbf{87.39} & 98.18 & \textbf{93.42} & \textbf{95.80} & \textbf{82.84} & \textbf{62.33} & \textbf{72.59} & \textbf{96.65} & \textbf{82.86} & \textbf{89.76} 
			\\
			
			\bottomrule
			
		\end{tabular}
		\vspace{0.5 em}\\
		\begin{tabular}{p{0.23\linewidth}|P{0.038\linewidth}P{0.038\linewidth}P{0.038\linewidth}|P{0.038\linewidth}P{0.038\linewidth}P{0.038\linewidth}|P{0.038\linewidth}P{0.038\linewidth}|P{0.038\linewidth}P{0.038\linewidth}|P{0.038\linewidth}P{0.038\linewidth}}
			& \multicolumn{6}{c|}{Task08 Hepatic Vessel} & \multicolumn{2}{c|}{Task06 Lung} & \multicolumn{2}{c|}{Task09 Spleen} & \multicolumn{2}{c}{Task10 Colon} \\
			\cline{2-13}
			&  \multicolumn{3}{c|}{DSC} &  \multicolumn{3}{c|}{NSD} & DSC & NSD &DSC & NSD  & DSC & NSD 
			\\
			Method & Organ  & Tumor  & Avg.  & Organ & Tumor & Avg. &  \multicolumn{2}{c|}{Tumor} & \multicolumn{2}{c|}{Organ} & \multicolumn{2}{c}{Tumor}\\ 
			\midrule
			Kim~\etal~\citep{kim2019scalable} & 62.34 & 68.63   & 65.49  & 83.22 & 78.43 & 80.83 & 63.10 & 62.51 & 91.92    & 94.83 & 49.32    & 62.21\\
			Trans VW~\citep{haghighi2021transferable}  & 65.80 & 71.44 & 68.62   & 84.01  & 80.15 & 82.08 &74.54    & 76.22 & 97.35    & 99.87 & 51.47    & 60.53\\
			C2FNAS\citep{yu2020c2fnas}  &  64.30 & 71.00   & 67.65  & 83.78 & 80.66 & 82.22  &70.44    & 72.22 & 96.28    & 97.66 & 58.90    & {72.56}\\
			Models Gen.~\citep{zhou2021models} & 65.80 & 71.44   & 68.62  & 84.01 & 80.15 & 82.08 & 74.54    & 76.22  & 97.35    & 99.87 & 51.47    & 60.53\\
			nnU-Net~\citep{isensee2021nnu}  & {66.46} & 71.78   & {69.12}  & 84.43 & 80.72 & 82.58 &73.97    & 76.02  & \textbf{97.43}    & \textbf{99.89} & 58.33    & 68.43\\
			DiNTS~\citep{he2021dints}  &  64.50 & 71.76   & 68.13  & 83.98 & 81.03 & 82.51  &74.75    & 77.02 & 96.98    & 99.83 & 59.21    & 70.34\\
			Swin UNETR~\citep{tang2022self} & 65.69 & {72.20} &68.95 & {84.83} & {81.62} & {83.23}  &{76.60} &{77.40} & 96.99 & 99.84 & {59.45} & 70.89\\
			\midrule
			\ourmodel\ & \textbf{67.15} & \textbf{75.86} & \textbf{71.51} & \textbf{84.84} & \textbf{85.23} & \textbf{85.04} & \textbf{80.01} & \textbf{81.25} & 97.27 & 99.87 & \textbf{63.14} & \textbf{75.15}
			\\
			\bottomrule
		\end{tabular}%
		\label{tab:msd_test}
	\end{table*}
	
	\begin{table*}[t]
		\caption
		{\textit{Five-fold cross-validation results on BTCV.} For a fair comparison, we did not use model ensemble during the evaluation. The dataset-specific training means train only on BTCV, while dataset-agnostic traning means training on multiple datasets. In dataset-agnostic training part, results of Swin UNETR and Multi-Talent are original from \citet{ulrich2023multitalent}. We re-implement TansUNet and UniSeg with same setting as \ourmodel. \ourmodel\ achieves the overall best performance. The performance is statistically significant at the P=0.05 level, with highlighting in a \significant{light red} box.}\vspace{2px}
		\centering
		\scriptsize
		
		\begin{tabular}{p{0.2\linewidth}P{0.035\linewidth}P{0.035\linewidth}P{0.035\linewidth}P{0.035\linewidth}P{0.035\linewidth}P{0.035\linewidth}P{0.035\linewidth}P{0.035\linewidth}P{0.035\linewidth}P{0.035\linewidth}P{0.035\linewidth}P{0.035\linewidth}|P{0.05\linewidth}}
			\toprule
			Methods & Spl  
			& RKid & LKid 
			& Gall  & Eso 
			& Liv & Sto 
			& Aor & IVC 
			& Veins   & Pan 
			& AG & Avg. \\ \midrule
			\multicolumn{14}{c}{\textbf{Dataset-specific Training}}
			\\
			TransUNet~\citep{chen2021transunet}    
			& 94.10 & 90.22                        
			& 88.84 & 65.49                    
			& 73.19 & 93.24                        
			& 80.85 & 87.47   
			& 80.48 & 71.47
			& 74.26 & 64.76
			& 79.16
			\\ 
			CoTr~\citep{xie2021cotr}     
			& 95.51 & 88.03                 
			& 89.19 & 68.49                        
			& 75.83 & 95.93                       
			& 81.84 & 89.01 
			& 82.32 & 73.39
			& 75.12 & 65.78
			& 80.48
			\\ 
			TransBTS~\citep{isensee2021nnu}     
			& 94.59 & 89.23                     
			& 90.47 & 68.50                       
			& 75.59 & 96.14                       
			& 83.72 & 88.85  
			& 82.28 & 74.25
			& 75.12 & 66.74 
			& 80.94
			\\ 
			nnFormer~\citep{zhou2021nnformer}     
			& 94.51 & 88.49                     
			& 93.39 & 65.51                        
			& 74.49 & 96.10                      
			& 83.83 & 88.91
			& 80.58 & 75.94
			& 77.71 & {68.19}
			& 81.22
			\\
			UNETR~\citep{hatamizadeh2022unetr}
			& 94.91 & 92.10                        
			& 93.12 & 76.98                        
			& 74.01 & 96.17                        
			& 79.98 & 89.74   
			& 81.20 & 75.05
			& 80.12 & 62.60
			& 81.43
			\\
			nnU-Net~\citep{isensee2021nnu}     
			& {95.92} & 88.28                        
			& 92.62 & 66.58                        
			& 75.71 & 96.49                        
			& 86.05 & 88.33   
			& 82.72 & \textbf{78.31}
			& 79.17 & 67.99
			& 82.01
			\\
			Swin UNETR~\citep{tang2022self}    
			& 95.44 & {93.38}                 
			& 93.40 & 77.12                    
			& 74.14 & 96.39                        
			& 80.12 & 90.02   
			& 82.93 & 75.08
			& 81.02 & 64.98
			& 82.06
			\\
			\midrule
			\multicolumn{14}{c}{\textbf{Dataset-agnostic Training}}
			\\
			TransUNet~\citep{chen2021transunet}
			& 94.90 & 90.89
			& 90.45 & 70.82
			& 80.31 & 95.01
			& 90.11 & 91.30
			& 84.21 & 74.53
			& 83.21 & 71.27
			& 84.75
			\\
			Swin UNETR~\citep{tang2022self}    
			& 91.04 & 86.93                 
			& 86.83 & 67.73                 
			& 77.87 & 95.14              
			& 86.77 & 89.92
			& 84.34 & 73.88
			& 82.27 & 65.34
			& 82.33
			\\
			Multi-Talent~\citep{ulrich2023multitalent}
			& 93.61 & 90.89
			& 90.47 & 72.12
			& \textbf{79.28} & 96.23
			& \textbf{92.83} & 91.60
			& \textbf{87.31} & 77.58
			& \textbf{84.92} & \textbf{72.16}
			& 85.75
			\\
			UniSeg~\citep{ye2023uniseg}
			&\textbf{96.99} &{94.11} 
			&92.21 &68.13 
			&78.15 & {96.43}
			&85.93 & 89.78 
			&86.69 & 76.12
			& 83.39 & 72.39
			& 85.02
			\\ 
			\ourmodel\ & 95.82 & \textbf{94.28} & \textbf{94.11} & \textbf{79.52} & {76.55} & \textbf{97.05} & {92.59} & \textbf{91.63} & {86.00} & 77.54 & {83.17} & {70.52} & \cellcolor{significant!40} \textbf{86.13}
			\\ \bottomrule
		\end{tabular}%
		\label{tab:btcv_benchmark}
	\end{table*}

	\begin{table*}[htbp]
		\centering
		\scriptsize
		\caption{\textit{Benchmark of different partially labeled methods on MOTS dataset.} \ourmodel* denotes the model trained exclusively on the MOTS training set, while \ourmodel~refers to our proposed model trained on the 14 integrated datasets in this work, excluding the 235 CT volumes present in the MOTS test set. The average performance is statistically significant at the P=0.05 level, with highlighting in a \significant{light red} box.}\vspace{2px}
		\resizebox{\textwidth}{!}{
			\begin{tabular}{l|cc|cc|cc|cc|c|c|c|c}
				\hline
				\multirow{2}{*}{Methods} & \multicolumn{2}{c|}{Task 1: Liver} & \multicolumn{2}{c|}{Task 2: Kidney} & \multicolumn{2}{c|}{Task 3: Hepatic} & \multicolumn{2}{c|}{Task 4: Pancreas} & Task 5: Colon & Task 6: Lung & Task 7: Spleen & Average \\
				\cline{2-12}
				& Organ & Tumor & Organ & Tumor & Organ & Tumor & Organ & Tumor & Tumor & Tumor & Organ & Dice \\
				\midrule
				Multi-Nets & 96.54 & 63.9 & 96.52 & 78.07 & 62.85 & 71.56 & 83.18 & 56.17 & 42.39  & 61.68 & 94.37 & 73.38 \\
				Cond-Input \citep{chen2017fast} & 96.67 & 65.61 & 96.97 & 83.17 & 65.14 & 74.98 & 84.24 & 63.32 & 46.03 & 69.67 & 95.15 & 76.45 \\
				Multi-Head \citep{chen2019med3d} & 96.77 & 64.33 & 96.84 & 82.39 & 65.21 & 74.62 & 84.59 & 64.11 & 46.65 & 68.63 & 95.47 & 76.33 \\
				Cond-Dec \citep{dmitriev2019learning} & 96.23 & 65.25 & 96.43 & 83.43 & 65.38 & 72.26 & 83.86 & 62.97 & 50.67  & 63.77 & 94.33 & 75.87 \\
				TAL \citep{fang2020multi} & 96.21 & 64.1 & 96.01 & 78.87 & 64.64 & 73.87 & 83.39 & 60.93 & 45.14  & 67.59 & 94.72 & 75.04 \\
				DoDNet \citep{zhang2021dodnet} & 96.86 & 65.99 & 97.31 & 83.45 & 65.80 & 77.07 & 85.39 & 60.22 & 47.66 & 72.65 & 93.94 & 76.94 \\
				TGNet \citep{wu2022tgnet}& 96.83 & 63.84 & 96.17 & 81.05 & 62.59 & 74.65 & 83.30 & 60.71 & 47.37 & 61.69 & 94.18 & 74.76\\
				CCQ \citep{liu2023ccq} & 96.71 & 64.32 & 96.68 & 79.82 & 62.50 & \textbf{76.94} & 83.18 & 60.54 & 54.77 & 70.51 & 94.57 & 76.41 \\
				TransDoDNet \citep{xie2023learning} & \textbf{97.01} & 66.27 & \textbf{97.03} & 83.57 & 65.43 & 76.78 & 84.88 & 64.63 & 58.64 & \textbf{70.82} & 95.82 & 78.26 \\
				\ourmodel* & 96.93 & 67.14 & 96.09 & 82.34 & 64.39 & {76.31} & 84.42 & 64.68 & 59.49 & 70.31 & 95.31 & 77.95 \\
				\ourmodel & 96.81 & \textbf{70.18} & 96.96 & \textbf{87.87} & \textbf{66.15} & 76.45 &\textbf{85.16} & \textbf{67.21} & \textbf{63.92} & 69.63 & \textbf{96.84} & \cellcolor{significant!40} \textbf{79.74}  \\
				\hline
			\end{tabular}
		}
		\label{tab:mots}
	\end{table*}

	\smallskip\noindent\textbf{Inference.} \major{The input 3D image is divided into smaller overlapping windows or patches based on the (96, 96, 96)px region of interest and 0.5 overlap ratio. Each window is passed through the pre-trained model, and the model makes a prediction for that particular window. Then, the predictions from overlapping windows are aggregated using gaussian weighted blending, which gives less weight to predictions on edges of windows, to form the final prediction for the entire input image. The obtained binary prediction is refined through three post-processing techniques. 1) Left-Right Split: For organs with bilateral symmetry, we fit a hyperplane through the vertebral and sagittal centers to split the left and right sides, subsequently remapping the side-related labels based on this hyperplane. 2) Non-largest Connected Component Suppression: For organs that occur only once, we identify three-dimensional connected components for the anatomy and remove non-largest connected components. 3) Anatomical Region Restriction: For certain categories with specific associated regions, we restrict the anatomy predictions to the corresponding body part. Since one pixel may belong to many classes at the same time, such as liver, liver tumor and Hepatic Vessel, it is optimal to store the prediction with binary mask for each classes.}
	
	\major{When merging the binary masks into a single mask, we follow a specific order to merge the individual binary masks. Firstly, the organ masks are combined, followed by the incorporation of vessel masks, and finally, the tumor masks are integrated. Moreover, in instances where the predicted masks of two organs overlap, we adjudicate the final result by comparing the associated confidence scores for each prediction.
	}

	\begin{figure*}[!h]
		\centerline{\includegraphics[width=1\linewidth]{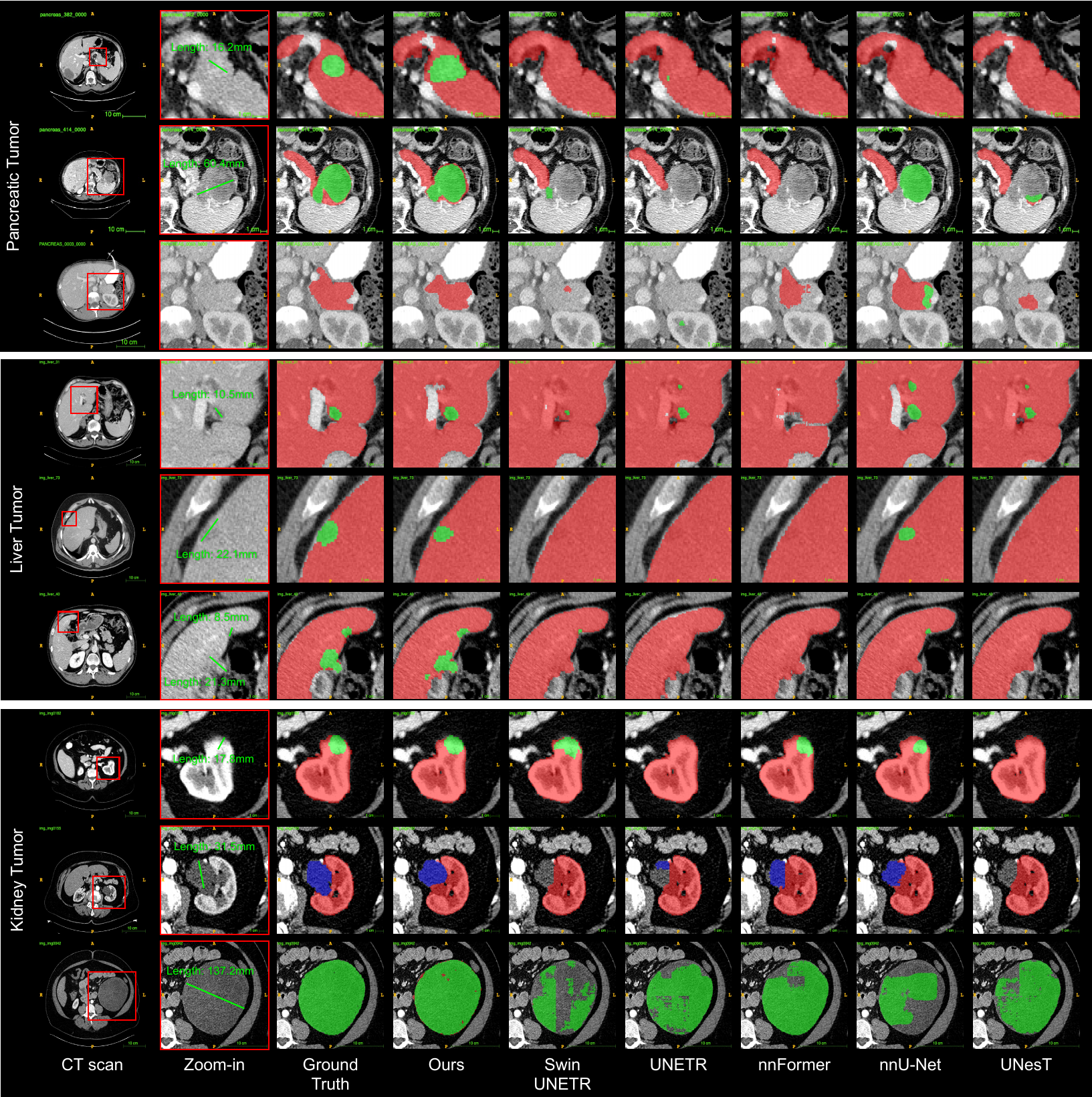}}
		\caption{
			\textit{Qualitative results of multi-tumor detection and segmentation.} We review the detection/segmentation results of each tumor type from smaller to larger sizes. Especially, \ourmodel\ generalizes well in organ segmentation and does not generate many false positives of tumors when it comes to a CT volume without tumors from other hospitals (Row 3).
		}
		\label{fig:qualitative_visualization}
	\end{figure*}
	
	\begin{table*}[!h]
		\caption
		{\textit{Quantitative results of multi-tumor detection.} The tumor detection performance analysis is based on CT volumes from the LiTS, KiTS, and MSD Pancreas datasets, containing tumors in the liver, kidney, and pancreas, respectively \citep{bilic2019liver, heller2019kits19, antonelli2021medical}. These volumes are utilized to compute the sensitivity of tumor detection. To assess specificity in an alternative manner, the CHAOS and Pancreas-CT datasets are employed \citep{valindria2018multi, roth2015deeporgan}. Importantly, it has been confirmed that the CHAOS dataset is devoid of liver or kidney tumors, while the Pancreas-CT dataset does not contain pancreatic tumors within the CT volumes. \ourmodel\ achieves a high harmonic mean, which signifies the capability to accurately identify tumor cases while minimizing false positives.}\vspace{2px}
		\centering
		\scriptsize
		\begin{tabular}{p{0.22\linewidth}|P{0.06\linewidth}P{0.06\linewidth}P{0.06\linewidth}|P{0.06\linewidth}P{0.06\linewidth}P{0.06\linewidth}|P{0.06\linewidth}P{0.06\linewidth}P{0.06\linewidth}}
			\toprule
			\multirow{2}{*}{Methods} & \multicolumn{3}{c|}{Liver Tumor} & \multicolumn{3}{c|}{Kidney Tumor} &\multicolumn{3}{c}{Pancreatic Tumor} \\
			& Sen. & Spec. & Harm.
			& Sen. & Spec. & Harm.
			& Sen. & Spec. & Harm.
			\\ \midrule
			nnU-Net~\citep{isensee2021nnu} & \textbf{94.44} & 75.00 & 83.60 & 96.88 & 85.00 & 90.55 & 95.18 & 88.75 & 91.85
			\\
			UNet++~\citep{zhou2019unet++}& \textbf{94.44} & 80.00 & 86.62 &N/A&N/A&N/A&N/A & N/A&N/A
			\\
			UNETR~\citep{hatamizadeh2022unetr} & 86.11 & \textbf{95.00} & 90.34 & 93.75 & \textbf{95.00} & \textbf{94.37} & 90.36 & 81.25 & 85.56
			\\
			Swin UNETR~\citep{tang2022self} & 91.67 & 85.00 & 88.21 & \textbf{97.91} & 70.00 & 81.63 & \textbf{97.59} & 87.50 & 92.26
			\\
			\midrule
			\ourmodel\ & 88.89 & \textbf{95.00} & \textbf{91.84} & 91.67 & \textbf{95.00} & 93.31 & 93.98 & \textbf{91.25} & \textbf{92.59}
			\\ \bottomrule
		\end{tabular}%
		\label{tab:high_specificity}
	\end{table*}
	
	\begin{table*}[t]
		\caption
		{\textit{Ablation study on language encoding.} We compare the language embedding among one-hot, BioBERT~\citep{yasunaga2022linkbert} and CLIP embedding in the test set of integrated data. The BioBERT method utilizes the prompt \textit{`A computerized tomography of a [CLS]'}. On the other hand, CLIP v1, v2, and v3 employ prompts such as \textit{`A photo of a [CLS]'}, \textit{`There is [CLS] in this computerized tomography'}, and \textit{`A computerized tomography of a [CLS]'}, respectively.}\vspace{2px}
		\centering
		\scriptsize
		\begin{tabular}{p{0.11\linewidth}P{0.055\linewidth}P{0.055\linewidth}P{0.055\linewidth}P{0.055\linewidth}P{0.055\linewidth}P{0.055\linewidth}P{0.055\linewidth}P{0.055\linewidth}P{0.055\linewidth}P{0.055\linewidth}P{0.055\linewidth}}
			\toprule
			\textbf{\textit{Encoding}} & spleen & kidneyR & kidneyL & gallblad. & esophagus & liver & stomach & aorta & postcava & PSV & pancreas
			\\ \midrule
			One-hot & 91.92 & 91.98 & 92.14 & 71.75 & 70.28 & 95.10 & 80.52 & 83.57 & 82.71 & 67.81 & 74.06
			\\
			BioBERT  & 94.65 & 93.26 & \textbf{92.98} & 75.14 & 72.32 & 95.09 & 87.68 & 91.05 & \textbf{83.91} & 67.83 & 80.51
			\\
			CLIP V1 & 92.35 & 91.83 & 91.89 & 72.45 & 71.38 & 90.23 & 73.07 & 86.77 & 78.17 & 74.00 & 74.91
			\\
			CLIP V2 & 93.05 & 92.14 & 91.42 & \textbf{75.88} & \textbf{75.56} & 94.75 & 75.79 & 91.15 & 80.64 & \textbf{78.90} & 78.94
			\\
			CLIP V3 & \textbf{94.69} & \textbf{94.09} & 92.77 & 73.45 & 72.87 & \textbf{95.71} & \textbf{89.19} & \textbf{92.19} & 83.44 & 59.20 & \textbf{86.09}
			\\ \midrule
			\vspace{1.0 em}\\
			\textbf{\textit{Embedding}} & adrenalR & adrenalL & duodenum & hepatic & lungR & lungL & colon & intestine & rectum & bladder & prostate
			\\ \midrule
			One-hot & 64.52 & 66.96 & 55.66 & 71.03 & \textbf{79.63} & 66.75 & 69.22 & 78.05 & 69.87 & 76.74 & 66.15
			\\
			BioBERT & 65.94 & 68.72 & \textbf{68.61} & 59.14 & 75.40 & 69.09 & 71.24  & \textbf{81.78} & 65.58 & 74.51 & 69.51
			\\
			CLIP V1 & 72.07 & 72.42 & 62.42 & \textbf{74.53} & 79.32 & 76.52 & 70.32 & 75.65 & 63.11 & 75.06 & 66.47
			\\
			CLIP V2 & \textbf{79.98} & \textbf{79.73} & 66.01 & 68.65 & 75.87 & \textbf{82.98} & \textbf{74.88} & 70.82 & 64.64 & 70.06 & 68.8
			\\
			CLIP V3 & 64.75 & 70.18 & 71.11 & 65.43 & 77.48 & 62.11 & 71.77 & 81.47 & \textbf{79.42} & \textbf{86.71} & \textbf{72.96}
			\\ \midrule
			\vspace{1.0 em}\\
			\textbf{\textit{Embedding}} & femurL & femurR & celiac & KiT & LiT & PT & HVT & LuT & CoT & KiC & Ave
			\\ \midrule
			One-hot & 70.27 & 60.23 & 78.92 & 63.84 & 68.02 & 55.48 & 52.31 & 53.87 & 48.39 & \textbf{35.81} & 70.42
			\\
			BioBERT & 74.39 & 79.07 & 80.69 & 57.41 & 63.44 & 39.70 & 57.88 & 58.57 & 54.19 & 20.33 & 71.55
			\\
			CLIP V1 & 74.61 & 72.53 & 79.28 & 56.62 & 76.24 & 61.05 & 56.49 & 73.60 & 55.03 & 32.87 & 73.49
			\\
			CLIP V2 & 69.98 & 75.73 & \textbf{84.04} & 67.04 & \textbf{82.09} & \textbf{77.75} & 67.45 & \textbf{75.38} & 55.55 & 35.79 & 75.66
			\\
			CLIP V3 & \textbf{84.94} & \textbf{89.45} & 77.55 & \textbf{68.72} & 74.87 & 65.46 & \textbf{73.53} & 73.12 & \textbf{60.66} & 30.44 & \textbf{76.11}
			\\ \bottomrule
		\end{tabular}%
		\label{tab:clip_ablation}
		\vspace{-0.2cm}
	\end{table*}

	\begin{figure*}[t]
		\centerline{\includegraphics[width=1.0\linewidth]{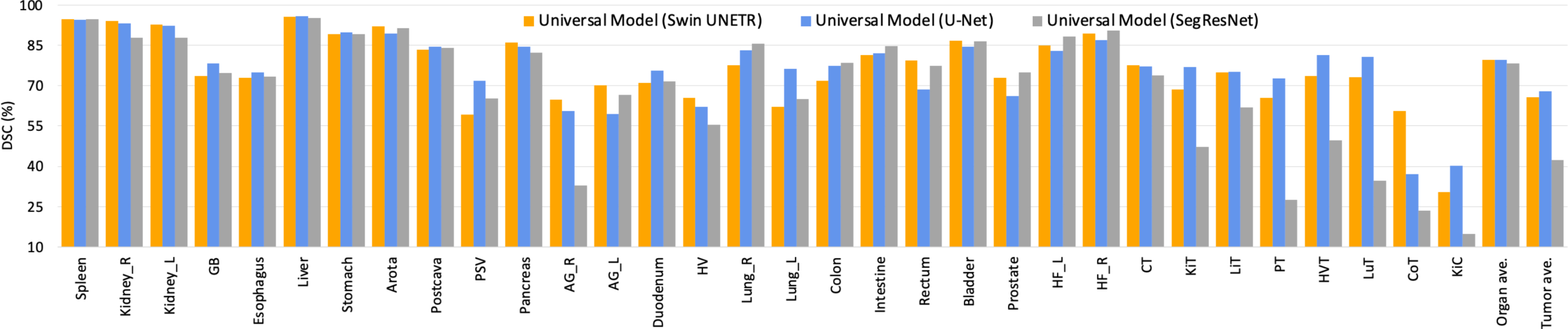}}
		\caption{
			\textbf{\textit{Ablation study on different segmentation backbones.}} \ourmodel\ can be expanded to Transformer-based (Swin UNETR) and CNN-based (U-Net, SegResNet-Tiny) backbone. These backbones achieve comparable results. The numbers of parameters of Swin UNETR, U-Net, and SegResNet-Tiny are 62.19M, 19.08M, and 4.7M, respectively. The order of classes is the same as \tableautorefname~\ref{tab:label_index}.
		}
		\label{fig:expansibility}
		\vspace{-0.5cm}
	\end{figure*}

	\begin{table*}[h!]
		\caption{\textbf{\textit{Generalizability.} Results on external datasets.} We evaluate \ourmodel\ and eight models on data from two external sources without additional fine-tuning or domain adaptation. {mDSC* is the average DSC score of the first seven organs.} Compared with dataset-specific models, our \ourmodel\ performs more robustly to CT volumes taken from a variety of scanners, protocols, and institutes. The average performance is statistically significant at the P=0.05 level, with highlighting in a \significant{light red} box.}\vspace{2px}
		\centering
		\scriptsize
		\begin{tabular}{p{0.28\linewidth}P{0.04\linewidth}P{0.04\linewidth}P{0.04\linewidth}P{0.04\linewidth}P{0.04\linewidth}P{0.04\linewidth}P{0.04\linewidth}P{0.04\linewidth}P{0.04\linewidth}P{0.04\linewidth}P{0.04\linewidth}}
			\toprule
			\textbf{\textit{3D-IRCADb}} & spleen & kidneyR & kidneyL
			& gallblad. & liver & stomach 
			& pancreas & lungR & lungL & mDSC* & mDSC
			\\ \midrule
			SegResNet~\citep{siddiquee2021redundancy} & 94.08 & 80.01 & 91.60 & 69.59 & 95.62 & \textbf{89.53} & 79.19 & N/A & N/A & 85.66 & N/A
			\\
			nnFormer~\citep{zhou2021nnformer} & 93.75 & 88.20 & 90.11 & 62.22 & 94.93 & 87.93 & 78.90  & N/A & N/A & 85.14 & N/A
			\\
			UNesT~\citep{yu2022unest} & 94.02 & 84.90 & \textbf{94.95} & 68.58 & 95.10 & 89.28 & 79.94 & N/A & N/A & 86.68 & N/A
			\\
			TransBTS~\citep{wang2021transbts} & 91.33 & 76.22 & 88.87 & 62.50 & 94.42 & 85.87 & 63.90 & N/A & N/A & 80.44 & N/A
			\\
			TransUNet~\citep{chen2021transunet} & 94.09 & 82.07 & 89.92 & 63.07 & 95.55 & 89.12 & 79.53 & N/A & N/A & 84.76 & N/A
			\\
			UNETR~\citep{hatamizadeh2022unetr} & 92.23 & 91.28 & 94.19 & 56.20 & 94.25 & 86.73 & 72.56 & 91.56 & 93.31 & 83.92 & 85.81
			\\
			Swin UNETR~\citep{tang2022self} & 93.51 & 66.34 & 90.63 & 61.05 & 94.73 & 87.37 & 73.77 & 93.72 & 92.17 & 81.05 & 83.69
			\\
			\midrule
			\ourmodel\ & \textbf{95.76} & \textbf{94.99} & 94.42 & \textbf{88.79} & \textbf{97.03} & 89.36 & \textbf{80.99} & \textbf{97.71} & \textbf{96.72} & \cellcolor{significant!40} \textbf{91.62} & \cellcolor{significant!40} \textbf{92.86}
			\\ \bottomrule
		\end{tabular}
		\vspace{0.5 em}\\
		\begin{tabular}{p{0.28\linewidth}P{0.04\linewidth}P{0.04\linewidth}P{0.04\linewidth}P{0.04\linewidth}P{0.04\linewidth}P{0.04\linewidth}P{0.04\linewidth}P{0.04\linewidth}P{0.04\linewidth}P{0.04\linewidth}P{0.04\linewidth}}
			\textbf{\textit{JHH}} & spleen & kidneyR & kidneyL
			& gallblad. & liver & stomach 
			& pancreas & aorta & postcava
			& vein & mDSC
			\\ \midrule
			SegResNet~\citep{siddiquee2021redundancy} & 93.11 & 89.92 & 87.84 & 74.62 & 95.37 & 87.90 & 76.33 & 84.05 & 79.36 & 57.13 & 82.56
			\\
			nnFormer~\citep{zhou2021nnformer} & 86.71 & 87.03 & 84.28 & 63.37 & 91.64 & 73.18 & 71.88 & 84.73 & 78.61 & 55.31 & 77.67
			\\
			UNesT~\citep{yu2022unest} & 93.82 & 90.42 & 89.04 & 76.40 & 95.30 & 89.65 & 78.97 & 84.36 & 79.61 & 59.70 & 83.73
			\\
			TransBTS~\citep{wang2021transbts} & 85.47 & 81.58 & 82.00 & 60.58 & 92.50 & 72.29 & 63.25 & 83.47 & 75.07 & 55.38 & 75.16
			\\
			TransUNet~\citep{chen2021transunet} & 94.63 & 89.86 & 89.61 & 77.28 & 95.85 & 88.95 & 79.98 & 85.06 & \textbf{81.02} & \textbf{59.76} & 84.20
			\\
			UNETR~\citep{hatamizadeh2022unetr} & 91.89 & 89.07 & 87.60 & 66.97 & 91.48 & 83.18 & 70.56 & 82.92 & 75.20 & 57.53 & 79.64
			\\
			Swin UNETR~\citep{tang2022self} & 92.23 & 84.34 & 82.95 & 74.06 & 94.91 & 82.28 & 71.17 & \textbf{85.50} & 79.18 & 55.11 & 80.17
			\\
			\midrule
			\ourmodel\ & \textbf{93.94} & \textbf{91.53} & \textbf{90.21} & \textbf{84.15} & \textbf{96.25} & \textbf{92.51} & \textbf{82.72} & 77.35 & 79.64 & 57.10 & \cellcolor{significant!40} \textbf{84.54}
			\\ \bottomrule
		\end{tabular}%
		\label{tab:generalizability}
	\end{table*}

	\subsection{Organ Segmentation Benchmark}
	\label{sec:strong_challenge_ranking}
	
	We secured the top \#1 solution in two prominent medical segmentation challenges: Medical Segmentation Decathlon (MSD)\footnote{\href{https://decathlon-10.grand-challenge.org/evaluation/challenge/leaderboard/}{decathlon-10.grand-challenge.org/evaluation/challenge/leaderboard/}} and Beyond The Cranial Vault (BTCV), surpassing the runners-up by a significant margin. \major{Notably, our \ourmodel\ demonstrates exceptional performance by providing solutions for six CT tasks, while predicting the results for four MRI tasks using open-sourced nnU-Net \citep{isensee2021nnu}.}
	
	\subsubsection{Medical Segmentation Decathlon (MSD)}
	\label{sec:benchmark_msd}
	
	To assess the effectiveness of our approach, we conducted a comprehensive evaluation of both the official test set and a 5-fold cross-validation on the Medical Segmentation Decathlon (MSD) dataset. Detailed comparisons of our model's performance across various metrics are presented in Table \ref{tab:msd_test}, providing a comprehensive overview of its strengths. \major{Specifically, for the liver, pancreas, hepatic vessel, and spleen organs, \ourmodel\ achieved DSC scores of 95.42\%, 82.84\%, 67.15\%, and 97.27\%, respectively. Furthermore, \ourmodel\ demonstrated superior performance in the segmentation of liver tumors, pancreas tumors, hepatic vessel tumors, lung tumors, and colon tumors, with DSC scores of 79.35\%, 62.33\%, 75.86\%, 80.01\%, and 63.14\%, respectively.}
	
	\subsubsection{Beyond The Cranial Vault (BTCV)}
	\label{sec:benchmark_btcv}
	
	\major{Table \ref{tab:btcv_benchmark} presents a comparison between \ourmodel\ and other methods under dataset-specific training and dataset-agnostic training on the Biomedical Translational Clinical Vision (BTCV) benchmark. \ourmodel\ surpassed the performance of dataset-specific training methods by a minimum of 3.5\% across various evaluation metrics. Furthermore, it outperformed dataset-agnostic training approaches, achieving an average Dice score superiority of 0.38\%.} Overall, \ourmodel\ consistently exhibits strong performance on both the MSD and BTCV benchmarks outperforming the performance of alternative methods. These results affirm the robustness and effectiveness of \ourmodel\ for organ segmentation across multiple tasks.
	
	\major{
		\subsubsection{Multi-Organ and Tumor Segmentation (MOTS)}
		\label{sec:benchmark_mots}
		To facilitate an equitable comparison of \ourmodel~with other partial label learning methods, we utilized the Multi-Organ and Tumor Segmentation (MOTS) dataset \citep{zhang2021dodnet} as a benchmark. The MOTS dataset comprises seven partially labeled sub-datasets, encompassing seven organ and tumor segmentation tasks. It consists of 1155 CT volumes, with 920 volumes designated for the training set and 235 for the test set. We conducted separate training procedures for \ourmodel~ on the MOTS training set and the 14 integrated datasets in this work (excluding the 235 CT volumes present in the MOTS test set). The results presented in Table~\ref{tab:mots} demonstrate the effectiveness of our proposed framework, achieving the second-best performance compared to other state-of-the-art methods. Moreover, when trained on a larger integrated dataset, \ourmodel~exhibited improved performance for most organs and tumors, validating the efficacy of training models on extensive integrated datasets.
	}

	\begin{figure*}[t]
		\centerline{\includegraphics[width=\linewidth]{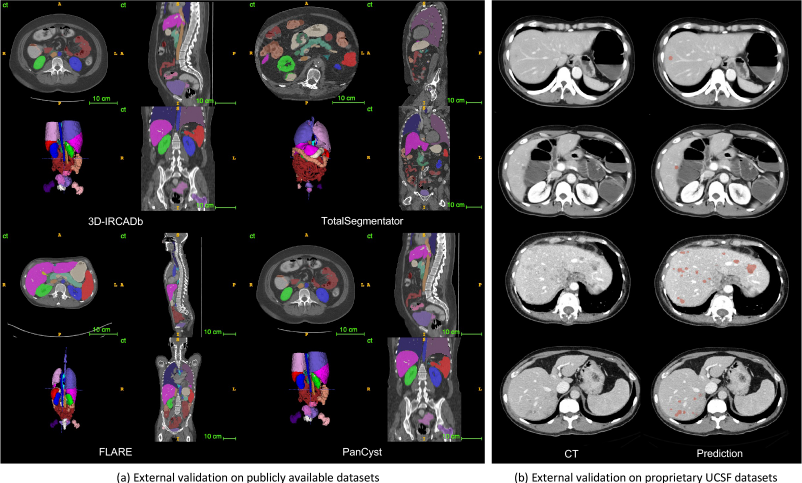}}
		\caption{
			\textit{(a) Pseudo-label visualization.} 25 organs and 6 tumors in four unseen datasets are visualized. \major{\textit{(b) External validation for liver tumor detection.} In Cases 1 and 2, \ourmodel\ successfully identified small new liver tumors, that have been overlooked during radiological evaluation. In Cases 3 and 4, where multiple liver tumors were present, \ourmodel\ detected them, resulting in improved diagnostic efficiency.} 
		}
		\label{fig:fig_generalization}
	\end{figure*}

	\subsection{Tumor Detection on Five Datasets}
	\label{sec:high_specificity}
	Usually, tumor detection task solely relies on DSC scores, which may not accurately reflect the performance. DSC scores are typically calculated only on abnormal CT volumes that contain tumors, which can lead to an overestimation of performance~\citep{isensee2021nnu,xia2022felix}. The presence of normal CT volumes without tumors may result in numerous false positives generated by the AI system~\citep{shen2021artificial}. To address this limitation, we evaluate the Sensitivity and Specificity at the patient level for the detection of three types of tumors. The harmonic mean of sensitivity and specificity is reported to provide a balanced assessment of the model's abilities. To obtain normal CT volumes without tumors for evaluation, we utilize the CHAOS and Pancreas-CT datasets, which provide pathological verification of the absence of tumors~\citep{valindria2018multi,roth2015deeporgan}. \tableautorefname~\ref{tab:high_specificity} show that \ourmodel\ achieves a harmonic mean of 91.84\% for liver tumors, 93.31\% for kidney tumors, and 92.59\% for pancreatic tumors, indicate that the proposed \ourmodel\ can accurately identify tumor cases while reducing false positives, highlighting its clinical significance.
	
	Moreover, we visualize several CT volumes from small to large size tumors in ~\figureautorefname~\ref{fig:qualitative_visualization} to demonstrate the superior capacity of \ourmodel. The visualization showcases the proficiency of \ourmodel\ in detecting both small and large size pancreatic tumors, while other methods may either ignore them or only detect a portion of the tumor. In addition, row presents a CT volume from another hospital, where \ourmodel\ demonstrates excellent generalization in pancreas organ segmentation and avoids generating false positives for tumors. This emphasizes the effectiveness of \ourmodel\ in diverse scenarios. Compared with dataset-specific models, the smaller number of false positives predicted by our \ourmodel\ underlines the necessity of assembling diverse datasets, benefiting from not only sufficient positive examples for training but also a larger number of negative examples as a control. 
	
	Overall, \ourmodel\ demonstrates superior performance in terms of specificity and harmonic mean compared to the evaluated methods, indicating its ability to achieve a balanced performance in tumor detection across different organs.

	\begin{table*}[h]
		\caption%
		{\textit{Transferability: Fine-tuning performance.} Comparing with MedicalNet~\citep{chen2019med3d}, Models Gen.~\citep{zhou2019models}, Swin UNETR~\citep{tang2022self}, UniMiSS~\citep{xie2022unimiss}, fine-tuning \ourmodel\ significantly outperforms learning from scratch on several downstream tasks (\ie, vertebrae, cardiac, muscles, organs and sub-classes of pancreatic tumor segmentation). Moreover, \ourmodel, trained by image segmentation as proxy task, can extract better visual representation---more related to segmentation tasks---than other pre-trained models developed in the medical domain. The average performance is statistically significant at the P=0.05 level, with highlighting in a \significant{light red} box.}\vspace{2px}
		\centering
		\scriptsize
		\begin{tabular}{p{0.1\linewidth}|P{0.05\linewidth}P{0.05\linewidth}P{0.05\linewidth}P{0.05\linewidth}P{0.05\linewidth}|P{0.05\linewidth}P{0.05\linewidth}P{0.05\linewidth}|P{0.05\linewidth}P{0.05\linewidth}P{0.05\linewidth}P{0.05\linewidth}}
			\toprule
			Method & Total\_vert. & Total\_card. & Total\_mus. & Total\_org. & Total\_ave & JHH\_card. & JHH\_org. & JHH\_ave & PDAC & Cyst & PanNet & Pancre\_ave
			\\ \midrule
			Scratch & 81.06&84.47&88.83&86.42 & 85.20
			& 71.63 & 89.08 & 80.36
			& 53.1 & 41.2 & 34.6 & 42.97
			\\
			MedicalNet&82.28&87.40&91.36&86.90 & 86.99
			& 58.07 & 77.68 & 67.88
			& N/A & N/A & N/A & N/A
			\\
			ModelsGen.&85.12&86.51&89.96&85.78 & 86.84
			& \textbf{74.25} & 88.64 & 81.45
			& N/A & N/A & N/A & N/A
			\\
			Swin UNETR&86.23&87.91&92.39&88.56 &88.77
			& 67.85 & 87.21 & 77.53
			& 53.4 & 41.6 & 35.4 & 43.47
			\\
			UniMiSS&85.12&88.96&92.86&88.51&88.86 
			& 69.33 & 82.53 & 75.93
			& N/A & N/A & N/A & N/A
			\\
			\midrule
			Ours &  \textbf{86.49} &  \textbf{89.57} & \textbf{94.43} &  \textbf{88.95} & \cellcolor{significant!40} \textbf{89.86} 
			& 72.06 & \textbf{89.37}  &\cellcolor{significant!40} \textbf{80.72}
			& \textbf{53.6} & \textbf{49.2} & \textbf{45.7} &\cellcolor{significant!40} \textbf{49.5}
			\\
			\bottomrule
		\end{tabular}%
		\label{tab:transfer_learning}
		\vspace{-0.2cm}
	\end{table*}

	\subsection{Ablation Study}
	\label{sec:ablation_study}
	
	\subsubsection{Language Encoding}
	In order to assess the effectiveness of language encoding, we conducted an experiment using different language models and varying prompt templates as shown in \tableautorefname~\ref{tab:clip_ablation}. The results indicate that the CLIP text encoder equipped with the prompt template ``A computerized tomography of a [CLS],'' achieves the highest performance with an average Dice score of 76.11\%. It is worth noting that other language models and prompt templates also outperformed the one-hot encoding technique, which provides compelling evidence regarding the efficacy of language encoding in our study.
	
	\subsubsection{Different Backbones}
	\ourmodel\ framework can be applied flexibly to other backbones. We conduct experiments in two CNN-based backbones, namely\ U-Net~\citep{ronneberger2015u} and SegResNet~\citep{myronenko20193d}. Over 25 different organs, the framework achieved an average Dice Similarity Coefficient (DSC) score of 79.65\% with U-Net and 78.30\% with SegResNet. Furthermore, for 6 different tumor types, the framework achieved an average DSC score of 67.95\% with U-Net and 42.20\% with SegResNet. These scores are comparable to the performance of the Swin UNETR backbone with the average dice score of 79.56\% and 65.79\% for organs and tumors as shown in \figureautorefname~\ref{fig:expansibility}. The results highlight the versatility and effectiveness of the \ourmodel\, as it demonstrates its ability to be expanded to different backbone architectures. The comparable performance achieved by these backbones further supports the flexibility and adaptability of \ourmodel\ in various medical imaging tasks. In the future, more advanced backbones will be added over time in our GitHub repository.
	
	\begin{figure*}[t]
		\centerline{\includegraphics[width=1\linewidth]{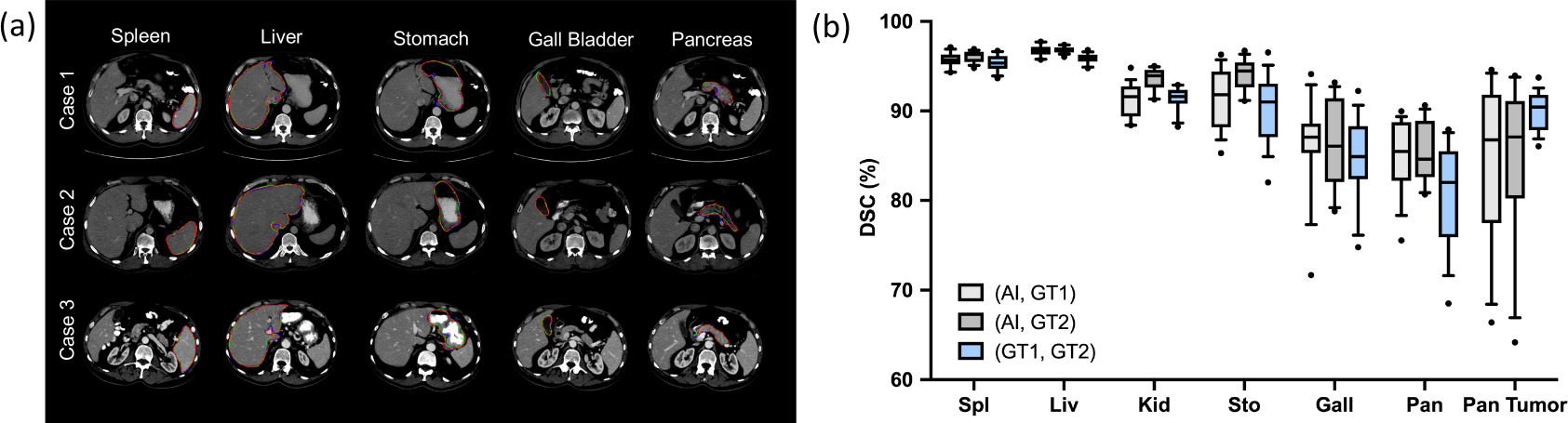}}
		\caption{
			\major{
			\textit{\textbf{(a) Contour line comparison among pseudo labels and two human experts.}} The \textcolor{red}{red} line represents the annotation from Doctor 1; \textcolor{green}{green} line indicates the annotation from Doctor 2; \textcolor{blue}{blue} line shows the results generated by \ourmodel. Examples of CT volumes annotated by our pseudo labels and two human experts with contour line comparison. The prediction results of these organs generated by the medical model are comparable with human experts.
			\textbf{\textit{(b) Intra-observer variability.}} We obtain similar performance between pseudo labels generated by \ourmodel\ (AI) and annotations performed by two human experts (Dr1,2) on 6 organs. Spleen (Spl), liver (Liv), kidneys (Kid), stomach (Sto), gallbladder (Gall), and pancreas (Pan) can be annotated by AI with a similar intra-observer variability to humans.}
		}
		\label{fig:pseudo_truth_evaluation}
	\end{figure*}

	\subsection{Generalization for External Datasets}
	\label{sec:generalizability}
	The generalizability of medical AI models is a crucial expectation, as they should perform well on new data from various hospitals, rather than being tailored to a single dataset \citep{mongan2020checklist, norgeot2020minimum, guo2021semantic}. In comparison to dataset-specific models, \ourmodel\ was trained on a significantly larger and more diverse collection of CT volumes. This broader training enables \ourmodel\ to demonstrate superior generalizability, as it can be directly tested on external data without the need for adaptation or fine-tuning. We conducted an assessment using both a public dataset, 3D-IRCADb, and a private dataset, JHH. These datasets were not included in the training phase and can be considered as external validation. The results, presented in \tableautorefname~\ref{tab:generalizability}, demonstrate that \ourmodel\ outperforms previous methods on both 3D-IRCADb and JHH, with a substantial improvement in DSC similarity coefficient (DSC) of 5\% and 4\%, respectively. 
	
	Additionally, we conducted a pseudo-label visualization of 25 organs in four unseen datasets, including 3D-IRCADb, TotalSegmentator, FLARE, and PanCyst. This visualization, depicted in Figure \ref{fig:fig_generalization} (a), provides further evidence of the performance on these unseen datasets achieved by \ourmodel. \major{For tumor detection, we opted for the private dataset from UCSF for external validation. \ourmodel\ served as an adjunctive tool in radiologist diagnostic practices. Upon evaluation by a radiologist, \ourmodel\ significantly reduced the false negative rate of liver tumors and improved diagnostic efficiency, particularly in the following clinical scenarios: 1. Detection of new small liver lesions, which are often overlooked but critical for initial tumor staging. 2. Assessment of changes in multiple tumors, which is a time-consuming and tedious task but essential for tumor prognosis. This visualization, depicted in Figure \ref{fig:fig_generalization} (b), illustrates the preliminary results of liver tumor detection by \ourmodel.}
	
	\subsection{Fine-tuning Results for Other Tasks}
	\label{sec:transfer_learning}
	\ourmodel\ possesses another significant characteristic, as it serves as a robust pre-training model for other tasks \citep{tang2024efficient}. By undergoing pre-training using an assembled dataset and subsequent fine-tuning on other datasets, \ourmodel\ achieves the highest DSC score compared to other pre-training methods. Specifically, in the TotalSegmentator dataset, \ourmodel\ achieves DSC scores of 86.49\%, 89.57\%, 94.43\%, and 88.95\% for four tasks, as presented in Table \ref{tab:transfer_learning}. This remarkable performance highlights the potential of \ourmodel\ in enhancing the generalization of medical imaging models by effectively capturing fine-grained information for segmentation purposes.
	\major{
		In addition, we investigated the cross-task ability of our model by transferring to the fine-grained segmentation of pancreatic tumors. We employed our proprietary dataset, which comprises 3,577 annotated pancreatic tumors, including detailed sub-types: 1,704 PDAC, 945 cysts, and 928 PanNets. The results shown in Table~\ref{tab:transfer_learning} also demonstrate that our model can transfer better to novel classes than self-supervised models, even in the more challenging sub-class tumor segmentation task.}

	\subsection{Comparing AI Predictions with Human Annotations}
	\label{sec:human}
	To compare the performance of AI model and human experts, we randomly select 17 CT volumes in BTCV. Two independent groups of radiologists from different institutes were involved in annotating six organs. The quality of the pseudo labels predicted by \ourmodel\ and the manual annotations performed by human experts were assessed. Figure \ref{fig:pseudo_truth_evaluation} (a) illustrates that the organ boundaries from both human experts and \ourmodel\ exhibit a high degree of alignment. 
	Additionally, it is worth noting that manual annotations performed by human experts exhibit inter-observer variance~\citep{ji2021learning}, particularly in segmentation tasks, which is attributed to the inherent blurriness and ambiguity of some organ boundaries. 
	We further calculate the mutual DSC scores (DSC) for six organs and one tumor between the AI prediction and the two human annotations (GT1 and GT2), denoted as AI$\leftrightarrow$GT1, AI$\leftrightarrow$GT2, and GT1$\leftrightarrow$GT2, in figure \ref{fig:pseudo_truth_evaluation} (b). \major{An intriguing observation was that the DSC for six organs between the AI model and human expert was slightly higher than the DSC among the human experts themselves. Notably, the AI model also achieved a smaller variance. However, the variance of the DSC for tumor segmentation between the AI model and human experts was substantially larger than that observed between the human experts, indicating that the model's tumor prediction capability still lags behind human expert proficiency. These findings underscore the model's proficiency in segmenting specific anatomical structures while concurrently underscoring the need for further refinements from human doctors to enhance its tumor segmentation during daily clinical application.}
	
	Furthermore, given that the six organs can be segmented by the AI model with a similar variance to human experts, it is encouraged for the research community to focus on creating annotations for more challenging organs and tumors.
	
	\section{Experiment II: Continual Learning}

    \subsection{Experimental Settings}
	\subsubsection{Datasets}
	In this study, we conducted two benchmark tests to evaluate the extensibility of the Continual \ourmodel. 
	
	\smallskip\noindent\textbf{Tumor Extension Benchmark.} The first benchmark setting was designed to simulate the tumor extension scenario encountered in clinical routine, and it utilized the BTCV dataset and LiTS dataset. The BTCV dataset consisted of 47 abdominal CT images that provided delineation for 13 different organs. Specifically, we utilized 13 classes from the BTCV dataset, namely spleen, right kidney, left kidney, gall bladder, esophagus, liver, stomach, aorta, inferior vena cava, portal vein and splenic vein, pancreas, right adrenal gland, and left adrenal gland, during the step 1 learning. For the step 2 learning, we employed the LiTS dataset, which comprised 130 contrast-enhanced abdominal CT volumes specifically for liver and liver tumor segmentation. In this stage, we focused on learning the liver tumor class from the LiTS dataset.
	
	\smallskip\noindent\textbf{Body Extension Benchamrk.} The second benchmark setting aimed to simulate extension scenarios involving different body regions, and it utilized the in-house JHH dataset. The JHH dataset contained annotated data for multiple classes, which we categorized into three groups: abdominal organs, gastrointestinal tract, and cardiovascular system. In the abdominal organs category, we selected seven classes for the step 1 learning, including spleen, right kidney, left kidney, gall bladder, liver, postcava, and pancreas. In the gastrointestinal tract category, three classes (stomach, colon, and intestine) were learned during the step 2 learning. Finally, for the step 3 learning, we focused on learning four classes from the cardiovascular system, namely aorta, portal vein and splenic vein, celiac truck, and superior mesenteric artery. This categorization was performed in accordance with the TotalSegmentator~\citep{wasserthal2022totalsegmentator}.

	\begin{table*}[!h]
		\centering 
		\scriptsize
		\caption{\textit{Results in tumor extension benchmark.} We present the DSC score of each class in two continual learning steps. The Organ Mean represents the average DSC score for 13 organs, and Mean stands for the average score for 14 classes. The average performance is statistically significant at the P=0.05 level, with highlighting in a \significant{light red} box.}\vspace{2px}
		\begin{tabular}{p{0.05\linewidth}|P{0.035\linewidth}P{0.035\linewidth}|P{0.035\linewidth}P{0.035\linewidth}|P{0.035\linewidth}P{0.035\linewidth}|P{0.035\linewidth}P{0.035\linewidth}|P{0.035\linewidth}P{0.035\linewidth}|P{0.035\linewidth}P{0.035\linewidth}|P{0.035\linewidth}P{0.035\linewidth}|P{0.035\linewidth}P{0.035\linewidth}}
			\toprule
			& \multicolumn{2}{c|}{Spleen} & \multicolumn{2}{c|}{R kidney}  & \multicolumn{2}{c|}{L kidney}  & \multicolumn{2}{c|}{Gall Bladder}  & \multicolumn{2}{c|}{Esophagus}  & \multicolumn{2}{c|}{Liver}  & \multicolumn{2}{c|}{Stomach}  & \multicolumn{2}{c}{Aorta}\\
			& Step 1 & Step 2& Step 1 & Step 2&  Step 1 & Step 2& Step 1 & Step 2& Step 1 & Step 2&Step 1 & Step 2&Step 1 & Step 2&Step 1 & Step 2
			\\ \midrule
			LwF & \multirow{3}{*}{0.955}  &0.941 &\multirow{3}{*}{0.819}& 0.745 &\multirow{3}{*}{0.927}& 0.845 &\multirow{3}{*}{0.844}& 0.796 &\multirow{3}{*}{0.718}& 0.690 &\multirow{3}{*}{0.969}& 0.935 &\multirow{3}{*}{0.896}& 0.852 &\multirow{3}{*}{0.801}& 0.847 \\
			ILT & & \textbf{0.950} & & 0.687 & & \textbf{0.913} & & 0.750 & & \textbf{0.692} & & \textbf{0.960} & & 0.857&& 0.774 \\
			PLOP&  & 0.942 & & 0.778 & & 0.908 & & \textbf{0.823} & & 0.690 & & 0.959 & & 0.883&& 0.803 \\ \midrule
			Ours &0.952 & 0.941 & 0.917 & \textbf{0.860} & 0.922 & 0.872 & 0.840 & 0.728 &0.720 & 0.690 & 0.966& 0.955  & 0.886 & \textbf{0.862}& 0.903& \textbf{0.909} \\
			\bottomrule
		\end{tabular}
		\vspace{0.5 em}\\
		\begin{tabular}{p{0.05\linewidth}|P{0.035\linewidth}P{0.035\linewidth}|P{0.035\linewidth}P{0.035\linewidth}|P{0.035\linewidth}P{0.035\linewidth}|P{0.035\linewidth}P{0.035\linewidth}|P{0.035\linewidth}P{0.035\linewidth}|P{0.035\linewidth}P{0.035\linewidth}|P{0.035\linewidth}P{0.035\linewidth}|P{0.035\linewidth}P{0.035\linewidth}}
			&\multicolumn{2}{c|}{IVC}  & \multicolumn{2}{c|}{Veins}  & \multicolumn{2}{c|}{Pancreas}  & \multicolumn{2}{c|}{R gland}  & \multicolumn{2}{c|}{L gland}  & \multicolumn{2}{c|}{Liver Tumor} &\multicolumn{2}{c|}{Organ Mean} &\multicolumn{2}{c}{Mean} \\
			& Step 1 & Step 2& Step 1 & Step 2&  Step 1 & Step 2& Step 1 & Step 2& Step 1 & Step 2&Step 1 & Step 2&Step 1 & Step 2&Step 1 & Step 2
			\\ \midrule
			LwF & \multirow{3}{*}{0.877}& 0.784& \multirow{3}{*}{0.735} &0.628 & \multirow{3}{*}{0.826}& 0.812 & \multirow{3}{*}{0.743}& 0.671 & \multirow{3}{*}{0.654}& 0.468 & \multirow{3}{*}{N/A} & 0.456 & \multirow{3}{*}{0.828}& 0.770 & \multirow{3}{*}{N/A} & 0.748  \\
			ILT  & & 0.849 & & \textbf{0.757} & & 0.812 & & 0.692 & & 0.527 && 0.335 && 0.786  && 0.754\\
			PLOP & & 0.819 & & 0.634 & & \textbf{0.822} & & \textbf{0.708} & & 0.612 && 0.362 && 0.799 && 0.768 \\ \hline
			Ours  &  0.902& \textbf{0.872} &0.812 & 0.742  & 0.845& 0.808  & 0.755& 0.703 & 0.758 & \textbf{0.683} &N/A & \textbf{0.466} & 0.860 & \textbf{0.817} &N/A & \cellcolor{significant!40} \textbf{0.792} \\
			\midrule
		\end{tabular}
		\label{tab:public_incremental_baseline}
	\end{table*}
	
	\subsubsection{Baselines}
	For a fair comparison, all the compared methods use the same Swin UNETR~\citep{hatamizadeh2022swin} as the backbone network, the state-of-the-art model in many medical image segmentation tasks. We compare with three popular continual learning baseline methods that apply knowledge distillation, including LwF~\citep{li2017learning}, ILT~\citep{michieli2019incremental} and PLOP~\citep{douillard2021plop}.

	\subsubsection{Implementation Details}
	The data augmentation, network structures, and evaluation metrics employed in this experiment are consistent with \S \ref{sec:integrated}.
	
	\smallskip\noindent\textbf{Optimization.} We train the proposed network architecture on new classes while utilizing pseudo labeling for the old classes. No additional distillation techniques are employed. The image-aware organ-specific heads, designed to be lightweight, are an integral part of our architecture. Each head comprises three convolution layers, with the first two layers having 8 kernels and the last layer having 1 kernel. For all compared models, we employ the AdamW optimizer and train them for 100 epochs using a cosine learning rate scheduler. During training, we utilize a batch size of 2 and a patch size of 96 × 96 × 96. The initial learning rate is set to $1e^{-4}$, and the weight decay is set to $1e^{-5}$. We conduct our experiments using MONAI version 1.1.0. The training process is performed on NVIDIA TITAN RTX and Quadro RTX 8000 GPUs.
	
	\subsection{Results for Extension to Novel Classes}
	The continual segmentation results for the tumor extension benchmark and body extension benchmark are presented in Tables \ref{tab:public_incremental_baseline} and \ref{tab:felix_incremental_baseline}, respectively. Notably, our model achieves the highest DSC score of 81.7\% and 46.6\% for all organs and tumors during step 2, as shown in Table \ref{tab:public_incremental_baseline}. Furthermore, our model outperforms other methods in the body extension benchmark, indicating its superior ability to adapt to new data and classes while minimizing forgetting of previously learned classes.
	
\begin{figure}[t]
		\centerline{\includegraphics[width=1\columnwidth]{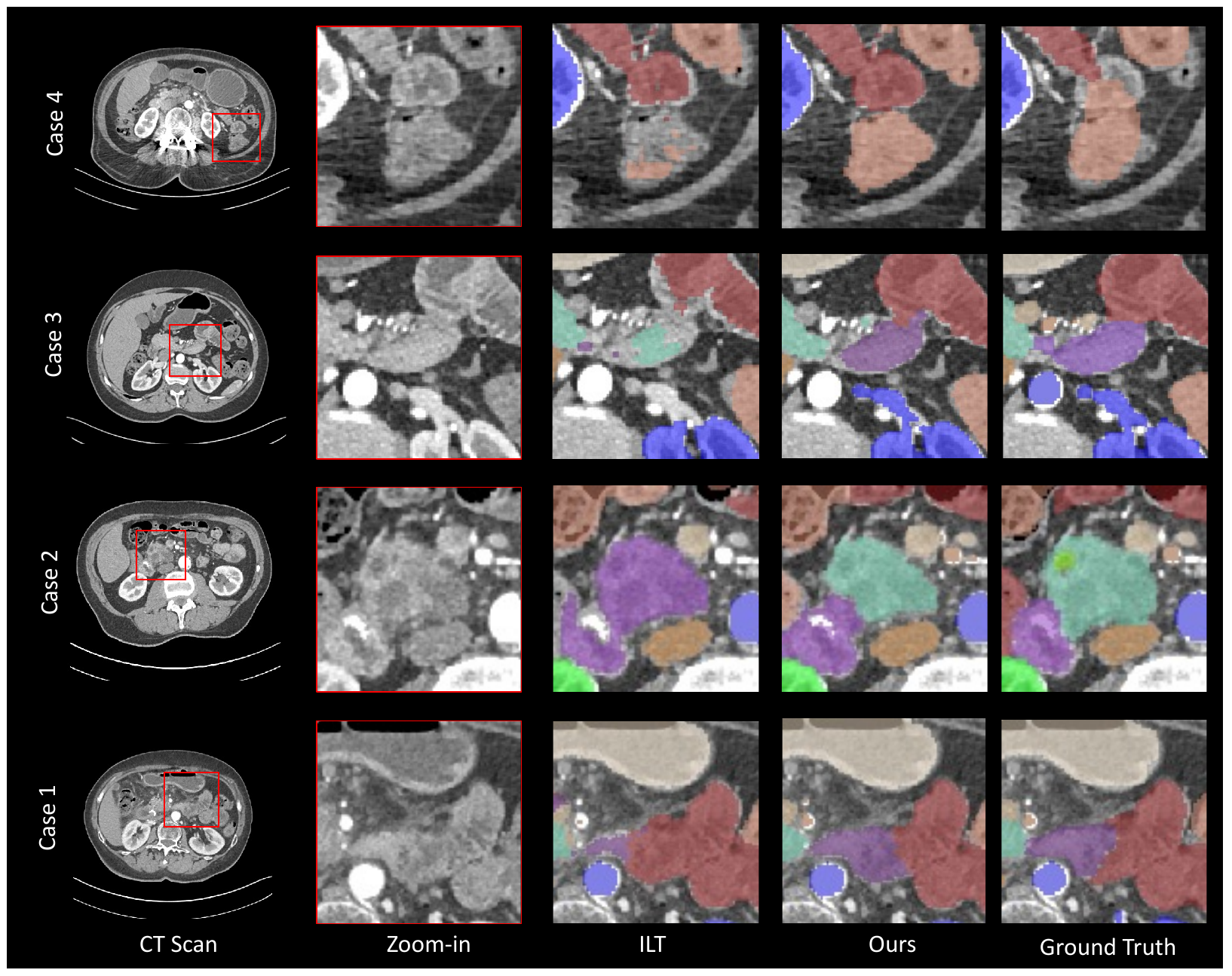}}
		\caption{
			\textit{The visualization for body extension benchmark.} We compare Continual \ourmodel\ with baseline model ILT across four case studies. Our model demonstrates effective adaptation to new organs while retaining knowledge of previously learned classes, thereby avoiding forgetting.
		}
		\label{fig:visual}
	\end{figure}

	To provide a qualitative assessment, we display the segmentation results of our proposed method alongside the best baseline method, ILT, for the body extension benchmark in Figure~\ref{fig:visual}. For instance, in case 2, ILT exhibits confusion between the stomach and pancreas. In contrast, our model accurately distinguishes these two organs due to the independent prediction enabled by the proposed LPG and CSH techniques, which prevent contradictions between different organs. The visualization effectively demonstrates that our method consistently achieves correct organ segmentation.
	
	\begin{table}[t]
		\scriptsize
		\centering
		\caption{\textit{Results in body extension benchmark and ablation study.} We present the average DSC score for three body parts in each step separately. The Ours\_1-hot replace the language embedding with one-hot embedding.}\vspace{2px}
		\begin{tabular}{p{0.14\linewidth}|P{0.06\linewidth}P{0.06\linewidth}P{0.08\linewidth}|P{0.08\linewidth}P{0.06\linewidth}|P{0.18\linewidth}}
			\toprule
			\multirowcell{2}[0pt][l]{Method} & \multicolumn{3}{c|}{JHH\_organ (7)} & \multicolumn{2}{c|}{JHH\_gastro (3)} & JHH\_cardiac (4) \\
			& Step~1 & Step~2 & Step~3 & Step~2 & Step~3 & Step~3 \\
			\midrule
			LwF & \textbf{0.891} & 0.777 & 0.767 & 0.530 & 0.486 & 0.360\\
			ILT & \textbf{0.891} & 0.775 & 0.776 & 0.653 & 0.480 & 0.484 \\
			PLOP & \textbf{0.891} & 0.780 & 0.777 & 0.427 & 0.464 & 0.318 \\
			\midrule
			Ours (1-hot) & 0.882 & 0.767 & 0.777 & 0.674 & 0.665 & 0.452 \\
			Ours (CLIP) & 0.887 & \textbf{0.783} & \textbf{0.787} & \textbf{0.695} & \textbf{0.692} & \textbf{0.636}\\
			\bottomrule
		\end{tabular}
		\label{tab:felix_incremental_baseline}
	\end{table}

	\subsection{Ablation Study}
	To assess the efficacy of the proposed model designs, we conducted an ablation study on the JHH dataset, as presented in Table \ref{tab:felix_incremental_baseline}. Specifically, we examined the performance impact of the organ-specific segmentation heads and the CLIP text embeddings. The first row in Table \ref{tab:felix_incremental_baseline} corresponds to the performance of the baseline Swin UNETR model trained with pseudo labeling (LwF). The fourth row (Ours\_1-hot) introduces the organ-specific segmentation heads but utilizes one-hot embeddings for each organ instead of the CLIP text embeddings. The third row represents the performance of the full method, incorporating both the organ-specific segmentation heads and the CLIP text embeddings. The results demonstrate a substantial improvement by adopting organ-specific heads as segmentation outputs. For instance, in step 2, we observe an improvement of 14.4\% and a remarkable 17.9\% enhancement in step 3 for gastrointestinal tracts. Furthermore, the utilization of CLIP text embeddings leads to further performance gains, with a margin of 18.4\% improvement in step 3 for the cardiovascular system. These findings validate the effectiveness of the proposed organ-specific segmentation heads and CLIP text embeddings in the continual organ segmentation task.
	
	\begin{figure}[t]
		\centerline{\includegraphics[width=\linewidth]{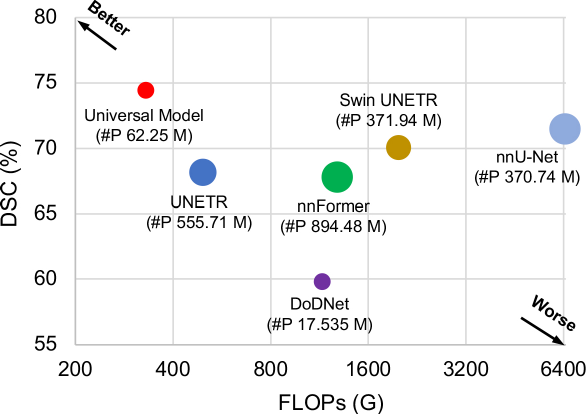}}
		\caption{
			\textit{Inference speed with FLOPs vs. DSC.} The plot illustrates the relationship between the average DSC similarity coefficient (DSC) score across the six Medical Segmentation Decathlon (MSD) tasks and the floating-point operations per second (FLOPs). The FLOPs are calculated based on input with a spatial size of $96 \times 96 \times 96$. The size of each circle in the plot corresponds to the number of parameters (`\#P'). \major{The used backbone of Universal Model is Swin UNETR.}
		}
		\label{fig:computational_efficiency}
	\end{figure}

	\subsection{Computational Complexity Analysis}
	Our work significantly reduces computational complexity in continual learning for segmentation. \citet{ji2023continual} introduced a state-of-the-art architecture for medical continual semantic segmentation that uses a pre-trained, frozen encoder with incrementally added decoders at each learning step. This approach, however, leads to substantial computational complexity in subsequent steps. For example, the Swin UNETR's decoder alone requires 466.08 GFLOPs, adding this amount with each new learning step.
	In contrast, our model adds only a few image-aware, organ-specific heads for new classes, each consuming just $9.44 \times 10^{-4}$ GFLOPs. Consequently, the computational complexity of our model remains nearly constant throughout the continual learning process, whereas Ji~\etal experiences linear growth in computational complexity with each step.
	
	Moreover, we also assessed the inference speed of various models, as this is crucial for clinical applications~\citep{chen2019augmented,esteva2021deep}. We utilized floating-point operations per second (FLOPS) as a metric to measure the inference speed. \figureautorefname~\ref{fig:computational_efficiency} presents a speed-performance plot, demonstrating that our \ourmodel\ achieves computational efficiency surpassing that of dataset-specific models, with a speed improvement of over 6 times. Remarkably, this efficiency is achieved while maintaining the highest average Dice similarity coefficient (DSC) score of 74\%.

	\section{Conclusion and Discussion}
	\label{sec:conclusion}
	In this work, we present a novel framework for multi-organ segmentation and tumor detection. This work integrates language embedding with segmentation models to enable a flexible and powerful segmentor, allowing the model to learn from an integrated dataset for high-performing organ segmentation and tumor detection, ranking first in both MSD and BTCV. More importantly, we demonstrate that language embedding can establish the anatomical relationship. Finally, the experiment results validate several clinically important merits of the CLIP-Driven Universal Model, such as compelling efficiency, generalizability, transferability, and extensibility. 
	
	While the universal model framework proposed in this study is primarily designed and evaluated using CT volumes, the general principles and techniques introduced herein could potentially be extended to other medical imaging modalities, such as magnetic resonance imaging (MRI) and ultrasound, with appropriate adaptations. MRI and ultrasound imaging exhibit unique characteristics, including bias fields, speckle noise, low contrast, and anisotropic resolution, which would necessitate specific preprocessing techniques and data augmentation strategies tailored to these modalities. Moreover, the backbone of our model is flexible and capable of employing any 2D or 3D network architecture, rendering it suitable for other medical imaging modalities. Furthermore, the language-driven parameter generator (LPG) and class-specific segment heads (CSH) leverage language embeddings from large language models like CLIP, which have been trained on diverse data, including medical images and text. Consequently, these language models could potentially capture semantic relationships between organs and abnormalities across various medical imaging modalities.
 
    Besides, there are some limitations of current works. We also propose some potential future directions for each limitation. 1) Multi-modal Integration: The dataset primarily includes abdominal CT volumes. Future models could benefit from integrating diverse imaging modalities like CT, MRI, and ultrasound into a unified framework. This integration could enhance the model's robustness and provide more comprehensive anatomical information for various clinical applications. 2) Language Embedding: The language embeddings used are derived from pre-trained models like CLIP, which are not optimized for the medical domain. Developing domain-specific language models tailored for medical imaging tasks could improve semantic encoding and overall model performance. 3) Unlabeled Region Utilization: Semi-supervised learning techniques, such as consistency loss and pseudo-labeling, could be incorporated to leverage hidden information in unlabeled regions. Integrating these techniques could improve the model's ability to learn from both labeled and unlabeled data, thereby enhancing its generalization capabilities.

	\section*{Acknowledgments}
	
	This work was supported by the Lustgarten Foundation for Pancreatic Cancer Research, the Patrick J. McGovern Foundation Award, and the National Natural Science Foundation of China (62001410). We thank Ali Hatamizadeh, Jieneng Chen, Junfei Xiao, Yongyi Lu, and Wenxuan Li for their constructive suggestions at several stages of the project.
	
	\bibliographystyle{model2-names.bst}\biboptions{authoryear}
	\bibliography{refs,zzhou}

\begin{thebibliography}{129}
\expandafter\ifx\csname natexlab\endcsname\relax\def\natexlab#1{#1}\fi
\providecommand{\url}[1]{\texttt{#1}}
\providecommand{\href}[2]{#2}
\providecommand{\path}[1]{#1}
\providecommand{\DOIprefix}{doi:}
\providecommand{\ArXivprefix}{arXiv:}
\providecommand{\URLprefix}{URL: }
\providecommand{\Pubmedprefix}{pmid:}
\providecommand{\doi}[1]{\href{http://dx.doi.org/#1}{\path{#1}}}
\providecommand{\Pubmed}[1]{\href{pmid:#1}{\path{#1}}}
\providecommand{\bibinfo}[2]{#2}
\ifx\xfnm\relax \def\xfnm[#1]{\unskip,\space#1}\fi
\bibitem[{Antonelli et~al.(2021)Antonelli, Reinke, Bakas, Farahani, Landman, Litjens, Menze, Ronneberger, Summers, van Ginneken et~al.}]{antonelli2021medical}
\bibinfo{author}{Antonelli, M.}, \bibinfo{author}{Reinke, A.}, \bibinfo{author}{Bakas, S.}, \bibinfo{author}{Farahani, K.}, \bibinfo{author}{Landman, B.A.}, \bibinfo{author}{Litjens, G.}, \bibinfo{author}{Menze, B.}, \bibinfo{author}{Ronneberger, O.}, \bibinfo{author}{Summers, R.M.}, \bibinfo{author}{van Ginneken, B.}, et~al., \bibinfo{year}{2021}.
\newblock \bibinfo{title}{The medical segmentation decathlon}.
\newblock \bibinfo{journal}{arXiv preprint arXiv:2106.05735} .
\bibitem[{Bai and Xia(2022)}]{bai2022end}
\bibinfo{author}{Bai, X.}, \bibinfo{author}{Xia, Y.}, \bibinfo{year}{2022}.
\newblock \bibinfo{title}{An end-to-end framework for universal lesion detection with missing annotations}, in: \bibinfo{booktitle}{2022 16th IEEE International Conference on Signal Processing (ICSP)}, \bibinfo{organization}{IEEE}. pp. \bibinfo{pages}{411--415}.
\bibitem[{Bilic et~al.(2023)Bilic, Christ, Li, Vorontsov, Ben-Cohen, Kaissis, Szeskin, Jacobs, Mamani, Chartrand et~al.}]{bilic2023liver}
\bibinfo{author}{Bilic, P.}, \bibinfo{author}{Christ, P.}, \bibinfo{author}{Li, H.B.}, \bibinfo{author}{Vorontsov, E.}, \bibinfo{author}{Ben-Cohen, A.}, \bibinfo{author}{Kaissis, G.}, \bibinfo{author}{Szeskin, A.}, \bibinfo{author}{Jacobs, C.}, \bibinfo{author}{Mamani, G.E.H.}, \bibinfo{author}{Chartrand, G.}, et~al., \bibinfo{year}{2023}.
\newblock \bibinfo{title}{The liver tumor segmentation benchmark (lits)}.
\newblock \bibinfo{journal}{Medical Image Analysis} \bibinfo{volume}{84}, \bibinfo{pages}{102680}.
\bibitem[{Bilic et~al.(2019)Bilic, Christ, Vorontsov, Chlebus, Chen, Dou, Fu, Han, Heng, Hesser et~al.}]{bilic2019liver}
\bibinfo{author}{Bilic, P.}, \bibinfo{author}{Christ, P.F.}, \bibinfo{author}{Vorontsov, E.}, \bibinfo{author}{Chlebus, G.}, \bibinfo{author}{Chen, H.}, \bibinfo{author}{Dou, Q.}, \bibinfo{author}{Fu, C.W.}, \bibinfo{author}{Han, X.}, \bibinfo{author}{Heng, P.A.}, \bibinfo{author}{Hesser, J.}, et~al., \bibinfo{year}{2019}.
\newblock \bibinfo{title}{The liver tumor segmentation benchmark (lits)}.
\newblock \bibinfo{journal}{arXiv preprint arXiv:1901.04056} .
\bibitem[{Brown et~al.(2020)Brown, Mann, Ryder, Subbiah, Kaplan, Dhariwal, Neelakantan, Shyam, Sastry, Askell et~al.}]{brown2020language}
\bibinfo{author}{Brown, T.}, \bibinfo{author}{Mann, B.}, \bibinfo{author}{Ryder, N.}, \bibinfo{author}{Subbiah, M.}, \bibinfo{author}{Kaplan, J.D.}, \bibinfo{author}{Dhariwal, P.}, \bibinfo{author}{Neelakantan, A.}, \bibinfo{author}{Shyam, P.}, \bibinfo{author}{Sastry, G.}, \bibinfo{author}{Askell, A.}, et~al., \bibinfo{year}{2020}.
\newblock \bibinfo{title}{Language models are few-shot learners}.
\newblock \bibinfo{journal}{Advances in neural information processing systems} \bibinfo{volume}{33}, \bibinfo{pages}{1877--1901}.
\bibitem[{Cai et~al.(2019)Cai, Xia, Yang, Xu, Yang and Roth}]{cai2019end}
\bibinfo{author}{Cai, J.}, \bibinfo{author}{Xia, Y.}, \bibinfo{author}{Yang, D.}, \bibinfo{author}{Xu, D.}, \bibinfo{author}{Yang, L.}, \bibinfo{author}{Roth, H.}, \bibinfo{year}{2019}.
\newblock \bibinfo{title}{End-to-end adversarial shape learning for abdomen organ deep segmentation}, in: \bibinfo{booktitle}{Machine Learning in Medical Imaging: 10th International Workshop, MLMI 2019, Held in Conjunction with MICCAI 2019, Shenzhen, China, October 13, 2019, Proceedings 10}, \bibinfo{organization}{Springer}. pp. \bibinfo{pages}{124--132}.
\bibitem[{Cardoso et~al.(2022)Cardoso, Li, Brown, Ma, Kerfoot, Wang, Murrey, Myronenko, Zhao, Yang et~al.}]{cardoso2022monai}
\bibinfo{author}{Cardoso, M.J.}, \bibinfo{author}{Li, W.}, \bibinfo{author}{Brown, R.}, \bibinfo{author}{Ma, N.}, \bibinfo{author}{Kerfoot, E.}, \bibinfo{author}{Wang, Y.}, \bibinfo{author}{Murrey, B.}, \bibinfo{author}{Myronenko, A.}, \bibinfo{author}{Zhao, C.}, \bibinfo{author}{Yang, D.}, et~al., \bibinfo{year}{2022}.
\newblock \bibinfo{title}{Monai: An open-source framework for deep learning in healthcare}, pp. \bibinfo{pages}{arXiv--2211}.
\bibitem[{Chambon et~al.(2022)Chambon, Bluethgen, Langlotz and Chaudhari}]{chambon2022adapting}
\bibinfo{author}{Chambon, P.}, \bibinfo{author}{Bluethgen, C.}, \bibinfo{author}{Langlotz, C.P.}, \bibinfo{author}{Chaudhari, A.}, \bibinfo{year}{2022}.
\newblock \bibinfo{title}{Adapting pretrained vision-language foundational models to medical imaging domains}.
\newblock \bibinfo{journal}{arXiv preprint arXiv:2210.04133} .
\bibitem[{Chen et~al.(2021a)Chen, Lu, Yu, Luo, Adeli, Wang, Lu, Yuille and Zhou}]{chen2021transunet}
\bibinfo{author}{Chen, J.}, \bibinfo{author}{Lu, Y.}, \bibinfo{author}{Yu, Q.}, \bibinfo{author}{Luo, X.}, \bibinfo{author}{Adeli, E.}, \bibinfo{author}{Wang, Y.}, \bibinfo{author}{Lu, L.}, \bibinfo{author}{Yuille, A.L.}, \bibinfo{author}{Zhou, Y.}, \bibinfo{year}{2021}a.
\newblock \bibinfo{title}{Transunet: Transformers make strong encoders for medical image segmentation}.
\newblock \bibinfo{journal}{arXiv preprint arXiv:2102.04306} .
\bibitem[{Chen et~al.(2023a)Chen, Xia, Yao, Yan, Zhang, Lu, Wang, Zhou, Qiu, Yu et~al.}]{chen2023cancerunit}
\bibinfo{author}{Chen, J.}, \bibinfo{author}{Xia, Y.}, \bibinfo{author}{Yao, J.}, \bibinfo{author}{Yan, K.}, \bibinfo{author}{Zhang, J.}, \bibinfo{author}{Lu, L.}, \bibinfo{author}{Wang, F.}, \bibinfo{author}{Zhou, B.}, \bibinfo{author}{Qiu, M.}, \bibinfo{author}{Yu, Q.}, et~al., \bibinfo{year}{2023}a.
\newblock \bibinfo{title}{Cancerunit: Towards a single unified model for effective detection, segmentation, and diagnosis of eight major cancers using a large collection of ct scans}, in: \bibinfo{booktitle}{Proceedings of the IEEE/CVF International Conference on Computer Vision}, pp. \bibinfo{pages}{21327--21338}.
\bibitem[{Chen et~al.(2024a)Chen, Qin, Lee, Yan and Li}]{chen2024learning}
\bibinfo{author}{Chen, K.}, \bibinfo{author}{Qin, T.}, \bibinfo{author}{Lee, V.H.F.}, \bibinfo{author}{Yan, H.}, \bibinfo{author}{Li, H.}, \bibinfo{year}{2024}a.
\newblock \bibinfo{title}{Learning robust shape regularization for generalizable medical image segmentation}.
\newblock \bibinfo{journal}{IEEE Transactions on Medical Imaging} .
\bibitem[{Chen et~al.(2019a)Chen, Gadepalli, MacDonald, Liu, Kadowaki, Nagpal, Kohlberger, Dean, Corrado, Hipp et~al.}]{chen2019augmented}
\bibinfo{author}{Chen, P.H.C.}, \bibinfo{author}{Gadepalli, K.}, \bibinfo{author}{MacDonald, R.}, \bibinfo{author}{Liu, Y.}, \bibinfo{author}{Kadowaki, S.}, \bibinfo{author}{Nagpal, K.}, \bibinfo{author}{Kohlberger, T.}, \bibinfo{author}{Dean, J.}, \bibinfo{author}{Corrado, G.S.}, \bibinfo{author}{Hipp, J.D.}, et~al., \bibinfo{year}{2019}a.
\newblock \bibinfo{title}{An augmented reality microscope with real-time artificial intelligence integration for cancer diagnosis}.
\newblock \bibinfo{journal}{Nature medicine} \bibinfo{volume}{25}, \bibinfo{pages}{1453--1457}.
\bibitem[{Chen et~al.(2024b)Chen, Chen, Song, Xiong, Yuille, Wei and Zhou}]{chen2024towards}
\bibinfo{author}{Chen, Q.}, \bibinfo{author}{Chen, X.}, \bibinfo{author}{Song, H.}, \bibinfo{author}{Xiong, Z.}, \bibinfo{author}{Yuille, A.}, \bibinfo{author}{Wei, C.}, \bibinfo{author}{Zhou, Z.}, \bibinfo{year}{2024}b.
\newblock \bibinfo{title}{Towards generalizable tumor synthesis}, in: \bibinfo{booktitle}{IEEE/CVF Conference on Computer Vision and Pattern Recognition}.
\newblock \URLprefix \url{https://github.com/MrGiovanni/DiffTumor}.
\bibitem[{Chen et~al.(2017)Chen, Xu and Koltun}]{chen2017fast}
\bibinfo{author}{Chen, Q.}, \bibinfo{author}{Xu, J.}, \bibinfo{author}{Koltun, V.}, \bibinfo{year}{2017}.
\newblock \bibinfo{title}{Fast image processing with fully-convolutional networks}, in: \bibinfo{booktitle}{Proceedings of the IEEE International Conference on Computer Vision}, pp. \bibinfo{pages}{2497--2506}.
\bibitem[{Chen et~al.(2019b)Chen, Ma and Zheng}]{chen2019med3d}
\bibinfo{author}{Chen, S.}, \bibinfo{author}{Ma, K.}, \bibinfo{author}{Zheng, Y.}, \bibinfo{year}{2019}b.
\newblock \bibinfo{title}{Med3d: Transfer learning for 3d medical image analysis}.
\newblock \bibinfo{journal}{arXiv preprint arXiv:1904.00625} .
\bibitem[{Chen et~al.(2021b)Chen, Sun, Bai, Han, Liu, Yao, Tang, Zhang, Lu, Huang et~al.}]{chen2021deep}
\bibinfo{author}{Chen, X.}, \bibinfo{author}{Sun, S.}, \bibinfo{author}{Bai, N.}, \bibinfo{author}{Han, K.}, \bibinfo{author}{Liu, Q.}, \bibinfo{author}{Yao, S.}, \bibinfo{author}{Tang, H.}, \bibinfo{author}{Zhang, C.}, \bibinfo{author}{Lu, Z.}, \bibinfo{author}{Huang, Q.}, et~al., \bibinfo{year}{2021}b.
\newblock \bibinfo{title}{A deep learning-based auto-segmentation system for organs-at-risk on whole-body computed tomography images for radiation therapy}.
\newblock \bibinfo{journal}{Radiotherapy and Oncology} \bibinfo{volume}{160}, \bibinfo{pages}{175--184}.
\bibitem[{Chen et~al.(2023b)Chen, Zheng, Li, Ma, Ma, Li and Fan}]{chen2023versatile}
\bibinfo{author}{Chen, X.}, \bibinfo{author}{Zheng, H.}, \bibinfo{author}{Li, Y.}, \bibinfo{author}{Ma, Y.}, \bibinfo{author}{Ma, L.}, \bibinfo{author}{Li, H.}, \bibinfo{author}{Fan, Y.}, \bibinfo{year}{2023}b.
\newblock \bibinfo{title}{Versatile medical image segmentation learned from multi-source datasets via model self-disambiguation}.
\newblock \bibinfo{journal}{arXiv preprint arXiv:2311.10696} .
\bibitem[{Conneau and Lample(2019)}]{conneau2019cross}
\bibinfo{author}{Conneau, A.}, \bibinfo{author}{Lample, G.}, \bibinfo{year}{2019}.
\newblock \bibinfo{title}{Cross-lingual language model pretraining}.
\newblock \bibinfo{journal}{Advances in neural information processing systems} \bibinfo{volume}{32}.
\bibitem[{Devlin et~al.(2018)Devlin, Chang, Lee and Toutanova}]{devlin2018bert}
\bibinfo{author}{Devlin, J.}, \bibinfo{author}{Chang, M.W.}, \bibinfo{author}{Lee, K.}, \bibinfo{author}{Toutanova, K.}, \bibinfo{year}{2018}.
\newblock \bibinfo{title}{Bert: Pre-training of deep bidirectional transformers for language understanding}.
\newblock \bibinfo{journal}{arXiv preprint arXiv:1810.04805} .
\bibitem[{Dmitriev and Kaufman(2019)}]{dmitriev2019learning}
\bibinfo{author}{Dmitriev, K.}, \bibinfo{author}{Kaufman, A.E.}, \bibinfo{year}{2019}.
\newblock \bibinfo{title}{Learning multi-class segmentations from single-class datasets}, in: \bibinfo{booktitle}{Proceedings of the IEEE/CVF Conference on Computer Vision and Pattern Recognition}, pp. \bibinfo{pages}{9501--9511}.
\bibitem[{Douillard et~al.(2021)Douillard, Chen, Dapogny and Cord}]{douillard2021plop}
\bibinfo{author}{Douillard, A.}, \bibinfo{author}{Chen, Y.}, \bibinfo{author}{Dapogny, A.}, \bibinfo{author}{Cord, M.}, \bibinfo{year}{2021}.
\newblock \bibinfo{title}{Plop: Learning without forgetting for continual semantic segmentation}, in: \bibinfo{booktitle}{Proceedings of the IEEE/CVF Conference on Computer Vision and Pattern Recognition}, pp. \bibinfo{pages}{4040--4050}.
\bibitem[{Eslami et~al.(2023)Eslami, Meinel and De~Melo}]{eslami2023pubmedclip}
\bibinfo{author}{Eslami, S.}, \bibinfo{author}{Meinel, C.}, \bibinfo{author}{De~Melo, G.}, \bibinfo{year}{2023}.
\newblock \bibinfo{title}{Pubmedclip: How much does clip benefit visual question answering in the medical domain?}, in: \bibinfo{booktitle}{Findings of the Association for Computational Linguistics: EACL 2023}, pp. \bibinfo{pages}{1151--1163}.
\bibitem[{Esteva et~al.(2021)Esteva, Chou, Yeung, Naik, Madani, Mottaghi, Liu, Topol, Dean and Socher}]{esteva2021deep}
\bibinfo{author}{Esteva, A.}, \bibinfo{author}{Chou, K.}, \bibinfo{author}{Yeung, S.}, \bibinfo{author}{Naik, N.}, \bibinfo{author}{Madani, A.}, \bibinfo{author}{Mottaghi, A.}, \bibinfo{author}{Liu, Y.}, \bibinfo{author}{Topol, E.}, \bibinfo{author}{Dean, J.}, \bibinfo{author}{Socher, R.}, \bibinfo{year}{2021}.
\newblock \bibinfo{title}{Deep learning-enabled medical computer vision}.
\newblock \bibinfo{journal}{NPJ digital medicine} \bibinfo{volume}{4}, \bibinfo{pages}{1--9}.
\bibitem[{Fang and Yan(2020)}]{fang2020multi}
\bibinfo{author}{Fang, X.}, \bibinfo{author}{Yan, P.}, \bibinfo{year}{2020}.
\newblock \bibinfo{title}{Multi-organ segmentation over partially labeled datasets with multi-scale feature abstraction}.
\newblock \bibinfo{journal}{IEEE Transactions on Medical Imaging} \bibinfo{volume}{39}, \bibinfo{pages}{3619--3629}.
\bibitem[{Gao et~al.(2021)Gao, Huang, Yang, Zhang, Shao, Tao, Chen, Metaxas, Li and Chen}]{gao2021focusnetv2}
\bibinfo{author}{Gao, Y.}, \bibinfo{author}{Huang, R.}, \bibinfo{author}{Yang, Y.}, \bibinfo{author}{Zhang, J.}, \bibinfo{author}{Shao, K.}, \bibinfo{author}{Tao, C.}, \bibinfo{author}{Chen, Y.}, \bibinfo{author}{Metaxas, D.N.}, \bibinfo{author}{Li, H.}, \bibinfo{author}{Chen, M.}, \bibinfo{year}{2021}.
\newblock \bibinfo{title}{Focusnetv2: Imbalanced large and small organ segmentation with adversarial shape constraint for head and neck ct images}.
\newblock \bibinfo{journal}{Medical Image Analysis} \bibinfo{volume}{67}, \bibinfo{pages}{101831}.
\bibitem[{Gao et~al.(2023)Gao, Li, Liu, Zhou, Zhang and Meta}]{gao2023training}
\bibinfo{author}{Gao, Y.}, \bibinfo{author}{Li, Z.}, \bibinfo{author}{Liu, D.}, \bibinfo{author}{Zhou, M.}, \bibinfo{author}{Zhang, S.}, \bibinfo{author}{Meta, D.N.}, \bibinfo{year}{2023}.
\newblock \bibinfo{title}{Training like a medical resident: Universal medical image segmentation via context prior learning}.
\newblock \bibinfo{journal}{arXiv preprint arXiv:2306.02416} .
\bibitem[{Germain et~al.(2014)Germain, Favelier, Cercueil, Denys, Kraus{\'e} and Guiu}]{germain2014liver}
\bibinfo{author}{Germain, T.}, \bibinfo{author}{Favelier, S.}, \bibinfo{author}{Cercueil, J.P.}, \bibinfo{author}{Denys, A.}, \bibinfo{author}{Kraus{\'e}, D.}, \bibinfo{author}{Guiu, B.}, \bibinfo{year}{2014}.
\newblock \bibinfo{title}{Liver segmentation: practical tips}.
\newblock \bibinfo{journal}{Diagnostic and interventional imaging} \bibinfo{volume}{95}, \bibinfo{pages}{1003--1016}.
\bibitem[{Guo et~al.(2021a)Guo, Wang, Zhou, Jiang and Patel}]{guo2021multi}
\bibinfo{author}{Guo, P.}, \bibinfo{author}{Wang, P.}, \bibinfo{author}{Zhou, J.}, \bibinfo{author}{Jiang, S.}, \bibinfo{author}{Patel, V.M.}, \bibinfo{year}{2021}a.
\newblock \bibinfo{title}{Multi-institutional collaborations for improving deep learning-based magnetic resonance image reconstruction using federated learning}, in: \bibinfo{booktitle}{Proceedings of the IEEE/CVF Conference on Computer Vision and Pattern Recognition}, pp. \bibinfo{pages}{2423--2432}.
\bibitem[{Guo et~al.(2021b)Guo, Liu and Yuan}]{guo2021semantic}
\bibinfo{author}{Guo, X.}, \bibinfo{author}{Liu, J.}, \bibinfo{author}{Yuan, Y.}, \bibinfo{year}{2021}b.
\newblock \bibinfo{title}{Semantic-oriented labeled-to-unlabeled distribution translation for image segmentation}.
\newblock \bibinfo{journal}{IEEE transactions on medical imaging} \bibinfo{volume}{41}, \bibinfo{pages}{434--445}.
\bibitem[{Haghighi et~al.(2021)Haghighi, Taher, Zhou, Gotway and Liang}]{haghighi2021transferable}
\bibinfo{author}{Haghighi, F.}, \bibinfo{author}{Taher, M.R.H.}, \bibinfo{author}{Zhou, Z.}, \bibinfo{author}{Gotway, M.B.}, \bibinfo{author}{Liang, J.}, \bibinfo{year}{2021}.
\newblock \bibinfo{title}{Transferable visual words: Exploiting the semantics of anatomical patterns for self-supervised learning}.
\newblock \bibinfo{journal}{IEEE Transactions on Medical Imaging} \URLprefix \url{https://github.com/fhaghighi/SemanticGenesis}.
\bibitem[{Hatamizadeh et~al.(2022a)Hatamizadeh, Nath, Tang, Yang, Roth and Xu}]{hatamizadeh2022swin}
\bibinfo{author}{Hatamizadeh, A.}, \bibinfo{author}{Nath, V.}, \bibinfo{author}{Tang, Y.}, \bibinfo{author}{Yang, D.}, \bibinfo{author}{Roth, H.R.}, \bibinfo{author}{Xu, D.}, \bibinfo{year}{2022}a.
\newblock \bibinfo{title}{Swin unetr: Swin transformers for semantic segmentation of brain tumors in mri images}, in: \bibinfo{booktitle}{Brainlesion: Glioma, Multiple Sclerosis, Stroke and Traumatic Brain Injuries: 7th International Workshop, BrainLes 2021, Held in Conjunction with MICCAI 2021, Virtual Event, September 27, 2021, Revised Selected Papers, Part I}, \bibinfo{organization}{Springer}. pp. \bibinfo{pages}{272--284}.
\bibitem[{Hatamizadeh et~al.(2022b)Hatamizadeh, Tang, Nath, Yang, Myronenko, Landman, Roth and Xu}]{hatamizadeh2022unetr}
\bibinfo{author}{Hatamizadeh, A.}, \bibinfo{author}{Tang, Y.}, \bibinfo{author}{Nath, V.}, \bibinfo{author}{Yang, D.}, \bibinfo{author}{Myronenko, A.}, \bibinfo{author}{Landman, B.}, \bibinfo{author}{Roth, H.R.}, \bibinfo{author}{Xu, D.}, \bibinfo{year}{2022}b.
\newblock \bibinfo{title}{Unetr: Transformers for 3d medical image segmentation}, in: \bibinfo{booktitle}{Proceedings of the IEEE/CVF Winter Conference on Applications of Computer Vision}, pp. \bibinfo{pages}{574--584}.
\bibitem[{He et~al.(2023)He, Nath, Yang, Tang, Myronenko and Xu}]{he2023swinunetr}
\bibinfo{author}{He, Y.}, \bibinfo{author}{Nath, V.}, \bibinfo{author}{Yang, D.}, \bibinfo{author}{Tang, Y.}, \bibinfo{author}{Myronenko, A.}, \bibinfo{author}{Xu, D.}, \bibinfo{year}{2023}.
\newblock \bibinfo{title}{Swinunetr-v2: Stronger swin transformers with stagewise convolutions for 3d medical image segmentation}, in: \bibinfo{booktitle}{International Conference on Medical Image Computing and Computer-Assisted Intervention}, \bibinfo{organization}{Springer}. pp. \bibinfo{pages}{416--426}.
\bibitem[{He et~al.(2021)He, Yang, Roth, Zhao and Xu}]{he2021dints}
\bibinfo{author}{He, Y.}, \bibinfo{author}{Yang, D.}, \bibinfo{author}{Roth, H.}, \bibinfo{author}{Zhao, C.}, \bibinfo{author}{Xu, D.}, \bibinfo{year}{2021}.
\newblock \bibinfo{title}{Dints: Differentiable neural network topology search for 3d medical image segmentation}, in: \bibinfo{booktitle}{Proceedings of the IEEE/CVF Conference on Computer Vision and Pattern Recognition}, pp. \bibinfo{pages}{5841--5850}.
\bibitem[{Heller et~al.(2020)Heller, McSweeney, Peterson, Peterson, Rickman, Stai, Tejpaul, Oestreich, Blake, Rosenberg et~al.}]{heller2020international}
\bibinfo{author}{Heller, N.}, \bibinfo{author}{McSweeney, S.}, \bibinfo{author}{Peterson, M.T.}, \bibinfo{author}{Peterson, S.}, \bibinfo{author}{Rickman, J.}, \bibinfo{author}{Stai, B.}, \bibinfo{author}{Tejpaul, R.}, \bibinfo{author}{Oestreich, M.}, \bibinfo{author}{Blake, P.}, \bibinfo{author}{Rosenberg, J.}, et~al., \bibinfo{year}{2020}.
\newblock \bibinfo{title}{An international challenge to use artificial intelligence to define the state-of-the-art in kidney and kidney tumor segmentation in ct imaging.}
\bibitem[{Heller et~al.(2019)Heller, Sathianathen, Kalapara, Walczak, Moore, Kaluzniak, Rosenberg, Blake, Rengel, Oestreich et~al.}]{heller2019kits19}
\bibinfo{author}{Heller, N.}, \bibinfo{author}{Sathianathen, N.}, \bibinfo{author}{Kalapara, A.}, \bibinfo{author}{Walczak, E.}, \bibinfo{author}{Moore, K.}, \bibinfo{author}{Kaluzniak, H.}, \bibinfo{author}{Rosenberg, J.}, \bibinfo{author}{Blake, P.}, \bibinfo{author}{Rengel, Z.}, \bibinfo{author}{Oestreich, M.}, et~al., \bibinfo{year}{2019}.
\newblock \bibinfo{title}{The kits19 challenge data: 300 kidney tumor cases with clinical context, ct semantic segmentations, and surgical outcomes}.
\newblock \bibinfo{journal}{arXiv preprint arXiv:1904.00445} .
\bibitem[{Hu et~al.(2023)Hu, Chen, Xiao, Sun, Chen, Yuille and Zhou}]{hu2023label}
\bibinfo{author}{Hu, Q.}, \bibinfo{author}{Chen, Y.}, \bibinfo{author}{Xiao, J.}, \bibinfo{author}{Sun, S.}, \bibinfo{author}{Chen, J.}, \bibinfo{author}{Yuille, A.L.}, \bibinfo{author}{Zhou, Z.}, \bibinfo{year}{2023}.
\newblock \bibinfo{title}{Label-free liver tumor segmentation}, in: \bibinfo{booktitle}{IEEE/CVF Conference on Computer Vision and Pattern Recognition}, pp. \bibinfo{pages}{7422--7432}.
\newblock \URLprefix \url{https://github.com/MrGiovanni/SyntheticTumors}.
\bibitem[{Hu et~al.(2022)Hu, Gan, Wang, Yang, Liu, Lu and Wang}]{hu2022scaling}
\bibinfo{author}{Hu, X.}, \bibinfo{author}{Gan, Z.}, \bibinfo{author}{Wang, J.}, \bibinfo{author}{Yang, Z.}, \bibinfo{author}{Liu, Z.}, \bibinfo{author}{Lu, Y.}, \bibinfo{author}{Wang, L.}, \bibinfo{year}{2022}.
\newblock \bibinfo{title}{Scaling up vision-language pre-training for image captioning}, in: \bibinfo{booktitle}{Proceedings of the IEEE/CVF conference on computer vision and pattern recognition}, pp. \bibinfo{pages}{17980--17989}.
\bibitem[{Huang et~al.(2023)Huang, Bianchi, Yuksekgonul, Montine and Zou}]{huang2023visual}
\bibinfo{author}{Huang, Z.}, \bibinfo{author}{Bianchi, F.}, \bibinfo{author}{Yuksekgonul, M.}, \bibinfo{author}{Montine, T.J.}, \bibinfo{author}{Zou, J.}, \bibinfo{year}{2023}.
\newblock \bibinfo{title}{A visual--language foundation model for pathology image analysis using medical twitter}.
\newblock \bibinfo{journal}{Nature medicine} \bibinfo{volume}{29}, \bibinfo{pages}{2307--2316}.
\bibitem[{Isensee et~al.(2021)Isensee, Jaeger, Kohl, Petersen and Maier-Hein}]{isensee2021nnu}
\bibinfo{author}{Isensee, F.}, \bibinfo{author}{Jaeger, P.F.}, \bibinfo{author}{Kohl, S.A.}, \bibinfo{author}{Petersen, J.}, \bibinfo{author}{Maier-Hein, K.H.}, \bibinfo{year}{2021}.
\newblock \bibinfo{title}{nnu-net: a self-configuring method for deep learning-based biomedical image segmentation}.
\newblock \bibinfo{journal}{Nature Methods} \bibinfo{volume}{18}, \bibinfo{pages}{203--211}.
\bibitem[{Jaus et~al.(2023)Jaus, Seibold, Hermann, Walter, Giske, Haubold, Kleesiek and Stiefelhagen}]{jaus2023towards}
\bibinfo{author}{Jaus, A.}, \bibinfo{author}{Seibold, C.}, \bibinfo{author}{Hermann, K.}, \bibinfo{author}{Walter, A.}, \bibinfo{author}{Giske, K.}, \bibinfo{author}{Haubold, J.}, \bibinfo{author}{Kleesiek, J.}, \bibinfo{author}{Stiefelhagen, R.}, \bibinfo{year}{2023}.
\newblock \bibinfo{title}{Towards unifying anatomy segmentation: automated generation of a full-body ct dataset via knowledge aggregation and anatomical guidelines}.
\newblock \bibinfo{journal}{arXiv preprint arXiv:2307.13375} .
\bibitem[{Ji et~al.(2021)Ji, Yu, Wu, Ma, Bian, Bi, Li, Liu, Cheng and Zheng}]{ji2021learning}
\bibinfo{author}{Ji, W.}, \bibinfo{author}{Yu, S.}, \bibinfo{author}{Wu, J.}, \bibinfo{author}{Ma, K.}, \bibinfo{author}{Bian, C.}, \bibinfo{author}{Bi, Q.}, \bibinfo{author}{Li, J.}, \bibinfo{author}{Liu, H.}, \bibinfo{author}{Cheng, L.}, \bibinfo{author}{Zheng, Y.}, \bibinfo{year}{2021}.
\newblock \bibinfo{title}{Learning calibrated medical image segmentation via multi-rater agreement modeling}, in: \bibinfo{booktitle}{Proceedings of the IEEE/CVF Conference on Computer Vision and Pattern Recognition}, pp. \bibinfo{pages}{12341--12351}.
\bibitem[{Ji et~al.(2022)Ji, Bai, Yang, Ge, Zhu, Zhang, Li, Zhang, Ma, Wan et~al.}]{ji2022amos}
\bibinfo{author}{Ji, Y.}, \bibinfo{author}{Bai, H.}, \bibinfo{author}{Yang, J.}, \bibinfo{author}{Ge, C.}, \bibinfo{author}{Zhu, Y.}, \bibinfo{author}{Zhang, R.}, \bibinfo{author}{Li, Z.}, \bibinfo{author}{Zhang, L.}, \bibinfo{author}{Ma, W.}, \bibinfo{author}{Wan, X.}, et~al., \bibinfo{year}{2022}.
\newblock \bibinfo{title}{Amos: A large-scale abdominal multi-organ benchmark for versatile medical image segmentation}.
\newblock \bibinfo{journal}{Neural Information Processing Systems (NeurIPS)} .
\bibitem[{Ji et~al.(2023)Ji, Guo, Wang, Yan, Lu, Xu, Wang, Ge, Gao, Ye et~al.}]{ji2023continual}
\bibinfo{author}{Ji, Z.}, \bibinfo{author}{Guo, D.}, \bibinfo{author}{Wang, P.}, \bibinfo{author}{Yan, K.}, \bibinfo{author}{Lu, L.}, \bibinfo{author}{Xu, M.}, \bibinfo{author}{Wang, Q.}, \bibinfo{author}{Ge, J.}, \bibinfo{author}{Gao, M.}, \bibinfo{author}{Ye, X.}, et~al., \bibinfo{year}{2023}.
\newblock \bibinfo{title}{Continual segment: Towards a single, unified and non-forgetting continual segmentation model of 143 whole-body organs in ct scans}, in: \bibinfo{booktitle}{Proceedings of the IEEE/CVF International Conference on Computer Vision}, pp. \bibinfo{pages}{21140--21151}.
\bibitem[{Jiang et~al.(2023)Jiang, Huang, Zhang, Zhang and Zhang}]{jiang2023zept}
\bibinfo{author}{Jiang, Y.}, \bibinfo{author}{Huang, Z.}, \bibinfo{author}{Zhang, R.}, \bibinfo{author}{Zhang, X.}, \bibinfo{author}{Zhang, S.}, \bibinfo{year}{2023}.
\newblock \bibinfo{title}{Zept: Zero-shot pan-tumor segmentation via query-disentangling and self-prompting}.
\newblock \bibinfo{journal}{arXiv preprint arXiv:2312.04964} .
\bibitem[{Kim et~al.(2019)Kim, Kim, Lim, Baek, Kim, Cho, Yoon and Kim}]{kim2019scalable}
\bibinfo{author}{Kim, S.}, \bibinfo{author}{Kim, I.}, \bibinfo{author}{Lim, S.}, \bibinfo{author}{Baek, W.}, \bibinfo{author}{Kim, C.}, \bibinfo{author}{Cho, H.}, \bibinfo{author}{Yoon, B.}, \bibinfo{author}{Kim, T.}, \bibinfo{year}{2019}.
\newblock \bibinfo{title}{Scalable neural architecture search for 3d medical image segmentation}, in: \bibinfo{booktitle}{International Conference on Medical Image Computing and Computer-Assisted Intervention}, \bibinfo{organization}{Springer}. pp. \bibinfo{pages}{220--228}.
\bibitem[{Lai et~al.(2024)Lai, Chen, Wang, Yuille and Zhou}]{lai2024pixel}
\bibinfo{author}{Lai, Y.}, \bibinfo{author}{Chen, X.}, \bibinfo{author}{Wang, A.}, \bibinfo{author}{Yuille, A.}, \bibinfo{author}{Zhou, Z.}, \bibinfo{year}{2024}.
\newblock \bibinfo{title}{From pixel to cancer: Cellular automata in computed tomography}.
\newblock \bibinfo{journal}{arXiv preprint arXiv:2403.06459} \URLprefix \url{https://github.com/MrGiovanni/Pixel2Cancer}.
\bibitem[{Landman et~al.(2017)Landman, Xu, Eugenio~Igelsias, Styner, Robin~Langerak and Klein}]{landman2017multiatlas}
\bibinfo{author}{Landman, B.}, \bibinfo{author}{Xu, Z.}, \bibinfo{author}{Eugenio~Igelsias, J.}, \bibinfo{author}{Styner, M.}, \bibinfo{author}{Robin~Langerak, T.}, \bibinfo{author}{Klein, A.}, \bibinfo{year}{2017}.
\newblock \bibinfo{title}{Multi-atlas labeling beyond the cranial vault-workshop and challenge} .
\bibitem[{Landman et~al.(2015)Landman, Xu, Igelsias, Styner, Langerak and Klein}]{landman2015miccai}
\bibinfo{author}{Landman, B.}, \bibinfo{author}{Xu, Z.}, \bibinfo{author}{Igelsias, J.}, \bibinfo{author}{Styner, M.}, \bibinfo{author}{Langerak, T.}, \bibinfo{author}{Klein, A.}, \bibinfo{year}{2015}.
\newblock \bibinfo{title}{Miccai multi-atlas labeling beyond the cranial vault--workshop and challenge}, in: \bibinfo{booktitle}{Proc. MICCAI Multi-Atlas Labeling Beyond Cranial Vault—Workshop Challenge}, p.~\bibinfo{pages}{12}.
\bibitem[{Lewandowsky and Li(1995)}]{lewandowsky1995catastrophic}
\bibinfo{author}{Lewandowsky, S.}, \bibinfo{author}{Li, S.C.}, \bibinfo{year}{1995}.
\newblock \bibinfo{title}{Catastrophic interference in neural networks: Causes, solutions, and data}, in: \bibinfo{booktitle}{Interference and inhibition in cognition}. \bibinfo{publisher}{Elsevier}, pp. \bibinfo{pages}{329--361}.
\bibitem[{Li et~al.(2023)Li, Chou, Sun, Qiao, Yuille and Zhou}]{li2023early}
\bibinfo{author}{Li, B.}, \bibinfo{author}{Chou, Y.C.}, \bibinfo{author}{Sun, S.}, \bibinfo{author}{Qiao, H.}, \bibinfo{author}{Yuille, A.}, \bibinfo{author}{Zhou, Z.}, \bibinfo{year}{2023}.
\newblock \bibinfo{title}{Early detection and localization of pancreatic cancer by label-free tumor synthesis}.
\newblock \bibinfo{journal}{MICCAI Workshop on Big Task Small Data, 1001-AI} \URLprefix \url{https://github.com/MrGiovanni/SyntheticTumors}.
\bibitem[{Li et~al.(2024)Li, Yuille and Zhou}]{li2024well}
\bibinfo{author}{Li, W.}, \bibinfo{author}{Yuille, A.}, \bibinfo{author}{Zhou, Z.}, \bibinfo{year}{2024}.
\newblock \bibinfo{title}{How well do supervised models transfer to 3d image segmentation?}, in: \bibinfo{booktitle}{International Conference on Learning Representations}.
\newblock \URLprefix \url{https://github.com/MrGiovanni/SuPreM}.
\bibitem[{Li and Hoiem(2017)}]{li2017learning}
\bibinfo{author}{Li, Z.}, \bibinfo{author}{Hoiem, D.}, \bibinfo{year}{2017}.
\newblock \bibinfo{title}{Learning without forgetting}.
\newblock \bibinfo{journal}{IEEE transactions on pattern analysis and machine intelligence} \bibinfo{volume}{40}, \bibinfo{pages}{2935--2947}.
\bibitem[{Liang et~al.(2021)Liang, Li, Zhang, Xiong, Zhou and Xie}]{liang2021incorporating}
\bibinfo{author}{Liang, X.}, \bibinfo{author}{Li, N.}, \bibinfo{author}{Zhang, Z.}, \bibinfo{author}{Xiong, J.}, \bibinfo{author}{Zhou, S.}, \bibinfo{author}{Xie, Y.}, \bibinfo{year}{2021}.
\newblock \bibinfo{title}{Incorporating the hybrid deformable model for improving the performance of abdominal ct segmentation via multi-scale feature fusion network}.
\newblock \bibinfo{journal}{Medical Image Analysis} \bibinfo{volume}{73}, \bibinfo{pages}{102156}.
\bibitem[{Liu et~al.(2024)Liu, Xu, Gao, Li, Wang, Chabin, Oguz and Grbic}]{liu2024cosst}
\bibinfo{author}{Liu, H.}, \bibinfo{author}{Xu, Z.}, \bibinfo{author}{Gao, R.}, \bibinfo{author}{Li, H.}, \bibinfo{author}{Wang, J.}, \bibinfo{author}{Chabin, G.}, \bibinfo{author}{Oguz, I.}, \bibinfo{author}{Grbic, S.}, \bibinfo{year}{2024}.
\newblock \bibinfo{title}{Cosst: Multi-organ segmentation with partially labeled datasets using comprehensive supervisions and self-training}.
\newblock \bibinfo{journal}{IEEE Transactions on Medical Imaging} .
\bibitem[{Liu et~al.(2021)Liu, Guo and Yuan}]{liu2021graph}
\bibinfo{author}{Liu, J.}, \bibinfo{author}{Guo, X.}, \bibinfo{author}{Yuan, Y.}, \bibinfo{year}{2021}.
\newblock \bibinfo{title}{Graph-based surgical instrument adaptive segmentation via domain-common knowledge}.
\newblock \bibinfo{journal}{IEEE Transactions on Medical Imaging} \bibinfo{volume}{41}, \bibinfo{pages}{715--726}.
\bibitem[{Liu et~al.(2023a)Liu, Zhang, Chen, Xiao, Lu, A~Landman, Yuan, Yuille, Tang and Zhou}]{liu2023clip}
\bibinfo{author}{Liu, J.}, \bibinfo{author}{Zhang, Y.}, \bibinfo{author}{Chen, J.N.}, \bibinfo{author}{Xiao, J.}, \bibinfo{author}{Lu, Y.}, \bibinfo{author}{A~Landman, B.}, \bibinfo{author}{Yuan, Y.}, \bibinfo{author}{Yuille, A.}, \bibinfo{author}{Tang, Y.}, \bibinfo{author}{Zhou, Z.}, \bibinfo{year}{2023}a.
\newblock \bibinfo{title}{Clip-driven universal model for organ segmentation and tumor detection}, in: \bibinfo{booktitle}{Proceedings of the IEEE/CVF International Conference on Computer Vision}, pp. \bibinfo{pages}{21152--21164}.
\newblock \URLprefix \url{https://github.com/ljwztc/CLIP-Driven-Universal-Model}.
\bibitem[{Liu et~al.(2022a)Liu, Deng, Wang, Hui, Li, Li, Luo, Sun, Quan, Yang et~al.}]{liu2022universal}
\bibinfo{author}{Liu, P.}, \bibinfo{author}{Deng, Y.}, \bibinfo{author}{Wang, C.}, \bibinfo{author}{Hui, Y.}, \bibinfo{author}{Li, Q.}, \bibinfo{author}{Li, J.}, \bibinfo{author}{Luo, S.}, \bibinfo{author}{Sun, M.}, \bibinfo{author}{Quan, Q.}, \bibinfo{author}{Yang, S.}, et~al., \bibinfo{year}{2022}a.
\newblock \bibinfo{title}{Universal segmentation of 33 anatomies}.
\newblock \bibinfo{journal}{arXiv preprint arXiv:2203.02098} .
\bibitem[{Liu et~al.(2023b)Liu, Gu, Wu, Liao, Qian and Chen}]{liu20233d}
\bibinfo{author}{Liu, P.}, \bibinfo{author}{Gu, C.}, \bibinfo{author}{Wu, B.}, \bibinfo{author}{Liao, X.}, \bibinfo{author}{Qian, Y.}, \bibinfo{author}{Chen, G.}, \bibinfo{year}{2023}b.
\newblock \bibinfo{title}{3d multi-organ and tumor segmentation based on re-parameterize diverse experts}.
\newblock \bibinfo{journal}{Mathematics} \bibinfo{volume}{11}, \bibinfo{pages}{4868}.
\bibitem[{Liu et~al.(2022b)Liu, Wang, Fan, Pan, Yin, Zhu, Du, Zhao, Xiao, Ding et~al.}]{liu2022learning}
\bibinfo{author}{Liu, P.}, \bibinfo{author}{Wang, X.}, \bibinfo{author}{Fan, M.}, \bibinfo{author}{Pan, H.}, \bibinfo{author}{Yin, M.}, \bibinfo{author}{Zhu, X.}, \bibinfo{author}{Du, D.}, \bibinfo{author}{Zhao, X.}, \bibinfo{author}{Xiao, L.}, \bibinfo{author}{Ding, L.}, et~al., \bibinfo{year}{2022}b.
\newblock \bibinfo{title}{Learning incrementally to segment multiple organs in a ct image}, in: \bibinfo{booktitle}{Medical Image Computing and Computer Assisted Intervention--MICCAI 2022: 25th International Conference, Singapore, September 18--22, 2022, Proceedings, Part IV}, \bibinfo{organization}{Springer}. pp. \bibinfo{pages}{714--724}.
\bibitem[{Liu et~al.(2023c)Liu, Wen and Yang}]{liu2023ccq}
\bibinfo{author}{Liu, X.}, \bibinfo{author}{Wen, B.}, \bibinfo{author}{Yang, S.}, \bibinfo{year}{2023}c.
\newblock \bibinfo{title}{Ccq: cross-class query network for partially labeled organ segmentation}, in: \bibinfo{booktitle}{Proceedings of the AAAI Conference on Artificial Intelligence}, pp. \bibinfo{pages}{1755--1763}.
\bibitem[{Liu et~al.(2022c)Liu, Han, Xue, Song, Liu, Tang and Zhu}]{liu2022improving}
\bibinfo{author}{Liu, Z.}, \bibinfo{author}{Han, K.}, \bibinfo{author}{Xue, K.}, \bibinfo{author}{Song, Y.}, \bibinfo{author}{Liu, L.}, \bibinfo{author}{Tang, Y.}, \bibinfo{author}{Zhu, Y.}, \bibinfo{year}{2022}c.
\newblock \bibinfo{title}{Improving ct-image universal lesion detection with comprehensive data and feature enhancements}.
\newblock \bibinfo{journal}{Multimedia Systems} , \bibinfo{pages}{1--12}.
\bibitem[{L{\"u}ddecke and Ecker(2022)}]{luddecke2022image}
\bibinfo{author}{L{\"u}ddecke, T.}, \bibinfo{author}{Ecker, A.}, \bibinfo{year}{2022}.
\newblock \bibinfo{title}{Image segmentation using text and image prompts}, in: \bibinfo{booktitle}{Proceedings of the IEEE/CVF Conference on Computer Vision and Pattern Recognition}, pp. \bibinfo{pages}{7086--7096}.
\bibitem[{Luo et~al.(2021)Luo, Liao, Xiao, Song, Zhang, Li, Wang and Zhang}]{luo2021word}
\bibinfo{author}{Luo, X.}, \bibinfo{author}{Liao, W.}, \bibinfo{author}{Xiao, J.}, \bibinfo{author}{Song, T.}, \bibinfo{author}{Zhang, X.}, \bibinfo{author}{Li, K.}, \bibinfo{author}{Wang, G.}, \bibinfo{author}{Zhang, S.}, \bibinfo{year}{2021}.
\newblock \bibinfo{title}{Word: Revisiting organs segmentation in the whole abdominal region}.
\newblock \bibinfo{journal}{arXiv preprint arXiv:2111.02403} .
\bibitem[{Ma et~al.(2021)Ma, Zhang, Gu, Zhu, Ge, Zhang, An, Wang, Wang, Liu et~al.}]{ma2021abdomenct}
\bibinfo{author}{Ma, J.}, \bibinfo{author}{Zhang, Y.}, \bibinfo{author}{Gu, S.}, \bibinfo{author}{Zhu, C.}, \bibinfo{author}{Ge, C.}, \bibinfo{author}{Zhang, Y.}, \bibinfo{author}{An, X.}, \bibinfo{author}{Wang, C.}, \bibinfo{author}{Wang, Q.}, \bibinfo{author}{Liu, X.}, et~al., \bibinfo{year}{2021}.
\newblock \bibinfo{title}{Abdomenct-1k: Is abdominal organ segmentation a solved problem}.
\newblock \bibinfo{journal}{IEEE Transactions on Pattern Analysis and Machine Intelligence} .
\bibitem[{Mahmood et~al.(2019)Mahmood, Borders, Chen, McKay, Salimian, Baras and Durr}]{mahmood2019deep}
\bibinfo{author}{Mahmood, F.}, \bibinfo{author}{Borders, D.}, \bibinfo{author}{Chen, R.J.}, \bibinfo{author}{McKay, G.N.}, \bibinfo{author}{Salimian, K.J.}, \bibinfo{author}{Baras, A.}, \bibinfo{author}{Durr, N.J.}, \bibinfo{year}{2019}.
\newblock \bibinfo{title}{Deep adversarial training for multi-organ nuclei segmentation in histopathology images}.
\newblock \bibinfo{journal}{IEEE transactions on medical imaging} \bibinfo{volume}{39}, \bibinfo{pages}{3257--3267}.
\bibitem[{Mattikalli et~al.(2022)Mattikalli, Mathai and Summers}]{mattikalli2022universal}
\bibinfo{author}{Mattikalli, T.}, \bibinfo{author}{Mathai, T.S.}, \bibinfo{author}{Summers, R.M.}, \bibinfo{year}{2022}.
\newblock \bibinfo{title}{Universal lesion detection in ct scans using neural network ensembles}, in: \bibinfo{booktitle}{Medical Imaging 2022: Computer-Aided Diagnosis}, \bibinfo{organization}{SPIE}. pp. \bibinfo{pages}{864--868}.
\bibitem[{Michieli and Zanuttigh(2019)}]{michieli2019incremental}
\bibinfo{author}{Michieli, U.}, \bibinfo{author}{Zanuttigh, P.}, \bibinfo{year}{2019}.
\newblock \bibinfo{title}{Incremental learning techniques for semantic segmentation}, in: \bibinfo{booktitle}{Proceedings of the IEEE/CVF international conference on computer vision workshops}, pp. \bibinfo{pages}{0--0}.
\bibitem[{Mongan et~al.(2020)Mongan, Moy and Kahn}]{mongan2020checklist}
\bibinfo{author}{Mongan, J.}, \bibinfo{author}{Moy, L.}, \bibinfo{author}{Kahn, C.E.}, \bibinfo{year}{2020}.
\newblock \bibinfo{title}{Checklist for artificial intelligence in medical imaging (claim): A guide for authors and reviewers}.
\newblock \bibinfo{journal}{Radiology: Artificial Intelligence} \bibinfo{volume}{2}, \bibinfo{pages}{e200029}.
\newblock \URLprefix \url{https://doi.org/10.1148/ryai.2020200029}, \DOIprefix\doi{10.1148/ryai.2020200029}, \href{http://arxiv.org/abs/https://doi.org/10.1148/ryai.2020200029}{\tt arXiv:https://doi.org/10.1148/ryai.2020200029}. \bibinfo{note}{pMID: 33937821}.
\bibitem[{Myronenko(2019)}]{myronenko20193d}
\bibinfo{author}{Myronenko, A.}, \bibinfo{year}{2019}.
\newblock \bibinfo{title}{3d mri brain tumor segmentation using autoencoder regularization}, in: \bibinfo{booktitle}{Brainlesion: Glioma, Multiple Sclerosis, Stroke and Traumatic Brain Injuries: 4th International Workshop, BrainLes 2018, Held in Conjunction with MICCAI 2018, Granada, Spain, September 16, 2018, Revised Selected Papers, Part II 4}, \bibinfo{organization}{Springer}. pp. \bibinfo{pages}{311--320}.
\bibitem[{Naga et~al.(2022)Naga, Mathai, Paul and Summers}]{naga2022universal}
\bibinfo{author}{Naga, V.}, \bibinfo{author}{Mathai, T.S.}, \bibinfo{author}{Paul, A.}, \bibinfo{author}{Summers, R.M.}, \bibinfo{year}{2022}.
\newblock \bibinfo{title}{Universal lesion detection and classification using limited data and weakly-supervised self-training}, in: \bibinfo{booktitle}{Workshop on Medical Image Learning with Limited and Noisy Data}, \bibinfo{organization}{Springer}. pp. \bibinfo{pages}{55--64}.
\bibitem[{Norgeot et~al.(2020)Norgeot, Quer, Beaulieu-Jones, Torkamani, Dias, Gianfrancesco, Arnaout, Kohane, Saria, Topol et~al.}]{norgeot2020minimum}
\bibinfo{author}{Norgeot, B.}, \bibinfo{author}{Quer, G.}, \bibinfo{author}{Beaulieu-Jones, B.K.}, \bibinfo{author}{Torkamani, A.}, \bibinfo{author}{Dias, R.}, \bibinfo{author}{Gianfrancesco, M.}, \bibinfo{author}{Arnaout, R.}, \bibinfo{author}{Kohane, I.S.}, \bibinfo{author}{Saria, S.}, \bibinfo{author}{Topol, E.}, et~al., \bibinfo{year}{2020}.
\newblock \bibinfo{title}{Minimum information about clinical artificial intelligence modeling: the mi-claim checklist}.
\newblock \bibinfo{journal}{Nature medicine} \bibinfo{volume}{26}, \bibinfo{pages}{1320--1324}.
\bibitem[{Oktay et~al.(2018)Oktay, Schlemper, Folgoc, Lee, Heinrich, Misawa, Mori, McDonagh, Hammerla, Kainz et~al.}]{oktay2018attention}
\bibinfo{author}{Oktay, O.}, \bibinfo{author}{Schlemper, J.}, \bibinfo{author}{Folgoc, L.L.}, \bibinfo{author}{Lee, M.}, \bibinfo{author}{Heinrich, M.}, \bibinfo{author}{Misawa, K.}, \bibinfo{author}{Mori, K.}, \bibinfo{author}{McDonagh, S.}, \bibinfo{author}{Hammerla, N.Y.}, \bibinfo{author}{Kainz, B.}, et~al., \bibinfo{year}{2018}.
\newblock \bibinfo{title}{Attention u-net: Learning where to look for the pancreas}.
\newblock \bibinfo{journal}{arXiv preprint arXiv:1804.03999} .
\bibitem[{Orbes-Arteaga et~al.(2019)Orbes-Arteaga, Varsavsky, Sudre, Eaton-Rosen, Haddow, S{\o}rensen, Nielsen, Pai, Ourselin, Modat et~al.}]{orbes2019multi}
\bibinfo{author}{Orbes-Arteaga, M.}, \bibinfo{author}{Varsavsky, T.}, \bibinfo{author}{Sudre, C.H.}, \bibinfo{author}{Eaton-Rosen, Z.}, \bibinfo{author}{Haddow, L.J.}, \bibinfo{author}{S{\o}rensen, L.}, \bibinfo{author}{Nielsen, M.}, \bibinfo{author}{Pai, A.}, \bibinfo{author}{Ourselin, S.}, \bibinfo{author}{Modat, M.}, et~al., \bibinfo{year}{2019}.
\newblock \bibinfo{title}{Multi-domain adaptation in brain mri through paired consistency and adversarial learning}, in: \bibinfo{booktitle}{Domain Adaptation and Representation Transfer and Medical Image Learning with Less Labels and Imperfect Data}. \bibinfo{publisher}{Springer}, pp. \bibinfo{pages}{54--62}.
\bibitem[{Ozdemir et~al.(2018)Ozdemir, Fuernstahl and Goksel}]{ozdemir2018learn}
\bibinfo{author}{Ozdemir, F.}, \bibinfo{author}{Fuernstahl, P.}, \bibinfo{author}{Goksel, O.}, \bibinfo{year}{2018}.
\newblock \bibinfo{title}{Learn the new, keep the old: Extending pretrained models with new anatomy and images}, in: \bibinfo{booktitle}{Medical Image Computing and Computer Assisted Intervention--MICCAI 2018: 21st International Conference, Granada, Spain, September 16-20, 2018, Proceedings, Part IV 11}, \bibinfo{organization}{Springer}. pp. \bibinfo{pages}{361--369}.
\bibitem[{Ozdemir and Goksel(2019)}]{ozdemir2019extending}
\bibinfo{author}{Ozdemir, F.}, \bibinfo{author}{Goksel, O.}, \bibinfo{year}{2019}.
\newblock \bibinfo{title}{Extending pretrained segmentation networks with additional anatomical structures}.
\newblock \bibinfo{journal}{International journal of computer assisted radiology and surgery} \bibinfo{volume}{14}, \bibinfo{pages}{1187--1195}.
\bibitem[{Park et~al.(2022)Park, Woo, Oh, Kweon and Lee}]{park2022per}
\bibinfo{author}{Park, K.}, \bibinfo{author}{Woo, S.}, \bibinfo{author}{Oh, S.W.}, \bibinfo{author}{Kweon, I.S.}, \bibinfo{author}{Lee, J.Y.}, \bibinfo{year}{2022}.
\newblock \bibinfo{title}{Per-clip video object segmentation}, in: \bibinfo{booktitle}{Proceedings of the IEEE/CVF Conference on Computer Vision and Pattern Recognition}, pp. \bibinfo{pages}{1352--1361}.
\bibitem[{Qin et~al.(2022)Qin, Yi, Lao and Li}]{qin2022medical}
\bibinfo{author}{Qin, Z.}, \bibinfo{author}{Yi, H.H.}, \bibinfo{author}{Lao, Q.}, \bibinfo{author}{Li, K.}, \bibinfo{year}{2022}.
\newblock \bibinfo{title}{Medical image understanding with pretrained vision language models: A comprehensive study}, in: \bibinfo{booktitle}{The Eleventh International Conference on Learning Representations}.
\bibitem[{Qu et~al.(2023)Qu, Zhang, Qiao, Liu, Tang, Yuille and Zhou}]{qu2023annotating}
\bibinfo{author}{Qu, C.}, \bibinfo{author}{Zhang, T.}, \bibinfo{author}{Qiao, H.}, \bibinfo{author}{Liu, J.}, \bibinfo{author}{Tang, Y.}, \bibinfo{author}{Yuille, A.}, \bibinfo{author}{Zhou, Z.}, \bibinfo{year}{2023}.
\newblock \bibinfo{title}{Abdomenatlas-8k: Annotating 8,000 abdominal ct volumes for multi-organ segmentation in three weeks}.
\newblock \bibinfo{journal}{Conference on Neural Information Processing Systems} \URLprefix \url{https://github.com/MrGiovanni/AbdomenAtlas}.
\bibitem[{Radford et~al.(2021)Radford, Kim, Hallacy, Ramesh, Goh, Agarwal, Sastry, Askell, Mishkin, Clark et~al.}]{radford2021learning}
\bibinfo{author}{Radford, A.}, \bibinfo{author}{Kim, J.W.}, \bibinfo{author}{Hallacy, C.}, \bibinfo{author}{Ramesh, A.}, \bibinfo{author}{Goh, G.}, \bibinfo{author}{Agarwal, S.}, \bibinfo{author}{Sastry, G.}, \bibinfo{author}{Askell, A.}, \bibinfo{author}{Mishkin, P.}, \bibinfo{author}{Clark, J.}, et~al., \bibinfo{year}{2021}.
\newblock \bibinfo{title}{Learning transferable visual models from natural language supervision}, in: \bibinfo{booktitle}{International Conference on Machine Learning}, \bibinfo{organization}{PMLR}. pp. \bibinfo{pages}{8748--8763}.
\bibitem[{Rao et~al.(2022)Rao, Zhao, Chen, Tang, Zhu, Huang, Zhou and Lu}]{rao2022denseclip}
\bibinfo{author}{Rao, Y.}, \bibinfo{author}{Zhao, W.}, \bibinfo{author}{Chen, G.}, \bibinfo{author}{Tang, Y.}, \bibinfo{author}{Zhu, Z.}, \bibinfo{author}{Huang, G.}, \bibinfo{author}{Zhou, J.}, \bibinfo{author}{Lu, J.}, \bibinfo{year}{2022}.
\newblock \bibinfo{title}{Denseclip: Language-guided dense prediction with context-aware prompting}, in: \bibinfo{booktitle}{Proceedings of the IEEE/CVF Conference on Computer Vision and Pattern Recognition}, pp. \bibinfo{pages}{18082--18091}.
\bibitem[{Rister et~al.(2020)Rister, Yi, Shivakumar, Nobashi and Rubin}]{rister2020ct}
\bibinfo{author}{Rister, B.}, \bibinfo{author}{Yi, D.}, \bibinfo{author}{Shivakumar, K.}, \bibinfo{author}{Nobashi, T.}, \bibinfo{author}{Rubin, D.L.}, \bibinfo{year}{2020}.
\newblock \bibinfo{title}{Ct-org, a new dataset for multiple organ segmentation in computed tomography}.
\newblock \bibinfo{journal}{Scientific Data} \bibinfo{volume}{7}, \bibinfo{pages}{1--9}.
\bibitem[{Ronneberger et~al.(2015)Ronneberger, Fischer and Brox}]{ronneberger2015u}
\bibinfo{author}{Ronneberger, O.}, \bibinfo{author}{Fischer, P.}, \bibinfo{author}{Brox, T.}, \bibinfo{year}{2015}.
\newblock \bibinfo{title}{U-net: Convolutional networks for biomedical image segmentation}, in: \bibinfo{booktitle}{International Conference on Medical Image Computing and Computer-Assisted Intervention}, \bibinfo{organization}{Springer}. pp. \bibinfo{pages}{234--241}.
\bibitem[{Roth et~al.(2015)Roth, Lu, Farag, Shin, Liu, Turkbey and Summers}]{roth2015deeporgan}
\bibinfo{author}{Roth, H.R.}, \bibinfo{author}{Lu, L.}, \bibinfo{author}{Farag, A.}, \bibinfo{author}{Shin, H.C.}, \bibinfo{author}{Liu, J.}, \bibinfo{author}{Turkbey, E.B.}, \bibinfo{author}{Summers, R.M.}, \bibinfo{year}{2015}.
\newblock \bibinfo{title}{Deeporgan: Multi-level deep convolutional networks for automated pancreas segmentation}, in: \bibinfo{booktitle}{International conference on medical image computing and computer-assisted intervention}, \bibinfo{organization}{Springer}. pp. \bibinfo{pages}{556--564}.
\bibitem[{Schoppe et~al.(2020)Schoppe, Pan, Coronel, Mai, Rong, Todorov, M{\"u}skes, Navarro, Li, Ert{\"u}rk et~al.}]{schoppe2020deep}
\bibinfo{author}{Schoppe, O.}, \bibinfo{author}{Pan, C.}, \bibinfo{author}{Coronel, J.}, \bibinfo{author}{Mai, H.}, \bibinfo{author}{Rong, Z.}, \bibinfo{author}{Todorov, M.I.}, \bibinfo{author}{M{\"u}skes, A.}, \bibinfo{author}{Navarro, F.}, \bibinfo{author}{Li, H.}, \bibinfo{author}{Ert{\"u}rk, A.}, et~al., \bibinfo{year}{2020}.
\newblock \bibinfo{title}{Deep learning-enabled multi-organ segmentation in whole-body mouse scans}.
\newblock \bibinfo{journal}{Nature communications} \bibinfo{volume}{11}, \bibinfo{pages}{5626}.
\bibitem[{Shen et~al.(2021)Shen, Shamout, Oliver, Witowski, Kannan, Park, Wu, Huddleston, Wolfson, Millet et~al.}]{shen2021artificial}
\bibinfo{author}{Shen, Y.}, \bibinfo{author}{Shamout, F.E.}, \bibinfo{author}{Oliver, J.R.}, \bibinfo{author}{Witowski, J.}, \bibinfo{author}{Kannan, K.}, \bibinfo{author}{Park, J.}, \bibinfo{author}{Wu, N.}, \bibinfo{author}{Huddleston, C.}, \bibinfo{author}{Wolfson, S.}, \bibinfo{author}{Millet, A.}, et~al., \bibinfo{year}{2021}.
\newblock \bibinfo{title}{Artificial intelligence system reduces false-positive findings in the interpretation of breast ultrasound exams}.
\newblock \bibinfo{journal}{Nature communications} \bibinfo{volume}{12}, \bibinfo{pages}{1--13}.
\bibitem[{Shi et~al.(2021)Shi, Xiao, Chen and Zhou}]{shi2021marginal}
\bibinfo{author}{Shi, G.}, \bibinfo{author}{Xiao, L.}, \bibinfo{author}{Chen, Y.}, \bibinfo{author}{Zhou, S.K.}, \bibinfo{year}{2021}.
\newblock \bibinfo{title}{Marginal loss and exclusion loss for partially supervised multi-organ segmentation}.
\newblock \bibinfo{journal}{Medical Image Analysis} \bibinfo{volume}{70}, \bibinfo{pages}{101979}.
\bibitem[{Siddiquee and Myronenko(2021)}]{siddiquee2021redundancy}
\bibinfo{author}{Siddiquee, M.M.R.}, \bibinfo{author}{Myronenko, A.}, \bibinfo{year}{2021}.
\newblock \bibinfo{title}{Redundancy reduction in semantic segmentation of 3d brain tumor mris}.
\newblock \bibinfo{journal}{arXiv preprint arXiv:2111.00742} .
\bibitem[{Silva-Rodr{\'\i}guez et~al.(2023)Silva-Rodr{\'\i}guez, Dolz and Ayed}]{silva2023towards}
\bibinfo{author}{Silva-Rodr{\'\i}guez, J.}, \bibinfo{author}{Dolz, J.}, \bibinfo{author}{Ayed, I.B.}, \bibinfo{year}{2023}.
\newblock \bibinfo{title}{Towards foundation models and few-shot parameter-efficient fine-tuning for volumetric organ segmentation}, in: \bibinfo{booktitle}{International Conference on Medical Image Computing and Computer-Assisted Intervention}, \bibinfo{organization}{Springer}. pp. \bibinfo{pages}{213--224}.
\bibitem[{Simpson et~al.(2019)Simpson, Antonelli, Bakas, Bilello, Farahani, Van~Ginneken, Kopp-Schneider, Landman, Litjens, Menze et~al.}]{simpson2019large}
\bibinfo{author}{Simpson, A.L.}, \bibinfo{author}{Antonelli, M.}, \bibinfo{author}{Bakas, S.}, \bibinfo{author}{Bilello, M.}, \bibinfo{author}{Farahani, K.}, \bibinfo{author}{Van~Ginneken, B.}, \bibinfo{author}{Kopp-Schneider, A.}, \bibinfo{author}{Landman, B.A.}, \bibinfo{author}{Litjens, G.}, \bibinfo{author}{Menze, B.}, et~al., \bibinfo{year}{2019}.
\newblock \bibinfo{title}{A large annotated medical image dataset for the development and evaluation of segmentation algorithms}.
\newblock \bibinfo{journal}{arXiv preprint arXiv:1902.09063} .
\bibitem[{Soler et~al.(2010)Soler, Hostettler, Agnus, Charnoz, Fasquel, Moreau, Osswald, Bouhadjar and Marescaux}]{soler20103d}
\bibinfo{author}{Soler, L.}, \bibinfo{author}{Hostettler, A.}, \bibinfo{author}{Agnus, V.}, \bibinfo{author}{Charnoz, A.}, \bibinfo{author}{Fasquel, J.}, \bibinfo{author}{Moreau, J.}, \bibinfo{author}{Osswald, A.}, \bibinfo{author}{Bouhadjar, M.}, \bibinfo{author}{Marescaux, J.}, \bibinfo{year}{2010}.
\newblock \bibinfo{title}{3d image reconstruction for comparison of algorithm database: A patient specific anatomical and medical image database}.
\newblock \bibinfo{journal}{IRCAD, Strasbourg, France, Tech. Rep} .
\bibitem[{Tang et~al.(2024)Tang, Liu, Zhou, Yu and Huo}]{tang2024efficient}
\bibinfo{author}{Tang, Y.}, \bibinfo{author}{Liu, J.}, \bibinfo{author}{Zhou, Z.}, \bibinfo{author}{Yu, X.}, \bibinfo{author}{Huo, Y.}, \bibinfo{year}{2024}.
\newblock \bibinfo{title}{Efficient 3d representation learning for medical image analysis}.
\newblock \bibinfo{journal}{World Scientific Annual Review of Artificial Intelligence} .
\bibitem[{Tang et~al.(2022)Tang, Yang, Li, Roth, Landman, Xu, Nath and Hatamizadeh}]{tang2022self}
\bibinfo{author}{Tang, Y.}, \bibinfo{author}{Yang, D.}, \bibinfo{author}{Li, W.}, \bibinfo{author}{Roth, H.R.}, \bibinfo{author}{Landman, B.}, \bibinfo{author}{Xu, D.}, \bibinfo{author}{Nath, V.}, \bibinfo{author}{Hatamizadeh, A.}, \bibinfo{year}{2022}.
\newblock \bibinfo{title}{Self-supervised pre-training of swin transformers for 3d medical image analysis}, in: \bibinfo{booktitle}{Proceedings of the IEEE/CVF Conference on Computer Vision and Pattern Recognition}, pp. \bibinfo{pages}{20730--20740}.
\bibitem[{Ulrich et~al.(2023)Ulrich, Isensee, Wald, Zenk, Baumgartner and Maier-Hein}]{ulrich2023multitalent}
\bibinfo{author}{Ulrich, C.}, \bibinfo{author}{Isensee, F.}, \bibinfo{author}{Wald, T.}, \bibinfo{author}{Zenk, M.}, \bibinfo{author}{Baumgartner, M.}, \bibinfo{author}{Maier-Hein, K.H.}, \bibinfo{year}{2023}.
\newblock \bibinfo{title}{Multitalent: A multi-dataset approach to medical image segmentation}, in: \bibinfo{booktitle}{International Conference on Medical Image Computing and Computer-Assisted Intervention}, \bibinfo{organization}{Springer}. pp. \bibinfo{pages}{648--658}.
\bibitem[{Valindria et~al.(2018)Valindria, Pawlowski, Rajchl, Lavdas, Aboagye, Rockall, Rueckert and Glocker}]{valindria2018multi}
\bibinfo{author}{Valindria, V.V.}, \bibinfo{author}{Pawlowski, N.}, \bibinfo{author}{Rajchl, M.}, \bibinfo{author}{Lavdas, I.}, \bibinfo{author}{Aboagye, E.O.}, \bibinfo{author}{Rockall, A.G.}, \bibinfo{author}{Rueckert, D.}, \bibinfo{author}{Glocker, B.}, \bibinfo{year}{2018}.
\newblock \bibinfo{title}{Multi-modal learning from unpaired images: Application to multi-organ segmentation in ct and mri}, in: \bibinfo{booktitle}{2018 IEEE winter conference on applications of computer vision (WACV)}, \bibinfo{organization}{IEEE}. pp. \bibinfo{pages}{547--556}.
\bibitem[{Wang et~al.(2021)Wang, Chen, Ding, Yu, Zha and Li}]{wang2021transbts}
\bibinfo{author}{Wang, W.}, \bibinfo{author}{Chen, C.}, \bibinfo{author}{Ding, M.}, \bibinfo{author}{Yu, H.}, \bibinfo{author}{Zha, S.}, \bibinfo{author}{Li, J.}, \bibinfo{year}{2021}.
\newblock \bibinfo{title}{Transbts: Multimodal brain tumor segmentation using transformer}, in: \bibinfo{booktitle}{International Conference on Medical Image Computing and Computer-Assisted Intervention}, \bibinfo{organization}{Springer}. pp. \bibinfo{pages}{109--119}.
\bibitem[{Wang et~al.(2022a)Wang, Lu, Li, Tao, Guo, Gong and Liu}]{wang2022cris}
\bibinfo{author}{Wang, Z.}, \bibinfo{author}{Lu, Y.}, \bibinfo{author}{Li, Q.}, \bibinfo{author}{Tao, X.}, \bibinfo{author}{Guo, Y.}, \bibinfo{author}{Gong, M.}, \bibinfo{author}{Liu, T.}, \bibinfo{year}{2022}a.
\newblock \bibinfo{title}{Cris: Clip-driven referring image segmentation}, in: \bibinfo{booktitle}{Proceedings of the IEEE/CVF Conference on Computer Vision and Pattern Recognition}, pp. \bibinfo{pages}{11686--11695}.
\bibitem[{Wang et~al.(2022b)Wang, Wu, Agarwal and Sun}]{wang2022medclip}
\bibinfo{author}{Wang, Z.}, \bibinfo{author}{Wu, Z.}, \bibinfo{author}{Agarwal, D.}, \bibinfo{author}{Sun, J.}, \bibinfo{year}{2022}b.
\newblock \bibinfo{title}{Medclip: Contrastive learning from unpaired medical images and text}, in: \bibinfo{booktitle}{Proceedings of the 2022 Conference on Empirical Methods in Natural Language Processing}, pp. \bibinfo{pages}{3876--3887}.
\bibitem[{Wasserthal et~al.(2023)Wasserthal, Breit, Meyer, Pradella, Hinck, Sauter, Heye, Boll, Cyriac, Yang et~al.}]{wasserthal2023totalsegmentator}
\bibinfo{author}{Wasserthal, J.}, \bibinfo{author}{Breit, H.C.}, \bibinfo{author}{Meyer, M.T.}, \bibinfo{author}{Pradella, M.}, \bibinfo{author}{Hinck, D.}, \bibinfo{author}{Sauter, A.W.}, \bibinfo{author}{Heye, T.}, \bibinfo{author}{Boll, D.T.}, \bibinfo{author}{Cyriac, J.}, \bibinfo{author}{Yang, S.}, et~al., \bibinfo{year}{2023}.
\newblock \bibinfo{title}{Totalsegmentator: Robust segmentation of 104 anatomic structures in ct images}.
\newblock \bibinfo{journal}{Radiology: Artificial Intelligence} \bibinfo{volume}{5}.
\bibitem[{Wasserthal et~al.(2022)Wasserthal, Meyer, Breit, Cyriac, Yang and Segeroth}]{wasserthal2022totalsegmentator}
\bibinfo{author}{Wasserthal, J.}, \bibinfo{author}{Meyer, M.}, \bibinfo{author}{Breit, H.C.}, \bibinfo{author}{Cyriac, J.}, \bibinfo{author}{Yang, S.}, \bibinfo{author}{Segeroth, M.}, \bibinfo{year}{2022}.
\newblock \bibinfo{title}{Totalsegmentator: robust segmentation of 104 anatomical structures in ct images}.
\newblock \bibinfo{journal}{arXiv preprint arXiv:2208.05868} .
\bibitem[{Wu et~al.(2022)Wu, Pang and Sowmya}]{wu2022tgnet}
\bibinfo{author}{Wu, H.}, \bibinfo{author}{Pang, S.}, \bibinfo{author}{Sowmya, A.}, \bibinfo{year}{2022}.
\newblock \bibinfo{title}{Tgnet: A task-guided network architecture for multi-organ and tumour segmentation from partially labelled datasets}, in: \bibinfo{booktitle}{2022 IEEE 19th International Symposium on Biomedical Imaging (ISBI)}, \bibinfo{organization}{IEEE}. pp. \bibinfo{pages}{1--5}.
\bibitem[{Xia et~al.(2022)Xia, Yu, Chu, Kawamoto, Park, Liu, Chen, Zhu, Li, Zhou et~al.}]{xia2022felix}
\bibinfo{author}{Xia, Y.}, \bibinfo{author}{Yu, Q.}, \bibinfo{author}{Chu, L.}, \bibinfo{author}{Kawamoto, S.}, \bibinfo{author}{Park, S.}, \bibinfo{author}{Liu, F.}, \bibinfo{author}{Chen, J.}, \bibinfo{author}{Zhu, Z.}, \bibinfo{author}{Li, B.}, \bibinfo{author}{Zhou, Z.}, et~al., \bibinfo{year}{2022}.
\newblock \bibinfo{title}{The felix project: Deep networks to detect pancreatic neoplasms}.
\newblock \bibinfo{journal}{medRxiv} .
\bibitem[{Xie et~al.(2022a)Xie, Hou, Ye and Shen}]{xie2022clims}
\bibinfo{author}{Xie, J.}, \bibinfo{author}{Hou, X.}, \bibinfo{author}{Ye, K.}, \bibinfo{author}{Shen, L.}, \bibinfo{year}{2022}a.
\newblock \bibinfo{title}{Clims: Cross language image matching for weakly supervised semantic segmentation}, in: \bibinfo{booktitle}{Proceedings of the IEEE/CVF Conference on Computer Vision and Pattern Recognition}, pp. \bibinfo{pages}{4483--4492}.
\bibitem[{Xie et~al.(2021)Xie, Zhang, Shen and Xia}]{xie2021cotr}
\bibinfo{author}{Xie, Y.}, \bibinfo{author}{Zhang, J.}, \bibinfo{author}{Shen, C.}, \bibinfo{author}{Xia, Y.}, \bibinfo{year}{2021}.
\newblock \bibinfo{title}{Cotr: Efficiently bridging cnn and transformer for 3d medical image segmentation}, in: \bibinfo{booktitle}{Medical Image Computing and Computer Assisted Intervention--MICCAI 2021: 24th International Conference, Strasbourg, France, September 27--October 1, 2021, Proceedings, Part III 24}, \bibinfo{organization}{Springer}. pp. \bibinfo{pages}{171--180}.
\bibitem[{Xie et~al.(2023)Xie, Zhang, Xia and Shen}]{xie2023learning}
\bibinfo{author}{Xie, Y.}, \bibinfo{author}{Zhang, J.}, \bibinfo{author}{Xia, Y.}, \bibinfo{author}{Shen, C.}, \bibinfo{year}{2023}.
\newblock \bibinfo{title}{Learning from partially labeled data for multi-organ and tumor segmentation}.
\newblock \bibinfo{journal}{IEEE Transactions on Pattern Analysis and Machine Intelligence} .
\bibitem[{Xie et~al.(2022b)Xie, Zhang, Xia and Wu}]{xie2022unimiss}
\bibinfo{author}{Xie, Y.}, \bibinfo{author}{Zhang, J.}, \bibinfo{author}{Xia, Y.}, \bibinfo{author}{Wu, Q.}, \bibinfo{year}{2022}b.
\newblock \bibinfo{title}{Unimiss: Universal medical self-supervised learning via breaking dimensionality barrier}, in: \bibinfo{booktitle}{European Conference on Computer Vision}, \bibinfo{organization}{Springer}. pp. \bibinfo{pages}{558--575}.
\bibitem[{Yan and Pei(2022)}]{yan2022clinical}
\bibinfo{author}{Yan, B.}, \bibinfo{author}{Pei, M.}, \bibinfo{year}{2022}.
\newblock \bibinfo{title}{Clinical-bert: Vision-language pre-training for radiograph diagnosis and reports generation}, in: \bibinfo{booktitle}{Proceedings of the AAAI Conference on Artificial Intelligence}, pp. \bibinfo{pages}{2982--2990}.
\bibitem[{Yan et~al.(2020a)Yan, Cai, Harrison, Jin, Xiao and Lu}]{yan2020universal}
\bibinfo{author}{Yan, K.}, \bibinfo{author}{Cai, J.}, \bibinfo{author}{Harrison, A.P.}, \bibinfo{author}{Jin, D.}, \bibinfo{author}{Xiao, J.}, \bibinfo{author}{Lu, L.}, \bibinfo{year}{2020}a.
\newblock \bibinfo{title}{Universal lesion detection by learning from multiple heterogeneously labeled datasets}.
\newblock \bibinfo{journal}{arXiv preprint arXiv:2005.13753} .
\bibitem[{Yan et~al.(2019)Yan, Tang, Peng, Sandfort, Bagheri, Lu and Summers}]{yan2019mulan}
\bibinfo{author}{Yan, K.}, \bibinfo{author}{Tang, Y.}, \bibinfo{author}{Peng, Y.}, \bibinfo{author}{Sandfort, V.}, \bibinfo{author}{Bagheri, M.}, \bibinfo{author}{Lu, Z.}, \bibinfo{author}{Summers, R.M.}, \bibinfo{year}{2019}.
\newblock \bibinfo{title}{Mulan: multitask universal lesion analysis network for joint lesion detection, tagging, and segmentation}, in: \bibinfo{booktitle}{International Conference on Medical Image Computing and Computer-Assisted Intervention}, \bibinfo{organization}{Springer}. pp. \bibinfo{pages}{194--202}.
\bibitem[{Yan et~al.(2020b)Yan, Huang, Xia, Gu, Yan, Wang and Tao}]{yan2020mri}
\bibinfo{author}{Yan, W.}, \bibinfo{author}{Huang, L.}, \bibinfo{author}{Xia, L.}, \bibinfo{author}{Gu, S.}, \bibinfo{author}{Yan, F.}, \bibinfo{author}{Wang, Y.}, \bibinfo{author}{Tao, Q.}, \bibinfo{year}{2020}b.
\newblock \bibinfo{title}{Mri manufacturer shift and adaptation: increasing the generalizability of deep learning segmentation for mr images acquired with different scanners}.
\newblock \bibinfo{journal}{Radiology: Artificial Intelligence} \bibinfo{volume}{2}, \bibinfo{pages}{e190195}.
\bibitem[{Yasunaga et~al.(2022)Yasunaga, Leskovec and Liang}]{yasunaga2022linkbert}
\bibinfo{author}{Yasunaga, M.}, \bibinfo{author}{Leskovec, J.}, \bibinfo{author}{Liang, P.}, \bibinfo{year}{2022}.
\newblock \bibinfo{title}{Linkbert: Pretraining language models with document links}.
\newblock \bibinfo{journal}{arXiv preprint arXiv:2203.15827} .
\bibitem[{Ye et~al.(2023a)Ye, Xie, Zhang, Chen, Wu and Xia}]{ye2023continual}
\bibinfo{author}{Ye, Y.}, \bibinfo{author}{Xie, Y.}, \bibinfo{author}{Zhang, J.}, \bibinfo{author}{Chen, Z.}, \bibinfo{author}{Wu, Q.}, \bibinfo{author}{Xia, Y.}, \bibinfo{year}{2023}a.
\newblock \bibinfo{title}{Continual self-supervised learning: Towards universal multi-modal medical data representation learning}.
\newblock \bibinfo{journal}{arXiv preprint arXiv:2311.17597} .
\bibitem[{Ye et~al.(2023b)Ye, Xie, Zhang, Chen and Xia}]{ye2023uniseg}
\bibinfo{author}{Ye, Y.}, \bibinfo{author}{Xie, Y.}, \bibinfo{author}{Zhang, J.}, \bibinfo{author}{Chen, Z.}, \bibinfo{author}{Xia, Y.}, \bibinfo{year}{2023}b.
\newblock \bibinfo{title}{Uniseg: A prompt-driven universal segmentation model as well as a strong representation learner}.
\newblock \bibinfo{journal}{arXiv preprint arXiv:2304.03493} .
\bibitem[{Yu et~al.(2020)Yu, Yang, Roth, Bai, Zhang, Yuille and Xu}]{yu2020c2fnas}
\bibinfo{author}{Yu, Q.}, \bibinfo{author}{Yang, D.}, \bibinfo{author}{Roth, H.}, \bibinfo{author}{Bai, Y.}, \bibinfo{author}{Zhang, Y.}, \bibinfo{author}{Yuille, A.L.}, \bibinfo{author}{Xu, D.}, \bibinfo{year}{2020}.
\newblock \bibinfo{title}{C2fnas: Coarse-to-fine neural architecture search for 3d medical image segmentation}, in: \bibinfo{booktitle}{Proceedings of the IEEE/CVF Conference on Computer Vision and Pattern Recognition}, pp. \bibinfo{pages}{4126--4135}.
\bibitem[{Yu et~al.(2022)Yu, Yang, Zhou, Cai, Gao, Lee, Li, Bao, Xu, Lasko et~al.}]{yu2022unest}
\bibinfo{author}{Yu, X.}, \bibinfo{author}{Yang, Q.}, \bibinfo{author}{Zhou, Y.}, \bibinfo{author}{Cai, L.Y.}, \bibinfo{author}{Gao, R.}, \bibinfo{author}{Lee, H.H.}, \bibinfo{author}{Li, T.}, \bibinfo{author}{Bao, S.}, \bibinfo{author}{Xu, Z.}, \bibinfo{author}{Lasko, T.A.}, et~al., \bibinfo{year}{2022}.
\newblock \bibinfo{title}{Unest: Local spatial representation learning with hierarchical transformer for efficient medical segmentation}.
\newblock \bibinfo{journal}{arXiv preprint arXiv:2209.14378} .
\bibitem[{Zeng et~al.(2023)Zeng, Xie, Lu, Lu, Wu and Xia}]{zeng2023segment}
\bibinfo{author}{Zeng, Q.}, \bibinfo{author}{Xie, Y.}, \bibinfo{author}{Lu, Z.}, \bibinfo{author}{Lu, M.}, \bibinfo{author}{Wu, Y.}, \bibinfo{author}{Xia, Y.}, \bibinfo{year}{2023}.
\newblock \bibinfo{title}{Segment together: A versatile paradigm for semi-supervised medical image segmentation}.
\newblock \bibinfo{journal}{arXiv preprint arXiv:2311.11686} .
\bibitem[{Zhang et~al.(2023a)Zhang, Chen, Wang, Hu, Liu and Li}]{zhang2023spatially}
\bibinfo{author}{Zhang, H.}, \bibinfo{author}{Chen, X.}, \bibinfo{author}{Wang, R.}, \bibinfo{author}{Hu, R.}, \bibinfo{author}{Liu, D.}, \bibinfo{author}{Li, G.}, \bibinfo{year}{2023}a.
\newblock \bibinfo{title}{Spatially covariant image registration with text prompts}.
\newblock \bibinfo{journal}{arXiv preprint arXiv:2311.15607} .
\bibitem[{Zhang et~al.(2021)Zhang, Xie, Xia and Shen}]{zhang2021dodnet}
\bibinfo{author}{Zhang, J.}, \bibinfo{author}{Xie, Y.}, \bibinfo{author}{Xia, Y.}, \bibinfo{author}{Shen, C.}, \bibinfo{year}{2021}.
\newblock \bibinfo{title}{Dodnet: Learning to segment multi-organ and tumors from multiple partially labeled datasets}, in: \bibinfo{booktitle}{Proceedings of the IEEE/CVF Conference on Computer Vision and Pattern Recognition}, pp. \bibinfo{pages}{1195--1204}.
\bibitem[{Zhang et~al.(2024)Zhang, Chen, Qu, Yuille and Zhou}]{zhang2024leveraging}
\bibinfo{author}{Zhang, T.}, \bibinfo{author}{Chen, X.}, \bibinfo{author}{Qu, C.}, \bibinfo{author}{Yuille, A.}, \bibinfo{author}{Zhou, Z.}, \bibinfo{year}{2024}.
\newblock \bibinfo{title}{Leveraging ai predicted and expert revised annotations in interactive segmentation: Continual tuning or full training?}, in: \bibinfo{booktitle}{IEEE International Symposium on Biomedical Imaging}, \bibinfo{organization}{IEEE}.
\newblock \URLprefix \url{https://github.com/MrGiovanni/ContinualLearning}.
\bibitem[{Zhang et~al.(2022)Zhang, Zhang, Wang, Yang, Huang, Yang, Wang and Han}]{zhang2022merging}
\bibinfo{author}{Zhang, W.}, \bibinfo{author}{Zhang, J.}, \bibinfo{author}{Wang, X.}, \bibinfo{author}{Yang, S.}, \bibinfo{author}{Huang, J.}, \bibinfo{author}{Yang, W.}, \bibinfo{author}{Wang, W.}, \bibinfo{author}{Han, X.}, \bibinfo{year}{2022}.
\newblock \bibinfo{title}{Merging nucleus datasets by correlation-based cross-training}.
\newblock \bibinfo{journal}{Medical Image Analysis} , \bibinfo{pages}{102705}.
\bibitem[{Zhang et~al.(2023b)Zhang, Li, Chen, Yuille, Liu and Zhou}]{zhang2023continual}
\bibinfo{author}{Zhang, Y.}, \bibinfo{author}{Li, X.}, \bibinfo{author}{Chen, H.}, \bibinfo{author}{Yuille, A.L.}, \bibinfo{author}{Liu, Y.}, \bibinfo{author}{Zhou, Z.}, \bibinfo{year}{2023}b.
\newblock \bibinfo{title}{Continual learning for abdominal multi-organ and tumor segmentation}, in: \bibinfo{booktitle}{International Conference on Medical Image Computing and Computer-Assisted Intervention}, \bibinfo{organization}{Springer}. pp. \bibinfo{pages}{35--45}.
\newblock \URLprefix \url{https://github.com/MrGiovanni/ContinualLearning}.
\bibitem[{Zhou et~al.(2021a)Zhou, Guo, Zhang, Yu, Wang and Yu}]{zhou2021nnformer}
\bibinfo{author}{Zhou, H.Y.}, \bibinfo{author}{Guo, J.}, \bibinfo{author}{Zhang, Y.}, \bibinfo{author}{Yu, L.}, \bibinfo{author}{Wang, L.}, \bibinfo{author}{Yu, Y.}, \bibinfo{year}{2021}a.
\newblock \bibinfo{title}{nnformer: Interleaved transformer for volumetric segmentation}.
\newblock \bibinfo{journal}{arXiv preprint arXiv:2109.03201} .
\bibitem[{Zhou et~al.(2019a)Zhou, Li, Bai, Wang, Chen, Han, Fishman and Yuille}]{zhou2019prior}
\bibinfo{author}{Zhou, Y.}, \bibinfo{author}{Li, Z.}, \bibinfo{author}{Bai, S.}, \bibinfo{author}{Wang, C.}, \bibinfo{author}{Chen, X.}, \bibinfo{author}{Han, M.}, \bibinfo{author}{Fishman, E.}, \bibinfo{author}{Yuille, A.L.}, \bibinfo{year}{2019}a.
\newblock \bibinfo{title}{Prior-aware neural network for partially-supervised multi-organ segmentation}, in: \bibinfo{booktitle}{Proceedings of the IEEE/CVF International Conference on Computer Vision}, pp. \bibinfo{pages}{10672--10681}.
\bibitem[{Zhou(2021)}]{zhou2021towards}
\bibinfo{author}{Zhou, Z.}, \bibinfo{year}{2021}.
\newblock \bibinfo{title}{Towards Annotation-Efficient Deep Learning for Computer-Aided Diagnosis}.
\newblock Ph.D. thesis. Arizona State University.
\bibitem[{Zhou et~al.(2022)Zhou, Gotway and Liang}]{zhou2022interpreting}
\bibinfo{author}{Zhou, Z.}, \bibinfo{author}{Gotway, M.B.}, \bibinfo{author}{Liang, J.}, \bibinfo{year}{2022}.
\newblock \bibinfo{title}{Interpreting medical images}, in: \bibinfo{booktitle}{Intelligent Systems in Medicine and Health}. \bibinfo{publisher}{Springer}, pp. \bibinfo{pages}{343--371}.
\bibitem[{Zhou et~al.(2019b)Zhou, Siddiquee, Tajbakhsh and Liang}]{zhou2019unet++}
\bibinfo{author}{Zhou, Z.}, \bibinfo{author}{Siddiquee, M.M.R.}, \bibinfo{author}{Tajbakhsh, N.}, \bibinfo{author}{Liang, J.}, \bibinfo{year}{2019}b.
\newblock \bibinfo{title}{Unet++: Redesigning skip connections to exploit multiscale features in image segmentation}.
\newblock \bibinfo{journal}{IEEE Transactions on Medical Imaging} \bibinfo{volume}{39}, \bibinfo{pages}{1856--1867}.
\newblock \URLprefix \url{https://github.com/MrGiovanni/UNetPlusPlus}.
\bibitem[{Zhou et~al.(2021b)Zhou, Sodha, Pang, Gotway and Liang}]{zhou2021models}
\bibinfo{author}{Zhou, Z.}, \bibinfo{author}{Sodha, V.}, \bibinfo{author}{Pang, J.}, \bibinfo{author}{Gotway, M.B.}, \bibinfo{author}{Liang, J.}, \bibinfo{year}{2021}b.
\newblock \bibinfo{title}{Models genesis}.
\newblock \bibinfo{journal}{Medical Image Analysis} \bibinfo{volume}{67}, \bibinfo{pages}{101840}.
\newblock \URLprefix \url{https://github.com/MrGiovanni/ModelsGenesis}.
\bibitem[{Zhou et~al.(2019c)Zhou, Sodha, Siddiquee, Feng, Tajbakhsh, Gotway and Liang}]{zhou2019models}
\bibinfo{author}{Zhou, Z.}, \bibinfo{author}{Sodha, V.}, \bibinfo{author}{Siddiquee, M.M.R.}, \bibinfo{author}{Feng, R.}, \bibinfo{author}{Tajbakhsh, N.}, \bibinfo{author}{Gotway, M.B.}, \bibinfo{author}{Liang, J.}, \bibinfo{year}{2019}c.
\newblock \bibinfo{title}{Models genesis: Generic autodidactic models for 3d medical image analysis}, in: \bibinfo{booktitle}{International Conference on Medical Image Computing and Computer-Assisted Intervention}, \bibinfo{organization}{Springer}. pp. \bibinfo{pages}{384--393}.
\newblock \URLprefix \url{https://github.com/MrGiovanni/ModelsGenesis}.
\bibitem[{Zlocha et~al.(2019)Zlocha, Glocker and Passerat-Palmbach}]{zlocha2019universal}
\bibinfo{author}{Zlocha, M.}, \bibinfo{author}{Glocker, B.}, \bibinfo{author}{Passerat-Palmbach, J.}, \bibinfo{year}{2019}.
\newblock \bibinfo{title}{Universal lesion detector: Deep learning for analysing medical scans} .

\end{thebibliography}
	
\end{document}